\documentclass[pre,twocolumn,superscriptaddress,nofootinbib,showpacs,showkeys]{revtex4}

\usepackage{graphicx}
\usepackage{dcolumn}

\usepackage{eucal}
\usepackage[dvips]{epsfig}
\usepackage{amssymb}

\usepackage{amsmath,amsthm}

\usepackage[]{color}
\usepackage[pdfencoding=auto,hyperindex]{hyperref}
\hypersetup{
breaklinks = {true},
colorlinks = {false},
linkcolor={black},
pdfpagemode = {None}, % prevents the automatic announcement of bookmarks
pdfborder = {0 0 1},
pdftitle = {Thermodynamic laws in isolated systems},
pdfsubject = {discussion of entropy, temperature, thermal equilibrium, and the Laws of Thermodynamics},
pdfauthor = {Stefan Hilbert, Jorn Dunkel, Peter Hanggi},
pdfkeywords = {entropy, temperature, thermal equilibrium, thermodynamics, statistical physics, microcanonical ensemble, isolated systems}
}

\allowdisplaybreaks[4]

\newcommand{\highlightchange}[1]{\textcolor{black}{#1}}

\newcommand{\vect}[1]{\boldsymbol{#1}}

\newcommand{\R}{\mathbb{R}}

\newcommand{\parder}[3][]{\frac{\partial^{#1} {#2}}{\partial {#3}^{#1}}}

\newcommand{\totder}[3][]{\frac{\mathrm{d}^{#1} {#2}}{\mathrm{d} {#3}^{#1}}}

\newcommand{\diff}[2][]{\mathrm{d}^{#1}{#2}}
\newcommand{\idiff}[2][]{\!\!\mathrm{d}^{#1}{#2}}

\newcommand{\Tr}{\mathrm{Tr}}
\newcommand{\EV}[1]{\left\langle{#1}\right\rangle}

\newcommand{\DiracDelta}{\delta}
\newcommand{\Heaviside}{\Theta}

\newcommand{\kB}{k_\mathrm{B}}

\newcommand{\PSDim}{D}

\newcommand{\DOp}{\rho}
\newcommand{\pdf}[1]{\ensuremath{\mathrm{p}_{#1}}}
\newcommand{\pdfE}[1]{\ensuremath{\pi_{#1}}}

\newcommand{\pq}{\xi}
\newcommand{\vpq}{\vect{\pq}}

\newcommand{\IDoS}{\Omega}
\newcommand{\DoS}{\omega}
\newcommand{\DDoS}{\nu}

\newcommand{\DLCF}{\Omega}
\newcommand{\ER}{E_\mathrm{R}}

\newcommand{\Ent}{S}
\newcommand{\Temp}{T}
\newcommand{\Heat}{Q}
\newcommand{\HeatC}{C}
\newcommand{\Press}{p}

\newcommand{\EntG}{\Ent_{\mathrm{G}}}
\newcommand{\EntB}{\Ent_{\mathrm{B}}}
\newcommand{\EntM}{\Ent_{\mathrm{M}}}
\newcommand{\EntC}{\Ent_{\mathrm{C}}}
\newcommand{\EntP}{\Ent_{\mathrm{P}}}

\newcommand{\TempG}{\Temp_{\mathrm{G}}}
\newcommand{\TempB}{\Temp_{\mathrm{B}}}
\newcommand{\TempM}{\Temp_{\mathrm{M}}}
\newcommand{\TempC}{\Temp_{\mathrm{C}}}
\newcommand{\TempP}{\Temp_{\mathrm{P}}}

\newcommand{\HeatCG}{\HeatC_{\mathrm{G}}}

\newcommand{\PressG}{\Press_{\mathrm{G}}}
\newcommand{\PressB}{\Press_{\mathrm{B}}}

\newcommand{\SubH}[1]{H_{#1}}
\newcommand{\SubE}[1]{E_{#1}}

\newcommand{\SubIDoS}[1]{\Omega_{#1}}
\newcommand{\SubDoS}[1]{\omega_{#1}}
\newcommand{\SubDDoS}[1]{\nu_{#1}}

\newcommand{\SubEnt}[1]{\Ent_{#1}}
\newcommand{\SubTemp}[1]{\Temp_{#1}}

\newcommand{\SubEntG}[1]{\Ent_{\mathrm{G}{#1}}}
\newcommand{\SubEntB}[1]{\Ent_{\mathrm{B}{#1}}}
\newcommand{\SubEntM}[1]{\Ent_{\mathrm{M}{#1}}}
\newcommand{\SubEntC}[1]{\Ent_{\mathrm{C}{#1}}}
\newcommand{\SubEntP}[1]{\Ent_{\mathrm{P}{#1}}}

\newcommand{\SubTempG}[1]{\Temp_{\mathrm{G}{#1}}}
\newcommand{\SubTempB}[1]{\Temp_{\mathrm{B}{#1}}}

\newcommand{\SubTempP}[1]{\Temp_{\mathrm{P}{#1}}}

\newcommand{\IDoSs}[1]{\IDoS_{\mathrm{s}{#1}}}
\newcommand{\Es}[1]{E_{\mathrm{s}{#1}}}
\newcommand{\es}[1]{\gamma_{#1}}

\newcommand{\SubEinit}[1]{E_{{#1}}}
\newcommand{\SubTempGinit}[1]{\Temp_{\mathrm{G}{#1}}}
\newcommand{\SubTempBinit}[1]{\Temp_{\mathrm{B}{#1}}}

\newcommand{\SubEML}[1]{E_{#1}^*}
\newcommand{\SubTempBML}[1]{\Temp^{*}_{\mathrm{B}{#1}}}

\newcommand{\IDoSinf}[1]{\IDoS_{\infty{#1}}}

\newcommand{\bs}{\boldsymbol}

\newcommand{\mcal}{\mathcal}

\newcommand{\mrm}{\mathrm}

\newcommand{\gd}{\delta}
\newcommand{\eps}{\epsilon}%\ge schon vergeben

\newcommand{\go}{\omega}

\newcommand{\gr}{\rho}

\newcommand{\p}{\partial}
\newcommand{\f}{\frac}

\newcommand{\lan}{\langle}
\newcommand{\ran}{\rangle}

\newcommand{\be}{\begin{eqnarray}}
\newcommand{\ee}{\end{eqnarray}}
\newcommand{\bse}{\begin{subequations}}
\newcommand{\ese}{\end{subequations}}

\begin{document}
%%%%%%%%%%%%%%%%%%%%%%%%%%%%%%%%%%%%%%%%%%%%%%%%%%%

\title{Thermodynamic laws in isolated systems}
%\title{Entropy, Temperature, Thermal Equilibrium, and the Laws of Thermodynamics}

\author{Stefan Hilbert}
%\email{hilbert@mpa-garching.mpg.de}
\email{stefan.hilbert@tum.de}
\affiliation{Exzellenzcluster Universe, Boltzmannstr. 2, D-85748 Garching, Germany}
%\affiliation{Max-Planck-Institut f{\"u}r Astrophysik, Karl-Schwarzschild-Str. 1, D-85748 Garching, Germany}

\author{Peter H{\"a}nggi}
\affiliation{Institute of Physics, University of Augsburg, Universit{\"a}tsstra{\ss}e 1, D-86135 Augsburg, Germany}
\affiliation{Nanosystems Initiative Munich, Schellingstr. 4, D-80799 M\"{u}nchen, Germany}

\author{J{\"o}rn Dunkel}
\affiliation{Department of Mathematics, Massachusetts Institute of Technology, 77 Massachusetts Avenue E17-412, Cambridge, MA 02139-4307, USA}

\date{\today}

\begin{abstract}
The recent experimental realization of exotic matter states in isolated quantum systems and the ensuing controversy about the existence of negative absolute temperatures demand a careful analysis of the conceptual foundations underlying microcanonical thermostatistics. Here, we provide a detailed comparison of the most commonly considered microcanonical entropy definitions, focussing specifically on whether they satisfy or violate the zeroth, first and second law of thermodynamics. Our analysis shows that, for a broad class of systems that includes all standard classical Hamiltonian systems, only the Gibbs volume entropy fulfills all three laws simultaneously. To avoid ambiguities, the discussion is restricted to exact results and analytically tractable examples.
\end{abstract}



\maketitle

% pacs / keywords:
% 05.20.-y 	Classical statistical mechanics
% 05.20.Gg 	Classical ensemble theory

% 05.30.-d  Quantum statistical mechanics
% 05.30.Ch  Quantum ensemble theory

% 05.70.-a  Thermodynamics

%%%%%%%%%%%%%%%%%%%%%%%%%%%%%%%%%%%%%%%%%%%%%%%%%
\section{Introduction}
%%%%%%%%%%%%%%%%%%%%%%%%%%%%%%%%%%%%%%%%%%%%%%%%%

Recent advances in experimental and observational techniques have made it possible to study in detail many-particle systems that, in good approximation, are thermally decoupled from their environment. Examples cover a wide range of length and energy scales, from isolated galactic clusters~\cite{2002Gross_Full} and nebulae to ultra-cold quantum gases~\cite{2013Braun} and spin systems~\cite{1951PurcellPound}. The thermostatistical description of such isolated systems relies on the microcanonical ensemble (MCE)~\cite{Gibbs,Khinchin,Becker,Huang}. Conceptually, the MCE is the most fundamental statistical equilibrium ensemble for it only assumes energy conservation, and because canonical and grand-canonical ensembles can be derived from the MCE (by considering the statistics of smaller subsystems~\cite{Becker}) but not \emph{vice versa}.\footnote{This statement is to be understood in a physical sense. Mathematically, the microcanonical density operator of many systems can be obtained from the canonical density operator \textit{via} an inverse Laplace transformation.} Although these facts are widely accepted, there still exists considerable confusion about the consistent treatment of entropy and the role of temperature in the MCE, as evidenced by the recent controversy regarding the (non)existence of negative absolute temperatures~\cite{2013Braun,2013Romero_PRE,2014DuHi_NatPhys,2014VilarRubi,2014FrenkelWarren,2014Schneider_Comment}.

\par
The debate has revealed some widespread misconceptions about the general meaning of temperature in isolated systems. For example, it is often claimed~\cite{2014VilarRubi,2014FrenkelWarren,2014Schneider_Comment}  that knowledge of the microcanonical temperatures suffices to predict the direction of heat flow. This statement, which is frequently mistaken as being equivalent to the second law, is true in many situations, but not in general, reflecting the fact that energy, not temperature, is the primary thermodynamic state variable of an isolated system  (see Sec.~\ref{sec:mce:non-unique_T}). These and other conceptual issues deserve careful and systematic clarification, as they affect the theoretically predicted efficiency bounds of quantum heat engines~\cite{1956Ramsey,2010Rapp,2013Braun} and the realizability of dark energy analogues in quantum systems~\cite{2013Braun}.

\par
The controversy about the existence of negative absolute temperatures revolves around the problem of identifying an entropy definition for isolated systems that is consistent with the laws of thermodynamics~\cite{Gibbs,1910Hertz_1,1910Hertz_2,1991Berdichevsky,2005Campisi,2007Campisi,2014DuHi_NatPhys}. Competing definitions include the \lq surface\rq{} entropy, which is often attributed to Boltzmann\footnote{According to Sommerfeld~\cite{Sommerfeld}, this attribution is probably not entirely correct historically; see also the historical remarks in Sec.~\ref{sec:mce:remarks_on_names} below.},  and the \lq volume\rq{} entropy derived by Gibbs (Chap. XIV in Ref.~\cite{Gibbs}). Although these and other entropy candidates often yield practically indistinguishable predictions for the thermodynamic properties of \lq normal\rq{} systems~\cite{KuboBook}, such as quasi-ideal gases with macroscopic particle numbers,  they can produce substantially different predictions for mesoscopic systems and \textit{ad hoc} truncated Hamiltonians with upper energy bounds~\cite{1991Berdichevsky,2014DuHi_NatPhys}. A related more subtle source of confusion is the precise formulation of the laws of thermodynamics and their interpretation in the context of isolated systems. Most authors seem to agree that a consistent thermostatistical formalism should respect the zeroth, first and second law, but often the laws themselves are stated in a heuristic or ambiguous form~\cite{2014FrenkelWarren,2014Schneider_Comment} that may lead to incorrect conclusions and spurious disputes. 

\par
Aiming to provide a comprehensive foundation for future discussions, we pursue here a two-step approach: Building on the work by Gibbs, Planck and others, we first identify formulations of the zeroth, first and second law that (i) are feasible in the context of isolated systems, (ii) permit a natural statistical interpretation within the MCE, and (iii) provide directly verifiable criteria. In the second step, we analyze whether or not the most commonly considered microcanonical entropy definitions comply with those laws.  In contrast to previous studies, which considered a narrow range of specific examples~\cite{2014DuHi_NatPhys,2014VilarRubi,2014FrenkelWarren,2014Schneider_Comment}, the focus here is on exact generic results that follow from general functional characteristics of the density of states (DoS) and, hence, hold true for a broad class of systems. Thereby, we deliberately refrain from imposing thermodynamic limits (TDLs). TDLs provide a useful technical tool for describing phase transitions in terms of formal singularities~\cite{1967GrRo,1969GrLe,1952YangLee,1952LeeYang} but they are not required on fundamental grounds. A dogmatic restriction~\cite{2014FrenkelWarren} of thermodynamic analysis to infinite systems is not only artificially prohibitive from a practical perspective but also mathematically unnecessary: Regardless of system size, the thermodynamics laws can be validated  based on general properties of the microcanonical DoS (non-negativity, behavior under convolutions, etc.). Therefore, within the qualifications specified in the next sections, the results below apply to finite and infinite  systems that may be extensive or non-extensive, including both long-range and short-range interactions. 

\par
We first introduce essential notation, \highlightchange{specify in detail the underlying assumptions,} and review the different microcanonical entropy definitions~ (Sec.~\ref{sec:mce}). The zeroth, first, and second law are discussed separately in Secs.~\ref{sec:zeroth_law},~\ref{sec:first_law} and~\ref{sec:second_law}. Exactly solvable examples that clarify practical implications are presented in Sec.~\ref{sec:examples}.  Some of these examples were selected to illustrate explicitly the behavior of the different entropy and temperature definitions  when two systems are brought into thermal contact. Others serve as counterexamples, showing that certain entropy definitions fail to satisfy basic consistency criteria. Section~\ref{sec:extensions:discrete} discusses a parameter-free smoothing procedure for systems with discrete spectra. The paper concludes with a discussion of the main results (Sec.~\ref{sec:summary}) \highlightchange{and avenues for future study (Sec.~\ref{sec:outlook}).}

\par
 \highlightchange{The main results of our paper can be summarized as follows: Among the considered entropy candidates,} only the Gibbs volume entropy, which implies a non-negative temperature and Carnot efficiencies~$\le 1$, satisfies all three thermodynamic laws exactly for the vast majority of physical systems (see Table~\ref{sec:summary:tab:comparison}).

%%%%%%%%%%%%%%%%%%%%%%%%%%%%%%%%%%%%%%%%%%%%%%%%%
\section{The microcanonical distribution and entropy definitions}
\label{sec:mce}
%%%%%%%%%%%%%%%%%%%%%%%%%%%%%%%%%%%%%%%%%%%%%%%%%

After recalling basic \highlightchange{assumptions and} definitions for the microcanonical ensemble (Sec.~\ref{sec:mce:setup}), we will summarize the most commonly considered microcanonical entropy definitions and their related temperatures in Sec.~\ref{sec:mce:entropy_candidates}. These entropy candidates will be tested in later, in Secs.~\ref{sec:zeroth_law},~\ref{sec:first_law} and~\ref{sec:second_law},  as to whether they satisfy the zeroth, first and second law of thermodynamics. Section~\ref{sec:mce:remarks_on_names} contains brief historical remarks on entropy naming conventions. Finally, in  Sec.~\ref{sec:mce:non-unique_T}, we demonstrate explicitly that, regardless of the adopted entropy definition,  the microcanonical temperature is, in general, not a unique (injective) function of the energy. This fact means that knowledge of temperature is, in general, not sufficient for predicting the direction of heat flow, implying that feasible versions of the second law must be formulated in terms of  entropy and energy (and not temperature).

%%%%%%%%%%%%%%%%%%%%%%%%%%%%%%%%%%%%%%%%%%%%%%%%%
\subsection{The microcanonical ensemble}
\label{sec:mce:setup}
%%%%%%%%%%%%%%%%%%%%%%%%%%%%%%%%%%%%%%%%%%%%%%%%%

We consider strictly isolated\footnote{\highlightchange{We distinguish isolated systems (no heat or matter exchange with environment), closed systems (no matter exchange, but heat exchange permitted), and open systems (heat and matter exchange possible).}} classical or quantum systems described by a Hamiltonian $H(\vpq; Z)$ where $\vpq$ denotes the microscopic states\footnote{\highlightchange{For classical systems, $\vpq$ comprises the canonical coordinates and momenta that specify points in phase space.} For quantum systems, $\vpq$ represents the labels (quantum numbers) of the energy eigenstates.} and $Z= (Z_1,\ldots)$ comprises external control parameters (volume, magnetic fields, etc.). It will be assumed throughout that the microscopic dynamics of the system conserves the system energy~$E$, that the energy is bounded from below, $E\ge 0$, and that the system is in a state in which its thermostatistical properties are described by the microcanonical ensemble with the microcanonical density operator \footnote{\highlightchange{For example, isolated systems with mixing dynamics~\cite{2003HuntMacKay} approach a state that can indeed be described by a MCE -- although, in many cases, weaker conditions like ergodicity are sufficient~\cite{1973LebowitzPenrose}. In contrast, systems with additional integrals of motion besides energy (e.g. angular momentum) may require a different thermostatistical description due to their nontrivial phase space topology.}}
\be
\label{eq:mce:pdf_mce}
\DOp(\vpq | E,Z)=\frac{\DiracDelta\bigl[E - H(\vpq, Z) \bigr]}{\DoS(E,Z)}.
\ee
The normalization constant is given by the density of states (DoS)
\be
\label{eq:mce:dos_mce}
\DoS(E,Z) = \Tr\left\{\DiracDelta\bigl[E - H(\vpq, Z) \bigr] \right\}.
\ee 
For classical systems, the trace $\Tr$ is defined as a phase-space integral (\highlightchange{normalized by symmetry factors and powers of the Planck constant}) and for quantum system by an integral over the basis vectors of the underlying Hilbert space. The DoS $\DoS$ is non-negative, $\DoS(E) \geq 0$, and with our choice of the ground-state energy, $\DoS(E) = 0$ for $E<0$.

\par
The energy derivative of the DoS will be denoted by
\begin{equation}
\label{eq:mce:ddos_mce}
\DDoS(E,Z) = \frac{\p \DoS(E,Z)}{\p E}.
\end{equation}
The integrated DoS, or dimensionless phase volume, is defined as
\begin{equation}
\label{eq:mce:idos_mce}
\IDoS(E,Z) = \Tr\left\{\Heaviside\bigl[E - H(\vpq, Z) \bigr] \right\},
\end{equation}
where $\Heaviside$ denotes the unit-step function. This definition implies that $\IDoS$ is a non-decreasing function of the energy $E$ that vanishes for $E<0$. For clarity, we shall assume throughout that $\IDoS$ is continuous and piecewise differentiable, so that, except at certain singular points, its partial derivatives are well-defined with $\DoS=\p\IDoS/\p E$ and $\DDoS=\p^2\IDoS/\p E^2$, and
\begin{equation}
\IDoS(E,Z) = \int_0^{E} \idiff{E'}\DoS(E', Z).
\end{equation}
These conditions are practically always fulfilled for classical Hamiltonian systems. For quantum systems with a discrete energy spectrum, additional smoothing procedures are required~(see discussion in Sec.~\ref{sec:extensions:discrete} below). 

\par
The expectation value of an observable $F(\vpq)$ with respect to the microcanonical density operator~$\DOp$ is defined by
\be
\EV{F}_{E,Z} =\Tr[\DOp{}\,F].
\ee 
As a special case, the probability density of some observable $F(\vpq)$ is given by 
\be\label{eq:mce:pdf_observable}
\pdf{F}(f|E,Z)=\EV{\DiracDelta(f-F)}_{E,Z}.
\ee
To avoid potential confusion, we would like to stress that, although we will compare different entropy functions $S(E,Z)$, all expectation values will always be defined with respect to the standard microcanonical density operator~$\DOp$, as defined in Eq.~\eqref{eq:mce:pdf_mce}. That is, expectation values $\EV{\,\cdot\,}_{E,Z}$ are always computed by averaging over microstates that are confined to the energy shell $E$. 

\par
For convenience, we adopt units such that the Boltzmann constant $\kB =1$. For a given entropy function $S(E,Z)$, the microcanonical temperature $T$ and the heat capacity $C$ are obtained according to the basic rules of thermodynamics by partial differentiation,
\begin{align}
T(E,Z) &= \left[\f{\p S(E,Z)}{\p E}\right]^{-1},
\label{eq:mce:df_temperature}
\\
C(E,Z) &= \left[\parder{T(E,Z)}{E}\right]^{-1}.
\label{eq:mce:df_heat_capacity}
\end{align}
It is important to emphasize that the primary thermodynamic state variables of an isolated system are $E$~and~$Z$.  More generally,  the primary thermodynamic state variables of an isolated system comprise the conserved quantities (\lq charges\rq) that characterize the underlying symmetries, and symmetry breaking parameters~\cite{Callen}, such as volume (broken translational invariance).  In contrast, the  temperature $T$ is a \emph{derived} quantity that, in general, does not uniquely characterize the thermodynamic state  (see detailed discussion in Sec.~\ref{sec:mce:non-unique_T} and Fig.~\ref{fig:wiggly_dos} below). 

\par
To simplify notation, when writing formulas that contain $\DoS, \IDoS, T$, etc., we will usually not explicitly state the functional dependence on $Z$ anymore, while keeping in mind that the Hamiltonian can contain several additional control parameters.

%%%%%%%%%%%%%%%%%%%%%%%%%%%%%%%%%%%%%%%%%%%%%%%%%
\subsection{Microcanonical entropy candidates}
\label{sec:mce:entropy_candidates}
%%%%%%%%%%%%%%%%%%%%%%%%%%%%%%%%%%%%%%%%%%%%%%%%%

%%%%%%%%%%%%%%%%%%%%%%%%%%%%%%%%%%%%%%%%%%%%%%%%%
\subsubsection{Gibbs entropy}
\label{sec:mce:Gibbs_entropy_and_temperature}
%%%%%%%%%%%%%%%%%%%%%%%%%%%%%%%%%%%%%%%%%%%%%%%%%

The Gibbs volume entropy is defined by (see Chap. XIV in Ref.~\cite{Gibbs})
\bse
\begin{equation}
\label{eq:mce:df_Gibbs_entropy}
\EntG(E) = \ln \IDoS (E).
\end{equation}
The associated Gibbs temperature~\cite{Gibbs,Becker,Khinchin}
\begin{equation}
\label{eq:mce:df_Gibbs_temperature}
\TempG(E) 
=\frac{\IDoS(E)}{\DoS(E)}
\end{equation}
\ese
is always non-negative, $\TempG(E)\ge 0$, and remains finite as long as $\DoS(E)>0$.

\par
For classical Hamiltonian systems with microstates (phase-space points) labelled by $\vpq=(\xi_1,\ldots,\xi_D)$, a continuous Hamiltonian $H(\bs \xi)$, a simple phase space $\{\vpq\}=\R^D$, and a phase space density 
described by the microcanonical density operator~\eqref{eq:mce:pdf_mce}, it is straightforward to prove that the Gibbs temperature satisfies the equipartition theorem~\cite{Khinchin}:
\begin{equation}
\label{eq:mce:equipartition_theorem}
\TempG(E) = \EV{\pq_i \parder{H}{\pq_i}}_{E}
\quad \forall\; i=1,\ldots,\PSDim.
\end{equation}
The proof of Eq.~\eqref{eq:mce:equipartition_theorem}, which is essentially a phase-space version of Stokes' theorem~\cite{Khinchin}, uses partial integrations, so that Eq.~\eqref{eq:mce:equipartition_theorem} holds for all standard classical Hamiltonian systems with simply connected closed energy manifolds $E=H(\bs \xi)$. However, the examples in Sec.~\ref{sec:examples:anharmonic_oscillator} show that Eq.~\eqref{eq:mce:equipartition_theorem} even holds for certain classical systems with bounded spectrum and non-monotonic DoS.

\par
A trivial yet important consequence of Eq.~\eqref{eq:mce:equipartition_theorem} is that any temperature satisfying the equipartition theorem must be identical to the Gibbs temperature, and thus, the associated entropy must be equal to the Gibbs entropy (plus any function independent of energy). This implies already that none of the other entropy definitions considered below leads to a temperature that satisfies the equipartition theorem (unless these other entropy definitions happen to coincide with the Gibbs entropy on some energy interval or in some limit). We shall return to Eq.~\eqref{eq:mce:equipartition_theorem} later, as it relates directly to the notion of \emph{thermal equilibrium} and the zeroth law.

%%%%%%%%%%%%%%%%%%%%%%%%%%%%%%%%%%%%%%%%%%%%%%%%%
\subsubsection{Boltzmann entropy}
\label{sec:mce:Boltzmann_entropy_and_temperature}
%%%%%%%%%%%%%%%%%%%%%%%%%%%%%%%%%%%%%%%%%%%%%%%%%

The perhaps most popular microcanonical entropy definition is the Boltzmann entropy
\bse
\begin{equation}
\label{eq:mce:df_Boltzmann_entropy}
\EntB(E) = \ln \left[\epsilon\, \DoS(E) \right],
\end{equation}
\ese
where~$\epsilon$ is a small energy constant required to make the argument of the logarithm dimensionless. The fact that the definition of~$\EntB$ requires an additional energy constant~$\epsilon$ is conceptually displeasing but bears no relevance for physical quantities that are related to derivatives of~$\EntB$. As we shall see below, however, the presence of $\epsilon$ will affect the validity of the second law.

\par
The associated Boltzmann temperature
\begin{equation}
\label{eq:mce:df_Boltzmann_temperature}
\TempB(E) =\frac{\DoS(E)}{\DDoS(E)}
\end{equation}
becomes negative when $\DoS(E)$ is a decreasing function of the energy~$E$, that is, when $\nu(E)=\p\DoS/\p E < 0$. The Boltzmann temperature and the Gibbs temperature are related by~\cite{2014DuHi_NatPhys}\footnote{We thank M. Campisi for directing our attention to this formula.}
\begin{equation}
\TempB(E) = \frac{\TempG(E)}{ 1 - \HeatCG^{-1}(E)},
\end{equation}
where $\HeatCG=(\p \TempG/\p E)^{-1}$ is the Gibbsian heat capacity measured in units of~$\kB$.
Thus, a small positive (generally non-extensive) heat capacity \mbox{$0< \HeatCG(E) < 1$} implies a negative Boltzmann temperature $\TempB(E) < 0$ and~\textit{vice versa}.\footnote{One should emphasize that $\TempB(E)$ is typically the effective canonical temperature of a small subsystem~\cite{2014DuHi_NatPhys}, appearing for example in one-particle momentum distributions and other reduced density operators. This fact, however, does not imply that $\TempB(E)$ is necessarily the absolute thermodynamic temperature of the whole system.}

%%%%%%%%%%%%%%%%%%%%%%%%%%%%%%%%%%%%%%%%%%%%%%%%%
\subsubsection{Modified Boltzmann entropy}
\label{sec:mce:modified_Boltzmann_entropy_and_temperature}
%%%%%%%%%%%%%%%%%%%%%%%%%%%%%%%%%%%%%%%%%%%%%%%%%

The energy constant $\epsilon$ in Eq.~\eqref{eq:mce:df_Boltzmann_entropy} is sometimes interpreted as a small uncertainty in the system energy~$E$. This interpretation suggests a modified microcanonical phase space probability density~\cite[][p.~115]{Gibbs}
\begin{equation}
\label{eq:pdf_gibbs}
\tilde\DOp(\vpq; E, \epsilon) = \frac{ \Heaviside\bigl( E + \epsilon - H \bigr)\;\Heaviside \bigl( H- E \bigr) }{\IDoS(E + \epsilon) - \IDoS(E)} .
\end{equation}
The Shannon information entropy of the modified density operator is given by
\bse
\begin{equation}
\begin{split}
\label{eq:mce:df_modified_Boltzmann_entropy}
\EntM(E,\eps) 
 &= -\Tr\left[\tilde\DOp\ln\tilde\DOp\right]
\\&= \ln \left[ \IDoS(E + \epsilon) - \IDoS(E) \right]. 	
\end{split}
\end{equation}
From $\EntM$, one can recover the Boltzmann entropy by expanding the argument of logarithm for $\eps\to 0$,
\be
\EntM (E) \approx \ln \left[ \epsilon \, \DoS(E) \right] = \EntB(E).
\ee
\ese
Note that this is~\emph{not} a systematic Taylor expansion of $\EntM$ itself, but rather of $\exp({\EntM})$.
The associated temperature 
\bse
\be
\label{eq:mce:df_modified_Boltzmann_temperature}
\TempM(E,\eps) = \frac{\IDoS(E + \epsilon) - \IDoS(E)}{\DoS(E + \epsilon) - \DoS(E)}
\ee
approaches for $\epsilon\to 0$ the Boltzmann temperature 
\be
\TempM(E) \approx \frac{\DoS(E)}{\DDoS(E)} =\TempB(E).
\ee
\ese
However, from a physical and mathematical point of view, the introduction of the finite energy uncertainty~$\epsilon$ is redundant, since according to the postulates of classical and quantum mechanics, systems can at least in principle be prepared in well-defined energy eigenstates that can be highly degenerate.  Moreover, from a more practical perspective, the explicit $\epsilon$-dependence of $\EntM$ and $\TempM$ means that any   thermodynamic formalism based on $\EntM$ involves $\epsilon$ as a second energy control parameter. The physically superfluous but technically required $\epsilon$-dependence disqualifies $\EntM$ from being a generic entropy definition for the standard microcanonical ensemble defined by Eq.~\eqref{eq:mce:pdf_mce}.

%%%%%%%%%%%%%%%%%%%%%%%%%%%%%%%%%%%%%%%%%%%%%%%%%
\subsubsection{Complementary Gibbs entropy}
\label{sec:mce:complementary_Gibbs_entropy_and_temperature}
%%%%%%%%%%%%%%%%%%%%%%%%%%%%%%%%%%%%%%%%%%%%%%%%%

If the total number of microstates is finite, \mbox{$\Omega_\infty\equiv \Omega(E\to\infty) < \infty$}, as for example in spin models with upper energy bound, then one can also define a complementary Gibbs entropy~\cite{2014Schneider_Comment}
\bse
\begin{equation}
\label{eq:mce:df_complementary_Gibbs_entropy}
\EntC(E) = \ln \left[ \IDoSinf{} - \IDoS(E) \right].
\end{equation}
The complementary Gibbs temperature
\begin{equation}
\label{eq:mce:df_complementary_Gibbs_temperature}
\TempC(E) = -\frac{\IDoSinf{} - \IDoS(E)}{\DoS(E)}.
\end{equation}
\ese
is always negative.

\par
In another universe, where~\mbox{$\Omega_\infty < \infty$} holds for all systems, the complementary Gibbs entropy provides an alternative thermodynamic description that, roughly speaking, mirrors the Gibbsian thermodynamics. In our Universe, however, many (if not all) physical systems are known to have a finite groundstate energy, but we are not aware of any experimental evidence for the existence of strict upper energy bounds.\footnote{
Model Hamiltonians with upper energy bounds (such as spin models) are usually truncations of more realistic (and often more complicated) Hamiltonians that are bounded from below but not from above.}

\par
From a practical perspective, a thermostatistical theory based on $\EntC$ is, by construction, restricted to systems with  upper energy bounds, thereby excluding many physically relevant systems such as classical and quantum gases. In particular, the  complementary Gibbs temperature is incompatible with conventional operational definitions of temperature that use Carnot efficiencies or  measurements with a gas thermometer.
Thus, even if one were content with the severe restriction to systems with upper energy bounds and considered the complementary Gibbs entropy~$\EntC$ as thermodynamic entropy, then all statements of conventional thermodynamics -- including the laws of thermodynamic themselves -- would have to be carefully examined and adjusted accordingly.

%%%%%%%%%%%%%%%%%%%%%%%%%%%%%%%%%%%%%%%%%%%%%%%%%
\subsubsection{Alternative entropy proposals}
\label{sec:mce:alternative_entropies}
%%%%%%%%%%%%%%%%%%%%%%%%%%%%%%%%%%%%%%%%%%%%%%%%%

Another interesting entropy definition is\footnote{We thank Oliver Penrose for bringing this possibility to our attention.}
\bse
\begin{equation}
\label{sec:mce:eq:df_Penrose_entropy}
\EntP(E) = \ln \IDoS(E) + \ln [ \IDoS_\infty - \IDoS(E) ] - \ln \IDoS_\infty.
\end{equation}
For systems with $\IDoS_\infty = \infty$, this alternative entropy becomes identical to the Gibbs entropy, assuming a sensible definition of $\lim_{E\to\infty} \EntP$. However, $\EntP$ differs from $\EntG$ for systems with bounded spectrum. The associated  temperature
\begin{equation}
\label{sec:mce:eq:df_Penrose_temperature}
\TempP(E) 
= 
\frac{1}{\DoS}\left[\frac{1}{ \IDoS}-\frac{1}{\IDoS_\infty - \IDoS}\right]^{-1}
\end{equation}
\ese
interpolates between $\TempG$ and $\TempC$ if $\IDoS_\infty < \infty$, and is equal to $\TempG$ otherwise. The example in Sec.~\ref{sec:examples:anharmonic_oscillator} demonstrates that, similar to the Boltzmann entropy, $\EntP$ also violates the classical equipartition theorem.

\par
In principle, one may also attempt to define entropies that have different analytic behaviors on different energy intervals~\cite{2014FrenkelWarren,2014Schneider_Comment}; for example, by constructing piecewise combinations of the Gibbs entropy  and the complementary Gibbs entropy, such as
\be
\label{sec:mce:eq:df_Schneider_entropy}
S_\mrm{G\vee C}(E)=\min(\EntG, \EntC).
\ee
However, constructions of this type seem unfeasible for more realistic model systems with an energy structure that goes beyond that of the simplest spin models (for example, when the DoS has more than one maximum)\footnote{The authors of Ref.~\cite{2014FrenkelWarren} speculate that suitably defined piecewise entropies could converge to the Boltzmann entropy in the TDL.}. In particular, the explicit dependence of $\EntP$ and $S_\mrm{G\vee C}$ on $\IDoS_\infty$ means that the equations of state obtained from these entropies,  depend on the choice of the upper energy cut-off, even though upper bounds are in fact artificial constraints resulting from \textit{ad hoc} truncations of the underlying Hamiltonians\footnote{All stable systems have a finite groundstate energy but, at least to our knowledge, no real physical system has a true upper energy bound. It does not seem reasonable to construct a thermodynamic formalism that gives different predictions depending on whether or not one neglects or includes higher-energy bands. For example, $\EntP$ can predict negative temperatures when only a single band is considered but these negative $\TempP$ regions disappear if one includes all higher bands. Similarly, $S_\mrm{G\vee C}$ can predict negative temperatures even for $\nu(E)=\omega'(E)>0$ depending on the choice of the energy cut-off.}. Another deficiency of piecewise entropies is that they predict spurious phase transitions arising from the non-analyticities at the interval boundaries. In view of such drawbacks, and due to the absence of a clearly formulated general definition that would be amenable to systematic analysis for a broader class of DoS functions, we do not study such piecewise entropies here.

%%%%%%%%%%%%%%%%%%%%%%%%%%%%%%%%%%%%%%%%%%%%%%%%%
\subsection{Historical remarks and naming conventions}
\label{sec:mce:remarks_on_names}
%%%%%%%%%%%%%%%%%%%%%%%%%%%%%%%%%%%%%%%%%%%%%%%%%

Boltzmann's tombstone famously carries the formula 
\be\label{eq:mcd:history_boltzmann}
S=k \log W,
\ee
even though it was probably Planck, and not Boltzmann, who established this equation (see Sommerfeld's discussion in the Appendix of Ref.~\cite{Sommerfeld}).
As described in many textbooks~(e.g., Ref.~\cite{Huang}), the entropy ~$\EntB$ defined in Eq.~\eqref{eq:mce:df_Boltzmann_entropy} is heuristically obtained from Eq.~\eqref{eq:mcd:history_boltzmann} by identifying $\log=\ln$ and interpreting $W=\epsilon \DoS(E)$ as the number of microstates accessible to a physical system at energy $E$. Perhaps for this reason, the entropy~\eqref{eq:mce:df_Boltzmann_entropy} is often called \lq Boltzmann entropy\rq{} nowadays.

\par
Upon dividing by Boltzmann's constant $k=\kB$, Eq.~\eqref{eq:mcd:history_boltzmann} coincides with Shannon's \emph{information} entropy
\be\label{eq:mcd:information_entropy}
I=-\sum_i p_i\log p_i
\ee  
for a uniform probability distribution $p_i=1/W$ (with $i=1,\ldots, W$) on a \emph{discrete} set of $W$ microstates\footnote{The fact that $\EntB$ can be connected to one of the many~\cite{1960Renyi,1978Wehrl,1991Wehrl} information entropies does \emph{not} imply that~$\EntB$ is equivalent to the phenomenological thermodynamic entropy and satisfies the laws of thermodynamics, even if such a connection might be appealing. Instead one has to verify whether or not $\EntB$ does indeed satisfy the thermodynamic laws for isolated systems.}. Boltzmann himself, while working on his $H$-theorem~\cite{1872Boltzmann} for classical $N$-particle systems, considered the continuum version of Eq.~\eqref{eq:mcd:information_entropy} for the reduced \emph{one-particle} distribution instead of the full $N$-particle distribution. Gibbs generalized Boltzmann's $H$-theorem to the $N$-particle distribution function\footnote{See Uffink~\cite{2007Uffink} for a detailed historical account.}.    

\par
In his comprehensive treatise on statistical mechanics~\cite{Gibbs}, Gibbs considered three different statistical entropy definitions and investigated whether or not they may serve as analogues for the phenomenological thermodynamic entropy (see Chap. XIV in Ref.~\cite{Gibbs}). The first definition, which amounts to \mbox{$S_N = -\Tr [\rho \ln \rho]$} in our notation, relates directly to his work on the generalized $H$-theorem. One should stress that Gibbs used this definition only when describing systems coupled to a heat bath within the framework of the canonical ensemble. Nowadays, $S_N$ is usually referred to as canonical Gibbs entropy in classical statistical mechanics, as von-Neumann entropy in quantum statistics,  or as Shannon entropy in information theory.  

\par
In the context of isolated systems, however, Gibbs investigated the two alternative entropy definitions~\eqref{eq:mce:df_Gibbs_entropy} and~\eqref{eq:mce:df_Boltzmann_entropy}. After performing a rigorous and detailed analysis, he concluded that, within the MCE, the definition~\eqref{eq:mce:df_Gibbs_entropy} provides a better anolog for the thermodynamic entropy. About a decade  later, in 1910, Gibbs' conclusion was corroborated by Hertz~\cite{1910Hertz_1,1910Hertz_2}, whose analysis focussed on adiabatic invariance. Hertz~\cite{1910Hertz_1} acknowledged explicitly that he could not add much new content to Gibbs' comprehensive treatment, but was merely trying to provide a more accessible approach to Gibbs' theory\footnote{Notwithstanding, Hertz's papers~\cite{1910Hertz_1,1910Hertz_2} received exceptional editorial support from Planck~\cite{2008Hoffmann} and highest praise from Einstein~\cite{1911Einstein}.}. It seems therefore appropriate to refer to the definition~\eqref{eq:mce:df_Gibbs_entropy} as the microcanonical \lq{}Gibbs entropy\rq{}, although some previous studies also used the term \lq{}Hertz entropy\rq{}.

%%%%%%%%%%%%%%%%%%%%%%%%%%%%%%%%%%%%%%%%%%%%%%%%%
\subsection{Non-uniqueness of microcanonical temperatures}
\label{sec:mce:non-unique_T}
%%%%%%%%%%%%%%%%%%%%%%%%%%%%%%%%%%%%%%%%%%%%%%%%%

%================================================
\begin{figure*}
\centerline{
\includegraphics[height=5cm]{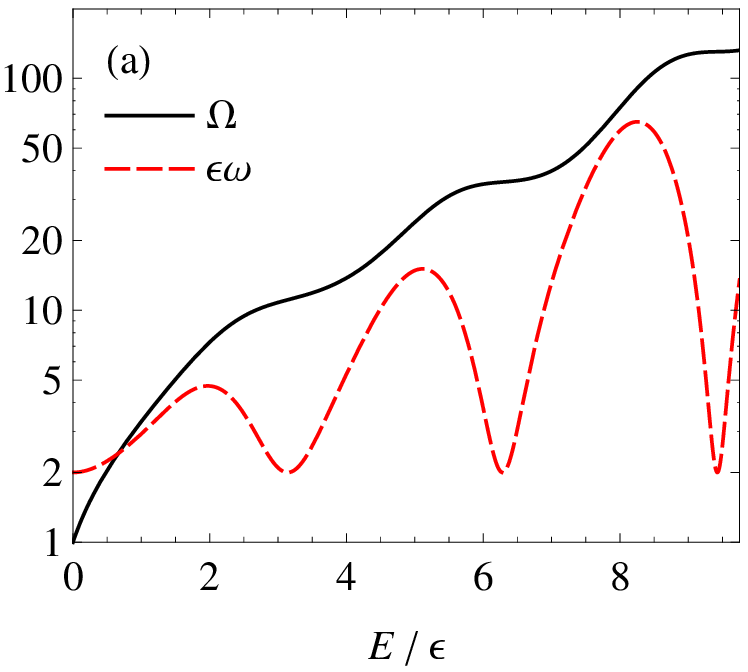}
\hfill
\includegraphics[height=5cm]{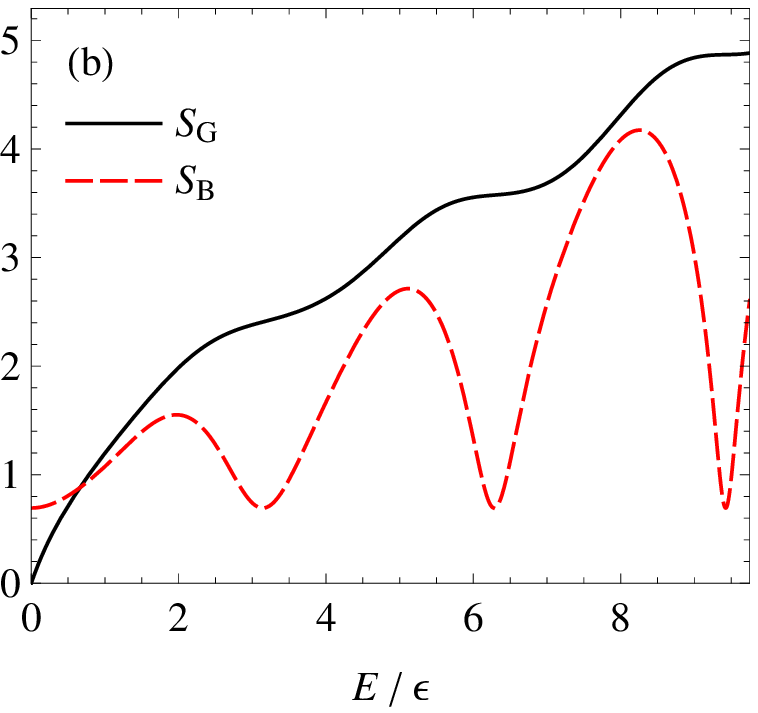}
\hfill
\includegraphics[height=5cm]{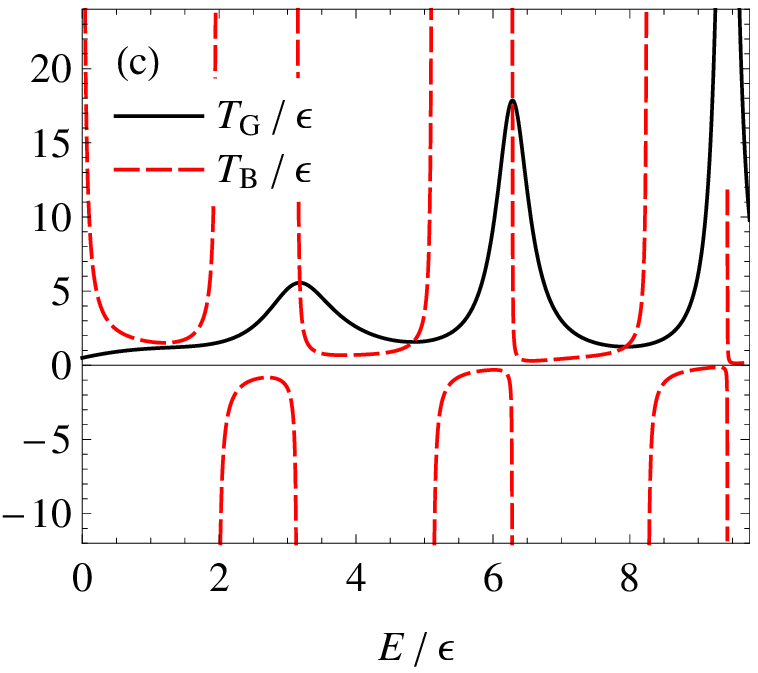}
}
\caption{
\label{fig:wiggly_dos}
Non-uniqueness of microcanonical temperatures illustrated for the integrated DoS from Eq.~\eqref{eq:wiggly_ddos}. Left (a): integrated DoS $\IDoS$ (black) and DoS $\DoS$ (red, dashed). Center (b): Gibbs entropy $\EntG$ (black) and Boltzmann entropy $\EntB$ (red, dashed). Right (c): Gibbs temperature~$\TempG$ (black) and Boltzmann temperature~$\TempB$ (red, dashed). This example shows that, in general, neither the Boltzmann nor the Gibbs temperature uniquely characterize the thermal state of an isolated system, as the same temperature value can correspond to very different energy values.
}
\end{figure*}
%================================================

It is often assumed that temperature tells us in which direction heat will flow when two bodies are placed in thermal contact. Although this heuristic rule-of-thumb works well in the case of \lq{}normal\rq{} systems that possess a monotonically increasing DoS $\DoS$, it is not difficult to show that, in general, neither the Gibbs temperature nor the Boltzmann temperature nor any of the other suggested alternatives are capable of specifying uniquely the direction of heat flow when two isolated systems become coupled. One obvious reason is simply that these microcanonical temperatures do~\emph{not} always uniquely characterize the state of an isolated system before it is coupled to another. To illustrate this explicitly, consider as a simple generic example a system with integrated DoS
\be\label{eq:wiggly_ddos}
\IDoS(E)=\exp \left[\frac{E}{2 \epsilon }-\frac{1}{4} \sin \left(\frac{2E}{\epsilon }\right)\right]+\frac{2 E}{\epsilon },
\ee
where $\eps$ is some energy scale. The associated DoS is non-negative and non-monotonic, $\DoS\equiv \p\IDoS/\p E\ge 0$ for all $E\ge 0$. 
As evident from Fig.~\ref{fig:wiggly_dos}, \emph{neither the Gibbs nor the Boltzmann temperature provide a unique thermodynamic characterization} in this case, as the same temperature value $\TempG$ or $\TempB$ can correspond to vastly different energy values. When coupling such a system to a second system, the direction of heat flow may be different for different initial energies of the first system, even if the corresponding initial temperatures of the first system may be the same. It is not difficult to see that qualitatively similar results are obtained for all continuous functions $\DoS(E)\ge 0$ that exhibit at least one local maximum and one local minimum on \highlightchange{$(0,\infty)$}. This ambiguity reflects the fact that the essential control parameter (thermodynamic state variable) of an isolated system is the energy $E$ and not the temperature. 

\par
\highlightchange{The above example shows that \emph{any} microcanonical temperature definition yielding an energy-temperature relation that is not always strictly one-to-one cannot tell us universally in which direction heat flows when two bodies are brought into thermal contact. In fact, one finds that the considered temperature definitions may even fail to  predict heat flows correctly for systems where the energy-temperature relation is one-to-one (see examples in Sec.~\ref{sec:examples:power_law_dos} and \ref{sec:examples:bounded_dos}). This indicates that, in general,} microcanonical temperatures do not specify the heat flow between two initially isolated systems and, therefore, temperature-based heat-flow arguments~\cite{2014FrenkelWarren,2014VilarRubi,2014Schneider_Comment} should not be used to judge entropy definitions.  One must instead analyze whether or not the different definitions respect the laws of thermodynamics.

%%%%%%%%%%%%%%%%%%%%%%%%%%%%%%%%%%%%%%%%%%%%%%%%%
\section{Zeroth law and thermal equilibrium}
\label{sec:zeroth_law}
%%%%%%%%%%%%%%%%%%%%%%%%%%%%%%%%%%%%%%%%%%%%%%%%%

%%%%%%%%%%%%%%%%%%%%%%%%%%%%%%%%
\subsection{Formulations}
\label{sec:zeroth_law:versions}
%%%%%%%%%%%%%%%%%%%%%%%%%%%%%%%%

In its most basic form, the zeroth law of thermodynamics states that:

\par
 (Z0) \emph{If two systems $\mcal{A}$ and $\mcal{B}$ are in thermal equilibrium with each other, and $\mcal{B}$ is in thermal equilibrium with a third system~$\mcal{C}$, then $\mcal{A}$ and~$\mcal{C}$ are also in thermal equilibrium with each other.}

 \par
 Clearly, for this statement to be meaningful, one needs to specify what is meant by \lq{}thermal equilibrium\rq{}\footnote{Equilibrium is a statement about the exchange of conserved quantities between systems. To avoid conceptual confusion, one should clearly distinguish between thermal equilibrium (no mean energy transfer), pressure equilibrium (no mean volume transfer), chemical equilibrium (no particle exchange on  average), etc. Complete thermodynamic equilibrium corresponds to a state where all the conserved fluxes between two coupled systems vanish.}.
We adopt here the following minimal definition:

\par
(E)\emph{Two systems are in thermal equilibrium, iff they are in contact so that they can exchange energy, and they have relaxed to a state in which there is no average net transfer of energy between them anymore. A system $\mcal{A}$ is in thermal equilibrium with itself, iff all its subsystems are in thermal equilibrium with each other. In this case,  $\mcal{A}$ is called a (thermal) equilibrium system.} 

\par
With this convention, the zeroth law (Z0), which demands transitivity of thermal equilibrium,  ensures that thermal equilibrium is an equivalence relation on the set of thermal equilibrium systems.  We restrict the discussion in this paper to thermal equilibrium systems as defined by (E). For brevity, we often write equilibrium instead of thermal equilibrium. 

\par
The basic form (Z0) of the zeroth law is a fundamental statement about energy flows, but it does not directly address entropy or temperature. Therefore, (Z0) cannot be used to distinguish entropy definitions.  A stronger version of the zeroth law is obtained by demanding that, in addition to (Z0), the following holds:

\par
(Z1) \emph{One can assign to every thermal equilibrium system a real-valued \lq{}temperature\rq{}~$T$, such that the temperature of any of its subsystems is equal to $T$.}

\par
The extension (Z1) implies that any two equilibrium systems that are also in thermal equilibrium with each other have the same temperature, which is useful for the interpretation of temperature measurements that are performed by bringing a thermometer in contact with another system.
%\highlightchange{The added strength of the extended version (Z0+Z1) we are particularly interested in here is that the temperature equality requirement (z1) imposes testable conditions on possible temperature candidates and their corresponding entropies.}
To differentiate between the weaker version (Z0) of the zeroth law from the stronger version (Z0+Z1), we will say that systems satisfying (Z0+Z1) are in \emph{temperature equilibrium}. 

\par
One may wonder whether there exist other feasible formulations of the zeroth law for isolated systems. For instance, since thermal equilibrium partitions the set of equilibrium systems into equivalence classes, one might be tempted to assume that temperature can be defined in such a way that it serves as a \emph{unique} label for these equivalence classes. If this were possible, then it would follow that any two systems that have the same temperature are in thermal equilibrium, even if they are not in contact. By contrast, the definition (E) adopted here implies that two systems cannot be in thermal equilibrium unless they are in contact, reflecting the fact that it seems meaningless to speak of thermal equilibrium if two systems are unable to exchange energy\footnote{If we uphold the definition (E), but still wish to uniquely label equivalence classes of systems in thermal equilibrium (and thus in thermal contact), we are confronted with the difficult task to always assign different temperatures to systems not in thermal contact.}.

\par
One may try to rescue the idea of using temperature to identify equivalence classes of systems in thermal equilibrium -- and, thus, of demanding that systems with the same temperature are in thermal equilibrium -- by broadening the definition of \lq{}thermal equilibrium\rq{} to include both \lq{}actual\rq{} thermal equilibrium, in sense of definition (E) above, and \lq{}potential\rq{} thermal equilibrium: Two systems are in potential thermal equilibrium, if they are not in thermal contact, but there would be no net energy transfer between the systems if they were brought into (hypothetical) thermal contact. One could then demand that two systems with the same temperature are in actual or potential thermal equilibrium. However, as already indicated in Sec.~\ref{sec:mce:non-unique_T}, and explicitly shown in \highlightchange{Sec.~\ref{sec:examples:bounded_dos}}, none of the considered temperature definitions satisfies this requirement either. The reason behind this general failure is that, for isolated systems, temperature as a secondary derived quantity does not always uniquely determine the thermodynamic state of the system, whereas potential heat flows and equilibria are determined by the \lq true\rq{} state variables $(E,Z)$. Thus, demanding that two systems with the same temperature must be in thermal equilibrium is not a feasible extension of the zeroth law\footnote{The situation is different for systems coupled to an infinite heat bath and described by the canonical ensemble. Then, by construction, the considered systems are in thermal equilibrium with a heat bath that sets both the temperature and the mean energy of the systems. If one assumes that any two systems with the same temperature couple to (and thus be in thermal equilibrium with) the same heat bath, then the basic form (Z0) of the zeroth law asserts that such systems are in thermal equilibrium with each other.}. 
%-- in contrast to Eq.~\eqref{eq:zeroth_law_ii}.

\par
In the remainder this section, we will analyze which of the different microcanonical entropy definitions is compatible with the condition (Z1). Before we can do this, however, we need to specify unambiguously what exactly is meant by \lq temperature of a subsystem\rq~in the context of the MCE. \highlightchange{To address this question systematically, we first recapitulate the meaning of thermal equilibrium in the MCE (Sec.~\ref{sec:zeroth_law:thermal_equilibrium}) and then discuss briefly subsystem energy and heat flow (Sec.~\ref{sec:zeroth_law:sub_energy}). These steps  will allow us to translate  (Z1) into testable statistical criterion for the subsystem temperature (Sec.~\ref{sec:zeroth_law:sub_temp}).}

%%%%%%%%%%%%%%%%%%%%%%%%%%%%%%%%
\subsection{Thermal equilibrium in the MCE}
\label{sec:zeroth_law:thermal_equilibrium}
%%%%%%%%%%%%%%%%%%%%%%%%%%%%%%%%

Consider an isolated  system consisting of two or more \emph{weakly} coupled subsystems. Assume the total energy of the compound system is conserved, so that its equilibrium state can be adequately described by the MCE. Due to the coupling, the energy values of the individual subsystems are not conserved by the microscopic dynamics and will fluctuate around certain average values. 
Since the microcanonical density operator of the compound system is stationary, the energy mean values of the subsystems are conserved, and there exists no net energy transfer, on average, between them. This means that part (Z0) of the zeroth law is always satisfied for systems in thermal contact, if the joint system is described by the MCE.

\par
To test whether or not part (Z1) also holds for a given entropy definition, it suffices to consider a compound system  that consists of two thermally coupled equilibrium systems. Let us therefore consider two initially isolated systems $\mcal{A}$ and $\mcal{B}$ with Hamiltonians $H_\mcal{A}(\vpq_\mcal{A})$ and $H_\mcal{B}(\vpq_\mcal{B})$, DoS $\SubDoS{\mcal{A}}\ge 0$ and $\SubDoS{\mcal{B}}\ge 0$ such that $\DoS_{\mcal{A},\mcal{B}}(E_{\mcal{A},\mcal{B}} < 0) = 0$, and denote the integrated DoS by $\Omega_\mcal{A}$ and $\Omega_\mcal{B}$. Before the coupling, the systems have fixed energies $\SubEinit{\mcal{A}}$ and~$\SubEinit{\mcal{B}}$, and each of the systems can be described by a microcanonical density operator,
\be
\DOp_i(\vpq_i|E_i)=\f{\DiracDelta\left[E_i - H_i(\vpq_i) \right]}{\DoS_i(E_i)},
\qquad 
i=\mcal{A},\mcal{B}.
\ee
In this pre-coupling state, one can compute for each system separately the various entropies $S_i(E_i)$ and temperatures $T_i(E_i)$ introduced in Sec.~\ref{sec:mce:entropy_candidates}.

\par
Let us further assume that the systems are brought into (weak) thermal contact and given a sufficiently long time to equilibrate. The two systems now form a joint systems $\mcal{AB}$ with microstates $\vpq=(\vpq_\mcal{A},\vpq_\mcal{B})$, Hamiltonian $H(\vpq)=H_\mcal{A}(\vpq_\mcal{A})+H_\mcal{B}(\vpq_\mcal{B})$ and conserved total energy~\mbox{$E= \SubEinit{\mcal{A}}+\SubEinit{\mcal{B}}=H(\vpq)$}. The microcanonical density operator of the new joint equilibrium  system  reads\footnote{Considering weak coupling, we formally neglect interaction terms in the joint Hamiltonian but assume nevertheless that the coupling interactions are still sufficiently strong to create mixing. }
\bse\label{eq:MCE_equilibrium}
\be
\DOp(\vpq|E)=\f{\DiracDelta\left[E - H(\vpq) \right]}{\DoS(E)},
\ee
where the joint DoS $\DoS$ is given by the convolution (see App.~\ref{app:sec:compound_system_density})
\begin{equation}
\label{eq:convolution_dos}
\begin{split}
\DoS(E) &=
\int_0^{\infty} \!\!\diff{\SubE{\mcal{A}}'}
\int_0^{\infty} \!\!\diff{\SubE{\mcal{B}}'}
\,
\SubDoS{\mcal{A}}(\SubE{\mcal{A}}')\,\SubDoS{\mcal{B}}(\SubE{\mcal{B}}')
\\&\quad\times
\DiracDelta(E - \SubE{\mcal{A}}' - \SubE{\mcal{B}}')
\\&=
\int_0^{E} \!\!\diff{\SubE{\mcal{A}}'}\, \SubDoS{\mcal{A}}(\SubE{\mcal{A}}')\,\SubDoS{\mcal{B}}(E - \SubE{\mcal{A}}')
.
\end{split}
\end{equation}
The associated integrated DoS $\Omega$ takes the form
\begin{equation}
\label{eq:convolution_idos}
\begin{split}
\IDoS(E) &=
\int_0^{\infty} \!\!\diff{\SubE{\mcal{A}}'}
\int_0^{\infty} \!\!\diff{\SubE{\mcal{B}}'}
\,
\SubDoS{\mcal{A}}(\SubE{\mcal{A}}')\,\SubDoS{\mcal{B}}(\SubE{\mcal{B}}')
\\&\quad\times
\Heaviside(E - \SubE{\mcal{A}}' - \SubE{\mcal{B}}')
\\&=
\int_0^{E} \!\!\diff{\SubE{\mcal{A}}'}\, \SubIDoS{\mcal{A}}(\SubE{\mcal{A}}')\,\SubDoS{\mcal{B}}(E-\SubE{\mcal{A}}')
.
\end{split}
\end{equation}
If $\lim_{\SubE{\mcal{A}}'\searrow 0}\SubDoS{\mcal{A}}(\SubE{\mcal{A}}')  = \SubDoS{\mcal{A}}(0^+) < \infty$, the differential DoS \mbox{$\DDoS=\p \DoS/\p E$} can be expressed as 
\begin{equation}
\label{eq:convolution_ddos}
\begin{split}
\DDoS(E) &= \int_0^{E} \!\!\diff{\SubE{\mcal{A}}'}\, \SubDDoS{\mcal{A}}(\SubE{\mcal{A}}')\,\SubDoS{\mcal{B}}(E-\SubE{\mcal{A}}')
\\&\quad +
 \SubDoS{\mcal{A}}(0^+)\, \SubDoS{\mcal{B}}(E)
.
\end{split}
\end{equation}
\ese
Note that Eq.~\eqref{eq:convolution_ddos} is not applicable if $\SubDoS{i}(\SubE{i}')$ diverges near $\SubE{i}' = 0$ for $i\in \{\mcal{A},\mcal{B}\}$.

\par
Since the joint system $\mcal{AB}$ is also described by the MCE, we can again directly compute any of the entropy definitions $S(E)$ introduced in Sec.~\ref{sec:mce:entropy_candidates} to obtain the associated temperature $T=(\p S/\p E)^{-1}$ of the compound system as function of the total energy $E$. 

%%%%%%%%%%%%%%%%%%%%%%%%%%%%%%%%
\subsection{Subsystem energies in the MCE}
\label{sec:zeroth_law:sub_energy}
%%%%%%%%%%%%%%%%%%%%%%%%%%%%%%%%

When in thermal contact, the subsystems with fixed external control parameters can permanently exchange energy, and their subsystem energies $E_i'=H_i(\vpq_i)$, with $i\in\/\mcal{A},\mcal{B}$, are fluctuating quantities. According to Eq.~\eqref{eq:mce:pdf_observable}, the probability distributions of the subsystem energies $E_i'$ for a given, fixed total energy $E$ are defined by:
\be
\label{eq:energy_dist_sub}
\pdfE{i}(E_i'|E)
= \pdf{H_i}(E_i'|E)
%=\EV{\delta(E_i-H_i)}_E 
=\Tr[\DOp{}\,\DiracDelta(E_i'-H_i)].
%=
%\Tr[\DiracDelta(E_i'-H_i(\vpq_i))\,\pdf(\vpq|E)].
%&=&
%\Tr_i[\DiracDelta(E_i'-H_i(\vpq_i))\,\mu_i(\vpq_i|E)]
\ee
From a calculation similar to that in Eq.~\eqref{eq:convolution_dos}, see App.~\ref{app:sec:compound_system_density}, one finds  for subsystem $\mcal{A}$  
\be\label{eq:energy_dist_sub_A}
\pdfE{\mcal{A}}(\SubE{\mcal{A}}'|E) =\f{\SubDoS{\mcal{A}}(\SubE{\mcal{A}}')\, \SubDoS{\mcal{B}}(E-\SubE{\mcal{A}}')}{\DoS(E)}.
\ee 
The energy density $\pdfE{\mcal{B}}(\SubE{\mcal{B}}'|E)$ of subsystem $\mcal{B}$ is obtained by exchanging labels~$\mcal{A}$ and~$\mcal{B}$ in Eq.~\eqref{eq:energy_dist_sub_A}.

\par
The conditional energy distribution $\pdfE{i}(E_i|E)$ can be used to compute expectation values $\EV{F}_E$ for quantities $F=F\bigl(H_i(\vpq_i)\bigr)$ that depend on the the system state $\vpq$ only through the subsystem energy $H_i$:
\begin{equation}
\label{eq:mean_for_functions_of_subsystem_energy}
  \EV{F(H_i)}_E = \int_0^{E}\!\!\diff[]{\SubE{i}'}\,\pdfE{i}(\SubE{i}'|E)\, F(\SubE{i}').
\end{equation}
For example, the mean energy of system $\mcal{A}$ after contact is given by
\begin{equation}
  \EV{H_\mcal{A}}_E = \int_0^{E}\!\!\diff[]{\SubE{\mcal{A}}'}\f{\SubDoS{\mcal{A}}(\SubE{\mcal{A}}')\, \SubDoS{\mcal{B}}(E-\SubE{\mcal{A}}')}{\DoS(E)} \SubE{\mcal{A}}'.
\end{equation}
Since the total energy $E={\SubEinit{\mcal{A}}+\SubEinit{\mcal{B}}}$ is conserved, the heat flow (mean energy transfer) between system $\mcal{A}$ and $\mcal{B}$ during thermalization can be computed as
\begin{equation}
\label{eq:heat_flow_definition}
  \Heat_{\mcal{A}\to\mcal{B}}(\SubEinit{\mcal{A}},\SubEinit{\mcal{B}}) = \SubEinit{\mcal{A}} - \EV{\SubH{\mcal{A}}}_{\SubEinit{\mcal{A}}+\SubEinit{\mcal{B}}}
.
\end{equation}
This equation implies that the heat flow is governed by the primary state variable energy rather than temperature.

%%%%%%%%%%%%%%%%%%%%%%%%%%%%%%%%
\subsection{Subsystem temperatures in the MCE}
\label{sec:zeroth_law:sub_temp}
%%%%%%%%%%%%%%%%%%%%%%%%%%%%%%%%

Verification of temperature amendment (Z1) requires an extension of the microcanonical temperature concept, as one needs to define subsystem temperatures first. The energy $\SubE{i}$  of a subsystem is subject to statistical fluctuations, precluding a direct application of the microcanonical entropy and temperature definitions. One can, however, compute subsystem entropies and temperatures for fixed subsystem energies $\SubE{i}$, by virtually decoupling the subsystem from the total system. In this case, regardless of the adopted definition, the  entropy of the decoupled subsystem is simply given by $S_i(\SubE{i})$, and the associated subsystem temperature $\SubTemp{i}(\SubE{i}) = [\p \SubEnt{i}(\SubE{i})/\p \SubE{i}]^{-1}$ is a function of the subsystem's energy $\SubE{i}$.

\par
We can then generalize the microcanonical subsystem temperature $\SubTemp{i}(\SubE{i})$, defined for a fixed subsystem energy $\SubE{i}$, by considering a suitably chosen microcanonical average~$\EV{\SubTemp{i}(\SubE{i})}_E$. A subsystem temperature average that is consistent with the general formula~\eqref{eq:mean_for_functions_of_subsystem_energy} reads
\begin{equation}
\label{eq:df_mean_sub_temp}
\EV{\SubTemp{i}(\SubE{i})}_E = \int_0^{E}\!\! \diff[]{\SubE{i}'}\,\pdfE{i}(\SubE{i}'|E)\, \SubTemp{i}(\SubE{i}').
\end{equation}
With this convention, the amendment (Z1) to the zeroth law takes the form
\be
\label{eq:zeroth_law_ii}
\EV{\SubTemp{i}(\SubE{i}')}_E \overset{!}{=} T(E),
\ee 
which can be tested for the various entropy candidates. 

\par
One should emphasize that Eq.~\eqref{eq:df_mean_sub_temp} implicitly assumes that the temperature $\SubTemp{i}(\SubE{i}')$ of the subsystem is well-defined for all energy values $\SubE{i}'$ in the integration range $[0,E]$, or at least for all $\SubE{i}'$ where $\pdfE{i}(\SubE{i}'|E) > 0$. The more demanding assumption that $\SubTemp{i}(\SubE{i}')$ is well defined for all $\SubE{i} \in [0,E]$ is typically not satisfied if the subsystem DoS has extended regions (band gaps) with $\SubDoS{i}(\SubE{i}')=0$  in the range $[0,E]$. The weaker assumption of a well defined subsystem temperature for energies with non-vanishing probability density may be violated, for example, for the Boltzmann temperature of subsystems exhibiting stationary points $E_i^*$ (e.g. maxima) with $\nu_i(\SubE{i}^*)=\SubDoS{i}'(\SubE{i}^*)=0$ in their DoS, in which case the mean subsystem Boltzmann temperature is ill-defined,  even if the Boltzmann temperature of the compound system is well defined and finite.

%%%%%%%%%%%%%%%%%%%%%%%%%%%%%%%%%%%%%%%%%%%%%%%%%
\subsubsection{Gibbs temperature}
\label{sec:zeroth_law:Gibbs_temperature}
%%%%%%%%%%%%%%%%%%%%%%%%%%%%%%%%%%%%%%%%%%%%%%%%%

We start by verifying Eq.~\eqref{eq:zeroth_law_ii} for the Gibbs entropy. To this end, we consider two systems $\mcal{A}$ and $\mcal{B}$ that become weakly coupled to form an isolated joint system~$\mcal{AB}$. The Gibbs temperatures of the subsystems before coupling are 
\begin{equation}
\label{eq:T_G_initial}
\SubTempGinit{i} =  \SubTempG{i}(\SubEinit{i}) = \frac{\SubIDoS{i}(\SubEinit{i})}{\SubDoS{i}(\SubEinit{i})},
\qquad i=\mcal{A},\mcal{B}.
\end{equation}
The Gibbs temperature of the combined system after coupling is
\be
\label{eq:T_G_total}
\TempG(E)=\f{\IDoS(E)}{\DoS(E)}
\ee
with $E = \SubEinit{\mcal{A}} + \SubEinit{\mcal{B}}$, and $\IDoS$ and $\DoS$ given in Eqs.~\eqref{eq:MCE_equilibrium}. Using the expression~\eqref{eq:mce:df_Gibbs_temperature}, the subsystem temperature $\SubTempG{\mcal{A}}$ for subsystem energy $\SubE{\mcal{A}}'$,
\begin{equation}
\label{eq:T_G_subsystem}
\SubTempG{\mcal{A}}(\SubE{\mcal{A}}') = \frac{\SubIDoS{\mcal{A}}(\SubE{\mcal{A}}')}{\SubDoS{\mcal{A}}(\SubE{\mcal{A}}')},
\end{equation}
which requires $\SubDoS{\mcal{A}}(\SubE{\mcal{A}}') > 0$ to be well defined. Assuming $\SubDoS{\mcal{A}}(\SubE{\mcal{A}}') > 0$ for all $\SubE{\mcal{A}}' \in (0,E)$ and making use of Eqs.~\eqref{eq:energy_dist_sub_A}, \eqref{eq:df_mean_sub_temp}, 
\eqref{eq:convolution_idos} and  \eqref{eq:T_G_total}, one finds that
\begin{equation}
\label{eq:zeroth_law_ii_Gibbs}
\begin{split}
&\quad 
\EV{\SubTempG{\mcal{A}}(\SubE{\mcal{A}}')}_E
\\&=
 \int_0^{E}\!\!\diff[]{\SubE{\mcal{A}}'}\,
 \frac{\SubDoS{\mcal{A}}(\SubE{\mcal{A}}') \, \SubDoS{\mcal{B}}(E - \SubE{\mcal{A}}')}{\DoS(E)}
 % \\&\quad\times
 \frac{\SubIDoS{\mcal{A}}(\SubE{\mcal{A}}')}{\SubDoS{\mcal{A}}(\SubE{\mcal{A}}')}\,
\\&=
\f{1}{\DoS(E)} 
% \\&\quad\times
\int_{0}^{E}\!\!\diff[]{\SubE{\mcal{A}}'}\,
\SubIDoS{\mcal{A}}(\SubE{\mcal{A}}')\, \SubDoS{\mcal{B}}(E - \SubE{\mcal{A}}')
\\&=
\TempG(E)
.
\end{split}
\end{equation}
By swapping labels $\mcal{A}$ and $\mcal{B}$, one obtains an analogous result for system $\mcal{B}$. Given that our choice of $\mcal{A}$ and $\mcal{B}$ was arbitrary, Eq.~\eqref{eq:zeroth_law_ii_Gibbs} implies that \emph{the Gibbs temperature satisfies part (Z1) of the zeroth law\footnote{\highlightchange{For clarity, consider any three thermally coupled subsystems $\mcal{A}_1,\mcal{A}_2,\mcal{A}_3$ and assume their DoS does not vanish for positive energies. In this case, Eq.~\eqref{eq:zeroth_law_ii_Gibbs} implies that $\EV{\SubTempG{\mcal{A}_1}}_E=T(E)$ and $\EV{\SubTempG{\mcal{A}_2}}_E=T(E)$ and $\EV{\SubTempG{\mcal{A}_3}}_E=T(E)$ and, therefore, $\EV{\SubTempG{\mcal{A}_1}}_E=\EV{\SubTempG{\mcal{A}_2}}_E=\EV{\SubTempG{\mcal{A}_3}}_E$, in agreement with the zeroth law.}} if the DoS  of the subsystems do not vanish for positive energies.}

\par
For classical Hamiltonian many-particle systems, the temperature equality~\eqref{eq:zeroth_law_ii_Gibbs} was, in fact, already discussed by Gibbs, see Chap.~X and his remarks below Eq. (487) in Chap. XIV in Ref.~\cite{Gibbs}. For such systems, one may arrive at the same conclusion by considering the equipartition theorem~\eqref{eq:mce:equipartition_theorem}. If the equipartition theorem holds, it ensures that $\EV{\pq_i \p H_\mcal{A} / \p \pq_i}_{H_\mcal{A}=\SubE{\mcal{A}}'} = \SubTempG{\mcal{A}}(\SubE{\mcal{A}}')$ for all microscopic degrees $i$ that are part of the subsystem $\mcal{A}$. Upon averaging over possible values of the subsystem energy $\SubE{\mcal{A}}'$, one obtains\footnote{We are assume, as before, weak coupling, $H=H_\mcal{A} +H_\mcal{B}$.} (see App.~\ref{app:sec:zero_equi})
\begin{equation} 
\label{eq:zeroth_law_equi}
\begin{split}
&\quad
\EV{ \SubTempG{\mcal{A}}(\SubE{\mcal{A}}')}_E 
\\
&=
\int_0^E\!\!\diff[]{\SubE{\mcal{A}}'} \, \pdfE{\mcal{A}}(\SubE{\mcal{A}}'|E) \EV{\!\pq_i \f{\p H_\mcal{A}}{\p \pq_i}\!}_{\!\!H_\mcal{A}=\SubE{\mcal{A}}'\!\!\!\!}
\\&
=
 \EV{\pq_i \f{\p H_\mcal{A}}{\p \pq_i}}_{E}
\\&= \TempG(E).
\end{split}
\end{equation}
Thus, for these classical systems and the Gibbs temperature, the zeroth law can be interpreted as a consequence of equipartition\footnote{For certain systems, such as those with energy gaps or upper energy bounds, $\SubDoS{\mcal{A}}(\SubE{\mcal{A}}')$ may vanish for a substantial part of the available energy range $0 < \SubE{\mcal{A}}' < E$. Then it may be possible that $\EV{\SubTempG{\mcal{A}}(\SubE{\mcal{A}}')}_E < \TempG(E)$; see Sec.~\ref{sec:examples:bounded_dos} for an example. For classical systems, this usually implies that the equipartition theorem~\eqref{eq:mce:equipartition_theorem} does not hold, and that at least one of the conditions for equipartition fails.}.

%%%%%%%%%%%%%%%%%%%%%%%%%%%%%%%%%%%%%%%%%%%%%%%%%
\subsubsection{Boltzmann temperature}
\label{sec:zeroth_law:Boltzmann_temperature}
%%%%%%%%%%%%%%%%%%%%%%%%%%%%%%%%%%%%%%%%%%%%%%%%%

We now perform a similar test for the Boltzmann entropy. The Boltzmann temperatures of the subsystems before coupling are 
\bse\label{eq:boltz_zero}
\begin{equation}
\SubTempB{i}(\SubEinit{i}) = \frac{\SubDoS{i}(\SubEinit{i})}{\SubDDoS{i}(\SubEinit{i})},
\qquad i=\mcal{A},\mcal{B},
\end{equation}
and the Boltzmann temperature of the combined system after coupling is
\be
\TempB(E)=\f{\DoS(E)}{\nu(E)}
\ee
\ese
with $\DoS$ and $\nu$ given in Eqs.~\eqref{eq:MCE_equilibrium}, and $E = \SubEinit{\mcal{A}} + \SubEinit{\mcal{B}}$. 
Assuming as before $\SubDoS{\mcal{A}}(\SubE{\mcal{A}})>0$ for all $0 < \SubE{\mcal{A}} < E$, we find 
\begin{equation}
\label{eq:zeroth_law_ii_no_Boltzmann}
%\label{eq:boltzmann_average}
\begin{split}
&\quad
\EV{ \SubTempB{\mcal{A}}(\SubE{\mcal{A}}')}_E 
\\&=
 \int_0^{E}\!\!\diff[]{\SubE{\mcal{A}}'}\,
 \frac{\SubDoS{\mcal{A}}(\SubE{\mcal{A}}') \SubDoS{\mcal{B}}(E - \SubE{\mcal{A}}')}{\DoS(E)}
 % \\&\quad\times
 \frac{\SubDoS{\mcal{A}}(\SubE{\mcal{A}}')}{\SubDDoS{\mcal{A}}(\SubE{\mcal{A}}')}\,
\\& \neq
\TempB(E).
\end{split}
\end{equation}
This shows that the mean Boltzmann temperature does \emph{not} satisfy the zeroth law (Z1).

\par
Instead, the first line in Eq.~\eqref{eq:zeroth_law_ii_no_Boltzmann}, combined with Eq.~\eqref{eq:convolution_ddos}, suggests that the Boltzmann temperature satisfies the following relation for the~\emph{inverse} temperature (see Chap.~X in Ref.~\cite{Gibbs} for a corresponding proof for classical $N$-particle systems):
\begin{equation}
\label{eq:zeroth_law_ii_inverse_Boltzmann}
\EV{\SubTempB{\mcal{A}}^{-1}(\SubE{\mcal{A}}')}^{-1} = \TempB(E),
\end{equation}
if  $\SubDoS{\mcal{A}}(\SubE{\mcal{A}}')>0$ for all $0 < \SubE{\mcal{A}}' < E$, and moreover $\SubDoS{\mcal{A}}(0) = 0$ and continuous\footnote{The second condition is crucial. In contrast, in certain cases, the first condition may be violated, while Eq.~\eqref{eq:zeroth_law_ii_inverse_Boltzmann} still holds.}.
Note that this equation is not consistent with the definition~\eqref{eq:mean_for_functions_of_subsystem_energy} of expectation values for the temperature itself, and therefore also disagrees with the zeroth law as stated in Eq.~\eqref{eq:zeroth_law_ii}. One may argue, however, that Eq.~\eqref{eq:zeroth_law_ii_inverse_Boltzmann} is consistent with the definition~\eqref{eq:mean_for_functions_of_subsystem_energy} for $\beta_\mathrm{B} = 1 / \TempB$.

\par
It is sometimes argued that the Boltzmann temperature characterizes the most probable energy state $E^*_i$ of a subsystem $i$ and that the corresponding temperature values ${\TempB}_{i}(E^*_i)$ coincides with the temperature of the compound system $\TempB(E)$. To investigate this statement, consider $i=\mcal{A}$ and recall that the probability $\pdfE{\mcal{A}}(\SubE{\mcal{A}}|E)$ of finding the first subsystem $\mcal{A}$ at energy $\SubE{\mcal{A}}$ becomes maximal either at a non-analytic point (e.g., a boundary value of the allowed energy range), or at a value $\SubEML{\mcal{A}}$ satisfying
\begin{equation}
 0 = 
 \f{\p \pdfE{\mcal{A}}(\SubE{\mcal{A}}|E)}{\p \SubE{\mcal{A}} }
 \biggr|_{\SubE{\mcal{A}} = \SubEML{\mcal{A}} } .
\end{equation}
Inserting $\pdfE{\mcal{A}}(\SubE{\mcal{A}}|E)$ from Eq.~\eqref{eq:energy_dist_sub_A}, one thus finds 
\begin{equation}
 \SubTempB{\mcal{A}}(\SubEML{\mcal{A}}) = \SubTempB{\mcal{B}}(E - \SubEML{\mcal{A}}).
\end{equation} 
Note, however, that in general
\begin{equation}
 \TempB(E) \neq \SubTempB{\mcal{A}}(\SubEML{\mcal{A}}) = \SubTempB{\mcal{B}}(E - \SubEML{\mcal{A}}),
\end{equation}
with the values~$\SubTempB{i}(\SubEML{i})$ usually depending on the specific decomposition into subsystems (see Sec.~\ref{sec:examples:polynomial_dos} for an example). This shows that the Boltzmann temperature $\TempB$ is in general~\emph{not} equal to the \lq most probable\rq~Boltzmann temperature~$\SubTempB{i}(\SubEML{i})$ of an arbitrarily chosen subsystem.

%%%%%%%%%%%%%%%%%%%%%%%%%%%%%%%%
\subsubsection{Other temperatures}
\label{sec:zeroth_law:other_temperatures}
%%%%%%%%%%%%%%%%%%%%%%%%%%%%%%%%
It is straightforward to verify through analogous calculations that, similar to the Boltzmann temperature, the temperatures derived from the other entropy candidates in Sec.~\ref{sec:mce:entropy_candidates} violate the zeroth law as stated in Eq.~\eqref{eq:zeroth_law_ii} for systems with non-vanishing $\SubDoS{i}(\SubE{i}>0)$. Only for certain systems with upper energy bounds, one finds that the complementary Gibbs entropy satisfies Eq.~\eqref{eq:zeroth_law_ii} for energies close to the highest permissible energy (see example in Sec.~\ref{sec:examples:bounded_dos}).

%%%%%%%%%%%%%%%%%%%%%%%%%%%%%%%%%%%%%%%%%%%%%%%%%
\section{First law}
\label{sec:first_law}
%%%%%%%%%%%%%%%%%%%%%%%%%%%%%%%%%%%%%%%%%%%%%%%%%

The first law of thermodynamics is the statement of energy conservation. That is, any change in the internal energy $dE$ of an isolated system is caused by heat transfer~$\DiracDelta Q$ from or into the system and external work $\DiracDelta A$ performed on or by the system,
\be
dE
&=&\DiracDelta Q+\DiracDelta A
\notag\\
&=&T \,dS - \sum_{n} p_n dZ_n,
\label{eq:first_law}
\ee
where the $p_n$ are the generalized pressure variables that characterize the energetic response of the system to changes in the control parameters $Z$. Specifically, pure work~$\DiracDelta A$ corresponds to an adiabatic variation of the parameters $Z=(Z_1,\ldots)$ of the Hamiltonian $H(\vpq;Z)$. Heat transfer~$\DiracDelta Q=T dS$ comprises all other forms of energy exchange (controlled injection or release of photons, etc.). Subsystems within the isolated system can permanently exchange heat although the total energy remains conserved in such internal energy redistribution processes.

\par
The formal differential relation~\eqref{eq:first_law} is trivially satisfied for~\emph{all} the entropies listed in Sec.~\ref{sec:mce:entropy_candidates}, if the generalized pressure variables are defined by:
\begin{equation}
\label{eq:consistency_formal}
\Press_j=
T \left(\f{\p S}{\p Z_j}\right)_{E,Z_n\neq Z_j}.
\end{equation}
Here, subscripts indicate quantities that are kept constant during differentiation. However, this formal definition does not ensure that the  abstract thermodynamic quantities $\Press_j$ have any relation to the relevant statistical quantities measured in an experiment. To obtain a meaningful theory,  the generalized pressure variables $\Press_j$ must be connected with the corresponding microcanonical expectation values. This requirement leads to the consistency relation
\begin{equation}
\label{eq:consistency_general}
\Press_j=
T \left(\f{\p S}{\p Z_j}\right)_{E,Z_n\neq Z_j}
%\overset{!}{=}
%-\left(\f{\p E}{\p Z_j}\right){S,Z_n\neq Z_j}
\overset{!}{=}
-\left\lan\f{\p H }{\p Z_j}\right\ran_E,
\end{equation}
which can be derived from the Hamiltonian or Heisenberg equations of motion (see, e.g., Supplementary Information of Ref.~\cite{2014DuHi_NatPhys}).
Equation~\eqref{eq:consistency_general} is physically relevant as it ensures that abstract thermodynamic observables agree with the statistical averages and measured quantities.

\par
As discussed in Ref.~\cite{2014DuHi_NatPhys}, any function of $\IDoS(E)$ satisfies Eq.~\eqref{eq:consistency_general}, implying that the Gibbs entropy, the complementary Gibbs entropy and the alternative proposals~$\Ent_P$ and $S_\mrm{G\vee C}$ are thermostatistically consistent with respect to this specific criterion. By contrast, the Boltzmann entropy $\EntB=\ln(\eps \DoS)$ violates Eq.~\eqref{eq:consistency_general} for finite systems of arbitrary size~\cite{2014DuHi_NatPhys}. The fact that, for isolated classical $N$-particle systems, the Gibbs entropy satisfies the thermodynamic relations for the empirical thermodynamic entropy exactly, whereas the Boltzmann entropy works only approximately, was already pointed out by Gibbs\footnote{Gibbs states on p.~179 in Ref.~\cite{Gibbs}: \lq\lq It would seem that in general averages are the most important, and that they lend themselves better to analytical transformations. This consideration would give preference to the system of variables in which $\log V$ [$=\EntG$ in our notation] is the analogue of entropy. Moreover, if we make $\phi$  [$=\EntB$ in our notation] the analogue of entropy, we are embarrassed by the necessity of making numerous exceptions for systems of one or two degrees of freedoms.\rq\rq
}  (Chap.~XIV in Ref.~\cite{Gibbs}) and Hertz~\cite{1910Hertz_1,1910Hertz_2}.
% (see Chap.~XIV in~Ref.~\cite{Gibbs}) and Hertz~\cite{1910Hertz_1,1910Hertz_2}.

\par
The above general statements can be illustrated with a very simple example already discussed by~Hertz~\cite{1910Hertz_2}. 
Consider a single classical molecule\footnote{Such an experiment could probably be performed nowadays using a suitably designed atomic trap.}
 moving with energy $E>0$ in the one-dimensional interval $[0,L]$.  This system is trivially ergodic with $\IDoS = a L E^{1/2}$ and $\DoS=a L/(2E^{1/2})$, where $a$ is a constant of proportionality that is irrelevant for our discussion. From the Gibbs entropy~$\EntG$, one obtains the temperature $\kB \TempG=2E>0$ and pressure $\PressG = 2E/L>0$, whereas the Boltzmann entropy~$\EntB$ yields $\kB \TempB=-2E<0 $ and  $\PressB =-2E/L<0$. Now, clearly, the kinetic force exerted by a molecule on the boundary is positive (outwards directed), which means that the pressure predicted by $\EntB$ cannot be correct. The failure of the Boltzmann entropy is a consequence of the general fact that, unlike the Gibbs entropy, $\EntB$  is not an adiabatic invariant~\cite{1910Hertz_1,1910Hertz_2}.  More generally, if one chooses to adopt non-adiabatic entropy definitions, but wants to maintain  the energy balance, then one must assume that heat and entropy is generated or destroyed in mechanically adiabatic processes. This, however, would imply that for mechanically adiabatic and reversible processes, entropy is not conserved, resulting in a violation of the second law.

%%%%%%%%%%%%%%%%%%%%%%%%%%%%%%%%%%%%%%%%%%%%%%%%%
%\pagebreak
\section{Second law}
\label{sec:second_law}
%%%%%%%%%%%%%%%%%%%%%%%%%%%%%%%%%%%%%%%%%%%%%%%%%

%%%%%%%%%%%%%%%%%%%%%%%%%%%%%%%%%%%%%%%%%%%%%%%%%
\subsection{Formulations}
\label{sec:second_law:versions}
%%%%%%%%%%%%%%%%%%%%%%%%%%%%%%%%%%%%%%%%%%%%%%%%%

The second law of thermodynamics concerns the non-decrease of entropy under rather general conditions. This law is sometimes stated in ambiguous form, and several authors appear to prefer different non-equivalent versions. Fortunately, in the case of isolated systems, it is relatively straightforward to identify a meaningful minimal version of the second law -- originally proposed by Planck~\cite[][]{PlanckBook1903} -- that imposes a testable constraint on the microcanonical entropy candidates. However, before focussing on Planck's formulation, let us briefly address two other rather popular versions that are \emph{not} feasible when dealing with isolated systems. 

\par
The perhaps simplest form of the second law states that the entropy of an isolated system never decreases. For isolated systems described by the MCE, this statement is meaningless, because the entropy $S(E,Z)$ of an isolated equilibrium system at fixed energy $E$ and fixed  control parameters $Z$ is constant regardless of the chosen entropy definition. 

\par
Another frequently encountered version of the second law, based on a simplification of Clausius' original statement~\cite{Clausius1854}, asserts that heat never flows spontaneously from a colder to a hotter body. As evident from the simple yet generic example in Sec.~\ref{sec:mce:non-unique_T}, the microcanonical temperature~$T(E)$ can be a non-monotonic or even oscillating function of energy and, therefore, temperature differences do not suffice to specify the direction of heat flow when two initially isolated systems are brought into thermal contact with each other.

\par
The deficiencies of the above formulations can be overcome by resorting to Planck's version of the second law. Planck postulated that the sum of entropies of all bodies taking any part in some process never decreases (p.~100 in Ref.~\cite[][]{PlanckBook1903}).\footnote{\highlightchange{Planck~\cite[][]{PlanckBook1903} regarded this as the most general version of the second law.}} This formulation is useful as it allows one to test the various microcanonical entropy definitions, e.g. in thermalization processes. More precisely, if $\mcal{A}$ and $\mcal{B}$ are two isolated systems with fixed energy values $\SubEinit{\mcal{A}}$ and $\SubEinit{\mcal{B}}$ and fixed entropies $\SubEnt{\mcal{A}}(\SubEinit{\mcal{A}})$ and $\SubEnt{\mcal{B}}(\SubEinit{\mcal{B}})$ before coupling, then the entropy of the compound system after coupling, $\Ent(\SubEinit{\mcal{A}}+\SubEinit{\mcal{B}})$ must be \highlightchange{equal or} larger than the sum of the initial entropies,
\be\label{eq:second_law}
\Ent(\SubEinit{\mcal{A}}+\SubEinit{\mcal{B}})
\ge 
\SubEnt{\mcal{A}}(\SubEinit{\mcal{A}}) +
\SubEnt{\mcal{B}}(\SubEinit{\mcal{B}}).
\ee
At this point, it may be useful recall that, before the coupling,  the two independent systems are described by the density operators $\gr_\mcal{A}=\gd(H_\mcal{A}-E_\mcal{A})/\go_\mcal{A}(E_\mcal{A})$ and \mbox{$\gr_\mcal{B}=\gd(H_\mcal{B}-E_\mcal{B})/\go_\mcal{B}(E_\mcal{B})$} corresponding the joint density operator  $\gr_{\mcal{A}\cup \mcal{B}}=\gr_\mcal{A}\cdot\gr_\mcal{B}$, whereas after the coupling their joint density operator is given by \mbox{$\gr_{\mcal{A}\mcal{B}}= \gd[(H_\mcal{A}+H_\mcal{B})-(E_\mcal{A}+E_\mcal{B})]/\go_\mcal{AB}(E_\mcal{A}+E_\mcal{B})$}. The transition from the product distribution $\gr_{\mcal{A}\cup \mcal{B}}$ to the coupled distribution $\gr_\mcal{AB}$ is what is formally meant by equilibration after coupling. 

\par
We next analyze whether the inequality~\eqref{eq:second_law} is fulfilled by the microcanonical entropy candidates introduced in Sec.~\ref{sec:mce:entropy_candidates}.

%%%%%%%%%%%%%%%%%%%%%%%%%%%%%%%%%%%%%%%%%%%%%%%%%
\subsection{Gibbs entropy}
\label{sec:second_law:Gibbs_entropy}
%%%%%%%%%%%%%%%%%%%%%%%%%%%%%%%%%%%%%%%%%%%%%%%%%

%================================================
\begin{figure}
%\centerline{\includegraphics[width=1\linewidth]{Gibbs_second_law}}
\centerline{\includegraphics[width=1\linewidth]{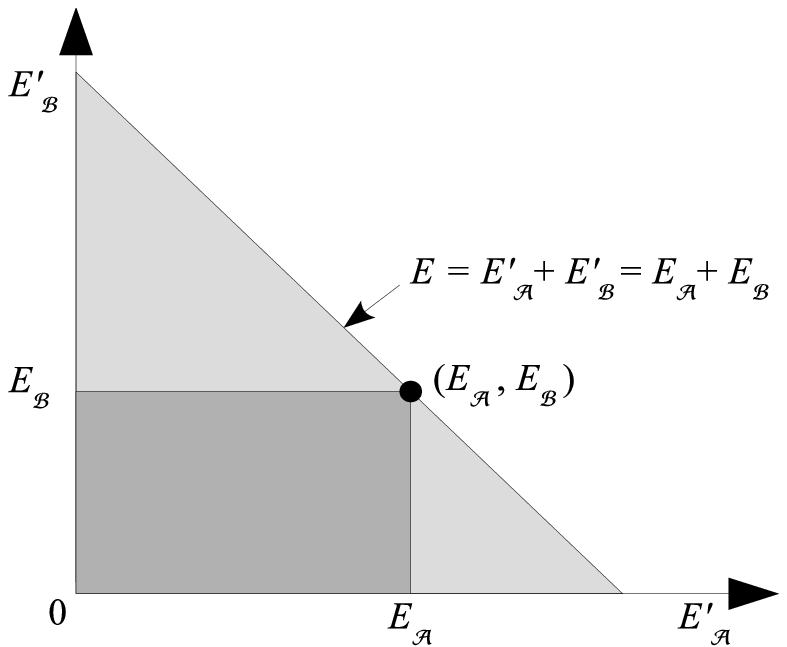}}
\caption{
\label{fig:Gibbs_second_law}
The phase volume $\IDoS(E)$ of a system composed of two subsystems $\mcal{A}$ and $\mcal{B}$ with initial energies $\SubEinit{\mcal{A}}$ and $\SubEinit{\mcal{B}}$ can be computed by integrating the product $\SubDoS{\mcal{A}}(\SubE{\mcal{A}}')\SubDoS{\mcal{B}}(\SubE{\mcal{B}}')$ of the subsystem densities $\SubDoS{\mcal{A}}$ and $\SubDoS{\mcal{B}}$ over the region bounded by the $\SubE{\mcal{A}}'+\SubE{\mcal{B}}' = \SubEinit{\mcal{A}}+\SubEinit{\mcal{B}} = E$-line in the $(\SubE{\mcal{A}}', \SubE{\mcal{B}}')$-plane (light and dark gray region). The product $\SubIDoS{\mcal{A}}(\SubEinit{\mcal{A}})\SubIDoS{\mcal{B}}(\SubEinit{\mcal{B}})$ of the phase volumes of the systems $\mcal{A}$ and $\mcal{B}$ before coupling is computed by integrating the same function $\SubDoS{\mcal{A}}(\SubE{\mcal{A}}')\SubDoS{\mcal{B}}(\SubE{\mcal{B}}')$, but now over the smaller region bounded by the lines of $\SubE{\mcal{A}}'=\SubEinit{\mcal{A}}$ and  $\SubE{\mcal{B}}'=\SubEinit{\mcal{B}}$ (dark gray region).
}
\end{figure}
%================================================

To verify Eq.~\eqref{eq:second_law} for the Gibbs entropy $\EntG=\ln\IDoS$, we have to compare the phase volume of the compound systems after coupling, $\IDoS(\SubEinit{\mcal{A}}+\SubEinit{\mcal{B}})$, with the phase volumes $\SubIDoS{\mcal{A}}(\SubEinit{\mcal{A}})$ and $\SubIDoS{\mcal{B}}(\SubEinit{\mcal{B}})$ of the subsystems before coupling. Starting from Eq.~\eqref{eq:convolution_idos}, we find (also see Fig.~\ref{fig:Gibbs_second_law})
\begin{equation}
\label{eq:Gibbs_second_law}
\begin{split}
&
\IDoS(\SubEinit{\mcal{A}}+\SubEinit{\mcal{B}})
 \\&=
\int_0^{\infty} \!\!\diff{\SubE{\mcal{A}}'}
\int_0^{\infty} \!\!\diff{\SubE{\mcal{B}}'}
\,
\SubDoS{\mcal{A}}(\SubE{\mcal{A}}')\SubDoS{\mcal{B}}(\SubE{\mcal{B}}')
\\&\quad\times
\Heaviside(\SubEinit{\mcal{A}}+\SubEinit{\mcal{B}} - \SubE{\mcal{A}}' - \SubE{\mcal{B}}')
 \\&\geq
\int_0^{\infty} \!\!\diff{\SubE{\mcal{A}}'}
\int_0^{\infty} \!\!\diff{\SubE{\mcal{B}}'}
\,
\SubDoS{\mcal{A}}(\SubE{\mcal{A}}')\SubDoS{\mcal{B}}(\SubE{\mcal{B}}')
\\&\quad\times
\Heaviside(\SubEinit{\mcal{A}} - \SubE{\mcal{A}}')
\Heaviside(\SubEinit{\mcal{B}} - \SubE{\mcal{B}}')
\\&=
\SubIDoS{\mcal{A}} (\SubE{\mcal{A}}) \,\SubIDoS{\mcal{B}} (\SubE{\mcal{B}})
.
\end{split}
\end{equation}
This result implies that the Gibbs entropy of the compound system is always at least as large as the sum of the Gibbs entropies of the subsystems before they were brought into thermal contact:
\begin{equation}
\EntG(\SubEinit{\mcal{A}} + \SubEinit{\mcal{B}}) \geq \SubEntG{\mcal{A}}(\SubEinit{\mcal{A}} ) + \SubEntG{\mcal{B}}(\SubEinit{\mcal{B}})
.
\end{equation}
Thus, the Gibbs entropy satisfies Planck's version of the second law.\footnote{\highlightchange{Note that, \emph{after} thermalization at fixed total energy $E$ and subsequent decoupling, the individual post-decoupling energies $\SubEinit{\mcal{A}}''$ and  $\SubEinit{\mcal{B}}''$  of the two subsystems  are \emph{not} exactly known (it is only known that $E=\SubEinit{\mcal{A}}''+\SubEinit{\mcal{B}}''$). That is, two ensembles of subsystems prepared by such a procedure are not in individual microcanonical states and their combined entropy remains at least $\EntG(\SubEinit{\mcal{A}}'' + \SubEinit{\mcal{B}}'')=\EntG(E)$. To reduce this entropy to a sum of microcanonical entropies, $\SubEntG{\mcal{A}}(\SubEinit{\mcal{A}}'' ) + \SubEntG{\mcal{B}}(\SubEinit{\mcal{B}}'')\le \EntG(E)$, an operator (Maxwell-type demon) would have to perform an additional energy measurement on one of the subsystems and only keep those systems in the ensemble that have exactly the same pairs of post-decoupling energies $\SubEinit{\mcal{A}}''$ and  $\SubEinit{\mcal{B}}''$. This information-based selection process transforms the original post-decoupling probability distributions into microcanonical density operators, causing a virtual entropy loss described by Eq.~\eqref{eq:second_law}.}} 

\par
Equality occurs only if the systems are energetically decoupled due to particular band structures and energy constraints that prevent actual energy exchange even in the presence of thermal coupling.  \highlightchange{The inequality is strict for an isolated system composed of two or more weakly coupled subsystems that can only energy exchange. However, the relative difference between $\EntG(\SubEinit{\mcal{A}} + \SubEinit{\mcal{B}})$ and $\SubEntG{\mcal{A}}(\SubEinit{\mcal{A}} ) + \SubEntG{\mcal{B}}(\SubEinit{\mcal{B}})$ may become small for \lq{}normal\rq{} systems (e.g., ideal gases and similar systems) in a suitably defined thermodynamic limit (see Sec.~\ref{sec:examples:power_law_dos:gibbs:second_law} for an example).}

%%%%%%%%%%%%%%%%%%%%%%%%%%%%%%%%%%%%%%%%%%%%%%%%%
\subsection{Boltzmann entropy}
\label{sec:second_law:Boltzmann_entropy}
%%%%%%%%%%%%%%%%%%%%%%%%%%%%%%%%%%%%%%%%%%%%%%%%%

To verify Eq.~\eqref{eq:second_law} for the Boltzmann entropy $\EntG=\ln(\eps\DoS)$, we have to compare the $\epsilon$-scaled DoS of the compound systems after coupling, $\epsilon \DoS(\SubEinit{\mcal{A}}+\SubEinit{\mcal{B}})$, with the product of the $\eps$-scaled DoS $\eps\SubDoS{\mcal{A}}(\SubEinit{\mcal{A}})$ and $\eps\SubDoS{\mcal{B}}(\SubEinit{\mcal{B}})$ before the coupling. But, according to Eq.~\eqref{eq:convolution_dos}, we have
\begin{equation}
\begin{split}
 &\epsilon \DoS(\SubEinit{\mcal{A}} + \SubEinit{\mcal{B}})
\\& =
 \eps
 \int_{0}^{\SubEinit{\mcal{A}} + \SubEinit{\mcal{B}}} \mspace{-15mu} \diff{\SubE{\mcal{A}}'} 
 \SubDoS{\mcal{A}}(\SubE{\mcal{A}}') \SubDoS{\mcal{B}}(\SubEinit{\mcal{A}} + \SubEinit{\mcal{B}} - \SubE{\mcal{A}}'),
\end{split}
\end{equation}
which, depending on $\epsilon$, can be larger or smaller than $\epsilon^2 \SubDoS{\mcal{A}}(\SubEinit{\mcal{A}}) \SubDoS{\mcal{B}}(\SubEinit{\mcal{B}})$. Thus, there is no strict relation between Boltzmann entropy of the compound system and the Boltzmann entropies of the subsystems before contact. That is, the Boltzmann entropy violates the Planck version of the second law for certain systems, as we will also demonstrate in Sec.~\ref{sec:examples:bounded_dos} with an example.

%%%%%%%%%%%%%%%%%%%%%%%%%%%%%%%%%%%%%%%%%%%%%%%%%
\subsection{Other entropy definitions}
\label{sec:second_law:other_entropies}
%%%%%%%%%%%%%%%%%%%%%%%%%%%%%%%%%%%%%%%%%%%%%%%%%

The modified Boltzmann entropy may violate the Planck version of the second law, if only a common energy width $\epsilon$ is used in the definition of the entropies. A more careful treatment reveals that the modified Boltzmann entropy satisfies:
\begin{multline}
\EntM(\SubEinit{\mcal{A}} + \SubEinit{\mcal{B}}, \epsilon_{\mcal{A}} + \epsilon_{\mcal{B}}) 
\\ \geq 
\SubEntM{\mcal{A}}(\SubEinit{\mcal{A}}, \epsilon_{\mcal{A}}) + \SubEntM{\mcal{B}}(\SubEinit{\mcal{B}}, \epsilon_{\mcal{B}})
.
\end{multline}
This shows that one has to properly propagate the uncertainties in the subsystem energies $\SubE{i}$ before coupling to the uncertainty in the total system energy $E$.

\par
A proof very similar to that for the Gibbs entropy shows that the complementary Gibbs entropy satisfies the Planck version of the second law (App.~\ref{app:sec:second}). The results for the Gibbs entropy and the complementary Gibbs entropy together imply that the alternative entropy $\EntP$ satisfies the Planck version as well.

%%%%%%%%%%%%%%%%%%%%%%%%%%%%%%%%%%%%%%%%%%%%%%%%%
%\pagebreak
\subsection{Adiabatic processes}
\label{sec:second_law:adiabatics}
%%%%%%%%%%%%%%%%%%%%%%%%%%%%%%%%%%%%%%%%%%%%%%%%%

So far, we have  focussed on whether the different microcanonical entropy definitions satisfy the second law during the thermalization of previously isolated systems after thermal coupling. Such thermalization processes are typically non-adiabatic and irreversible. Additionally,  one can also consider mechanically adiabatic processes performed on an isolated system, in order to assess whether a given microcanonical entropy definition obeys the second law.

\par
As already mentioned in Sec.~\ref{sec:first_law}, any entropy defined as a function of the integrated DoS $\IDoS(E)$ is an adiabatic invariant. Entropies of this type do not change in a mechanically adiabatic process, in agreement with the second law, ensuring that processes that are adiabatic in the mechanical sense (corresponding to \lq{}slow\rq{} changes of external control parameters $Z$) are also adiabatic in the thermodynamic sense ($dS = 0)$. Entropy definitions with this property include, for example, the Gibbs entropy~\eqref{eq:mce:df_Gibbs_entropy}, the complementary Gibbs entropy~\eqref{eq:mce:df_complementary_Gibbs_entropy}, and the alternative entropy~\eqref{sec:mce:eq:df_Penrose_entropy}.

\par
By contrast, the Boltzmann entropy~\eqref{eq:mce:df_Boltzmann_entropy} is a function of $\DoS(E)$ and, therefore, not an adiabatic invariant. As a consequence, $\EntB$ can change in a reversible mechanically adiabatic (quasi-static) process, which implies  that either during the forward process or its reverse the Boltzmann entropy decreases, in violation of the second law.
%For non-adiabatic processes both $\EntG$ and $\EntB$ will be in general not be constant.

%%%%%%%%%%%%%%%%%%%%%%%%%%%%%%%%%%%%%%%%%%%%%%%%%
\section{Examples}
\label{sec:examples}
%%%%%%%%%%%%%%%%%%%%%%%%%%%%%%%%%%%%%%%%%%%%%%%%%

The generic examples presented in this part illustrate the general results from above in more detail\footnote{Readers satisfied by the above general derivations may want to skip this section.}.
Section~\ref{sec:examples:power_law_dos} demonstrates that the Boltzmann temperature violates part (Z1) of zeroth law and fails to predict the direction of heat flows for systems with power-law DoS, whereas the Gibbs temperature does not. The example of a system with polynomial DoS in Sec.~\ref{sec:examples:polynomial_dos} illustrates that choosing the most probable Boltzmann temperature as subsystem temperature also violates the zeroth law (Z1). Section~\ref{sec:examples:bounded_dos} focuses on the thermal coupling of systems with bounded DoS, including an example for which the Boltzmann entropy violates the second law.
Subsequently, we still discuss in Sec.~\ref{sec:examples:classical_bounded} two classical Hamiltonian systems, where the equipartition formula~\eqref{eq:mce:equipartition_theorem} for the Gibbs temperature holds even for a bounded spectrum.

%%%%%%%%%%%%%%%%%%%%%%%%%%%%%%%%%%%%%%%%%%%%%%%%%
\subsection{Power-law densities}
\label{sec:examples:power_law_dos}
%%%%%%%%%%%%%%%%%%%%%%%%%%%%%%%%%%%%%%%%%%%%%%%%%

As the first example, we consider thermal contact between systems that have a power-law DoS. This class of systems includes important model systems such as ideal gases or harmonic oscillators\footnote{It is sometimes argued that thermodynamics must not be applied to small classical systems. We do not agree with this view as the Gibbs formalism works consistently even in these cases. As an example, consider an isolated the 1D harmonic pendulum with integrated DoS $\IDoS\propto E$. In this case, the Gibbs formalism yields $\kB\TempG=E$. For a macroscopic pendulum with a typical energy of, say, $E\sim 1$J this gives  a temperature of $\TempG \sim 10^{23}$K, which may seem prohibitively large. However, this results makes sense, upon recalling that an isolated pendulum moves, by definition, in a vacuum. If we let a macroscopically large number of gas molecules, which was kept at room temperature,  enter into the vacuum,  the mean kinetic energy of the pendulum will decrease very rapidly due to friction (i.e., heat will flow from the \lq hot\rq{} oscillator to the \lq cold\rq{} gas molecules) until the pendulum performs only miniscule thermal oscillations (\lq{}Brownian motions\rq{}) in agreement with the ambient gas temperature. Thus, $\TempG$ corresponds to the hypothetical gas temperature that would be required to maintain the same average pendulum amplitude or, equivalently, kinetic energy as in the vacuum. For a macroscopic pendulum, this temperature, must of course be extremely high. In essence, $\TempG \sim 10^{23}$K just tells us that it is practically impossible to drive macroscopic pendulum oscillations through molecular thermal fluctuations.}.

\par
Here, we show explicitly that the Gibbs temperature satisfies the zeroth law for systems with power-law DoS, whereas the Boltzmann temperature violates this law. Furthermore, we will demonstrate that for this class, the Gibbs temperature before thermal coupling  determines the direction of heat flow during coupling in accordance with naive expectation. By contrast, the Boltzmann temperature before coupling does not uniquely specify the heat flow direction during coupling.

\par
Specifically, we consider (initially) isolated systems $i=\mcal{A},\mcal{B},\ldots$, with energies $\SubE{i}$ and integrated DoS
\bse
\label{eq:examples:power_law_dos}
\begin{align}
\SubIDoS{i}(\SubE{i}) &= 
  \IDoSs{i}
  \begin{cases}
    \left({\SubE{i}}/{\Es{}}\right)^{\es{i}}, & 0 < \SubE{i}, \\
    0, & \text{otherwise,}
  \end{cases}
\end{align}
DoS
\begin{align}
\SubDoS{i}(\SubE{i}) &= 
 \es{i}\frac{\IDoSs{i}}{\Es{}}
  \begin{cases}
    \left({\SubE{i}}/{\Es{}}\right)^{\es{i} - 1}, & 0 < \SubE{i}, \\
    0, & \text{otherwise}
  \end{cases}
\end{align}
 and differential DoS
\begin{align}
\SubDDoS{i}(\SubE{i}) &= 
  \es{i}(\es{i}\! - \! 1)\frac{\IDoSs{\,i}}{\Es{}^2}\!
 \begin{cases}
    \left({\SubE{i}}/{\Es{}}\right)^{\es{i} - 2},& 0 < \SubE{i}, \\
    0 , & \text{otherwise.}
  \end{cases}
\end{align}
\ese
The parameter $\Es{}>0$ defines a characteristic energy scale, $\IDoSs{i}=\SubIDoS{i}(\Es{})$ is an amplitude parameter, and~\mbox{$\es{i} > 0$} denotes the power law index of the integrated DoS. For example, $\es{i} = N D / 2$ for an ideal gas of $N$ particles in $D$ dimensions, or $\es{i} = N D$ for $N$ weakly coupled $D$-dimensional harmonic oscillators.

\par
For $\SubE{i} \ge 0$, the Gibbs temperature of system $i$ is given by
\begin{equation}
 \SubTempG{i}(\SubE{i})  = \frac{\SubE{i}}{\es{i}} 
.
\end{equation} 
The Gibbs temperature is always non-negative, as already mentioned in the general discussion.

\par
For comparison, the Boltzmann temperature reads
\begin{equation}
 \SubTempB{i}(\SubE{i}) =\frac{\SubE{i}}{\es{i} - 1}  
.
\end{equation} 
For $\es{i} < 1$, the Boltzmann temperature is negative. A simple example for such a system with negative Boltzmann temperature is a single particle in a one-dimensional box (or equivalently any single one of the momentum degrees of freedom in an ideal gas), for which $\SubIDoS{i}(\SubE{i}) \propto \sqrt{\SubE{i}}$, corresponding to $\es{} = 1/2$ \cite[][]{2006DuHi,2014DuHi_NatPhys}.

\par
For $\es{i} = 1$, the Boltzmann temperature is infinite. Examples for this case include systems of two particles in a one-dimensional box, one particle in a two-dimensional box, or a single one-dimensional harmonic oscillator \cite[][]{2006DuHi,2014DuHi_NatPhys}.

\par
Since the integrated DoS is unbounded for large energies, $\IDoSinf{i}=\infty$, the complementary Gibbs entropy $\SubEntC{i}$ is not well defined. Furthermore, the entropy $\SubEntP{i}$ is identical to the Gibbs entropy, i.e. $\SubEntP{i}(\SubE{i}) = \SubEntG{i}(\SubE{i})$ for all $\SubE{i}$.

\par
Assume now two initially isolated systems $\mcal{A}$ and $\mcal{B}$ with an integrated DoS of the form~\eqref{eq:examples:power_law_dos} and initial energies $\SubEinit{\mcal{A}}$ and $\SubEinit{\mcal{B}}$ are brought into thermal contact. The energy of the resulting compound system $\SubE{\mcal{AB}} = \SubEinit{\mcal{A}} + \SubEinit{\mcal{B}}$. The integrated DoS $\SubIDoS{\mcal{AB}}$ of the compound system $\mcal{AB}$ follows a power law~\eqref{eq:examples:power_law_dos} with
\bse
\begin{align}
 \es{\mcal{AB}} &= \es{\mcal{A}} + \es{\mcal{B}},
 \\
 \IDoSs{\mcal{AB}} &= 
 \frac{\Gamma(\es{\mcal{A}}+1) \Gamma(\es{\mcal{B}}+1)}{\Gamma(\es{\mcal{AB}}+1)}
 \IDoSs{\mcal{A}} \IDoSs{\mcal{B}}
 .
\end{align}
\ese
Here, $\Gamma$ denotes the Gamma-function.

\par
The probability density of the energy $\SubE{i}$ of subsystem $i\in\{\mcal{A},\mcal{B}\}$ after thermalization reads
\begin{equation}
\label{eq:examples:power_law_dos:energy_pdf}
 \begin{split}
 \pdfE{i}(\SubE{i}| \SubE{\mcal{AB}}) &=
 \frac{\Gamma(\es{\mcal{AB}})}{ \Gamma(\es{\mcal{A}})\Gamma(\es{\mcal{B}})}
 \\&\quad\times
  \frac{\SubE{i}^{\es{i}-1} \bigl(\SubE{\mcal{AB}} - \SubE{i}\bigr)^{\es{\mcal{AB}} - \es{i} - 1}}{\SubE{\mcal{AB}}^{\es{\mcal{AB}}-1}}
  .
 \end{split}
\end{equation} 
The mean energy $\EV{\SubE{i}}_{\SubE{\mcal{AB}}}$ of system $i$  after thermalization is given by:
\begin{equation}
 \begin{split}
 \EV{\SubE{i}}_{\SubE{\mcal{AB}}} &=
 \frac{\es{i}}{\es{\mcal{AB}}} \SubE{\mcal{AB}}
  .
 \end{split}
\end{equation} 
The larger the index $\es{i}$, the bigger the share in energy for system $i$.

%%%%%%%%%%%%%%%%%%%
\subsubsection{Gibbs temperature predicts heat flow}
\label{sec:examples:power_law_dos:gibbs_heat_flow}
%%%%%%%%%%%%%%%%%%%

The Gibbs temperature of the compound system after thermalization is given by
\begin{equation}
 \begin{split}
  \SubTempG{\mcal{AB}} &=  \frac{\SubE{\mcal{AB}}}{\es{\mcal{AB}}}
= 
\frac{\es{\mcal{A}} \SubTempGinit{\mcal{A}} + \es{\mcal{B}} \SubTempGinit{\mcal{B}}}{\es{\mcal{A}} +  \es{\mcal{B}}}
.
 \end{split}
\end{equation} 
Thus, the Gibbs temperature $\SubTempG{\mcal{AB}}$ is a weighted mean of the temperatures $\SubTempGinit{\mcal{A}} = \SubTempG{\mcal{A}}(\SubEinit{\mcal{A}})$ and $\SubTempGinit{\mcal{B}}=\SubTempG{\mcal{B}}(\SubEinit{\mcal{B}})$ of the systems $\mcal{A}$ and $\mcal{B}$ before coupling. In simple words: \lq{}hot\rq{} (large $T$) and \lq{}cold\rq{} (small $T$) together yield \lq{}warm\rq{} (some intermediate $T$), as one might naively expect from everyday experience. In particular, if $\SubTempGinit{\mcal{A}} = \SubTempGinit{\mcal{B}}$, then $\SubTempG{\mcal{AB}} = \SubTempGinit{\mcal{A}} = \SubTempGinit{\mcal{B}}$.
This is however not universal, but rather a special property of systems with power-law densities, as already pointed out by Gibbs \cite[pp.~171]{Gibbs}.

The above equations imply that the energy before coupling is given by $\SubEinit{\mcal{A}} = \es{\mcal{A}} \SubTempGinit{\mcal{A}}$, and the energy after coupling by $\EV{\SubE{\mcal{A}}}_{\SubE{\mcal{AB}}} = \es{\mcal{A}} \SubTempG{\mcal{AB}}$. Since the final temperature $\SubTempG{\mcal{AB}}$ is a weighted mean of the initial temperatures $\SubTempGinit{\mcal{A}}$ and $\SubTempGinit{\mcal{B}}$,
\begin{align}
	\SubTempGinit{\mcal{A}} &\gtreqqless \SubTempGinit{\mcal{B}} &\Leftrightarrow && \SubEinit{\mcal{A}} &\gtreqqless \EV{\SubE{\mcal{A}}'}_{\SubE{\mcal{AB}}} 
.	
\end{align}
This means that for systems with power law densities, the difference in the initial Gibbs temperatures fully determines the direction of the heat flow~\eqref{eq:heat_flow_definition} between the systems during thermalization.

%%%%%%%%%%%%%%%%%%%
\subsubsection{Gibbs temperature satisfies the zeroth law}
\label{sec:examples:power_law_dos:gibbs:zeroth_law}
%%%%%%%%%%%%%%%%%%%

In Sec.~\ref{sec:zeroth_law}, we already presented a general proof that the Gibbs temperature obeys the zeroth law~\eqref{eq:zeroth_law_ii} for a wide class of systems. An explicit calculation confirms this for the Gibbs temperature of power-law density systems after coupling:
\begin{equation}
\begin{split}
	\EV{\SubTempG{\mcal{A}}(\SubE{\mcal{A}}')}_{\SubE{\mcal{AB}}} \! = \EV{\SubTempG{\mcal{B}}(\SubE{\mcal{B}}')}_{\SubE{\mcal{AB}}}  \!
	&=
	\SubTempG{\mcal{AB}}(\SubE{\mcal{AB}})
.
\end{split}
\end{equation}

%%%%%%%%%%%%%%%%%%%
\subsubsection{Gibbs temperature satisfies the second law}
\label{sec:examples:power_law_dos:gibbs:second_law}
%%%%%%%%%%%%%%%%%%%

\highlightchange{
In Sec.~\ref{sec:second_law} we already presented a general proof that the Gibbs entropy satisfies the second law~\eqref{eq:second_law}. Here, we illustrate this finding by an explicit example calculation. We also show that, although the inequality~\eqref{eq:second_law} is always strict for power law systems at finite energies, the relative difference become small in a suitable limit.
}

\par
\highlightchange{
For a given total energy $\SubE{\mcal{AB}} = \SubE{\mcal{A}} + \SubE{\mcal{B}}$, the sum $\SubEntG{\mcal{A}}(\SubE{\mcal{A}}) + \SubEntG{\mcal{B}}(\SubE{\mcal{B}})$ of Gibbs entropies of system~$\mcal{A}$ and~$\mcal{B}$ before coupling becomes maximal for energies
\begin{align}
\label{eq:examples:power_law_dos:e_maximal_entropy_before_coupling}
 \SubE{\mcal{A}}^=  &= \frac{\es{\mcal{A}} \SubE{\mcal{AB}}}{\es{\mcal{AB}}}
&\text{and}&&
 \SubE{\mcal{B}}^=  &= \frac{\es{\mcal{B}} \SubE{\mcal{AB}}}{\es{\mcal{AB}}}
.
\end{align}
These coincide with the energies, at which the subsystem Gibbs temperatures before coupling equal the Gibbs temperature of the compound system after coupling (but may differ slightly from the most probable energies during coupling, see Sec.~\ref{sec:examples:power_law_dos:boltzmann_zeroth_law} below), and for which there is no net heat flow during thermalization. Thus, for $\SubE{\mcal{AB}}>0$, we have
\begin{equation}
\label{eq:examples:power_law_dos:second_law}
\begin{split}
&\SubEntG{\mcal{AB}}(\SubE{\mcal{AB}})
\\&=
\ln\left[
\frac{\Gamma(\es{\mcal{A}} + 1) \Gamma(\es{\mcal{B}} + 1)}{\Gamma(\es{\mcal{AB}} + 1)} \IDoSs{\mcal{A}} \IDoSs{\mcal{B}}
\!\left(\frac{\SubE{\mcal{AB}}}{\Es{}}\right)^{\!\!\es{\mcal{AB}}}
\right]
\\&>
\SubEntG{\mcal{A}}(\SubE{\mcal{A}}^=) + \SubEntG{\mcal{B}}(\SubE{\mcal{B}}^=)
\\&=
\ln\left[
\frac{\es{\mcal{A}}^{\es{\mcal{A}}}\es{\mcal{B}}^{\es{\mcal{B}}}}{\es{\mcal{AB}}^{\es{\mcal{AB}}}} \IDoSs{\mcal{A}} \IDoSs{\mcal{B}}
\!\left(\frac{\SubE{\mcal{AB}}}{\Es{}}\right)^{\!\!\es{\mcal{AB}}}
\right]
\\&\geq
\SubEntG{\mcal{A}}(\SubE{\mcal{A}}) + \SubEntG{\mcal{B}}(\SubE{\mcal{B}})
.
\end{split}
\end{equation}
}

\par
\highlightchange{
The inequality \eqref{eq:examples:power_law_dos:second_law} shows that for finite energies, the total entropy always increases during coupling. However, for equal temperatures before coupling and large $\es{i}$ (e.g. large particle numbers in an ideal gas), the relative increase becomes small:
\begin{equation}
\label{eq:examples:power_law_dos:second_law_ratio}
0=\lim_{\es{i}\to \infty}
\frac{\SubEntG{\mcal{AB}}(\SubE{\mcal{AB}}) - \SubEntG{\mcal{A}}(\SubE{\mcal{A}}^=) - \SubEntG{\mcal{B}}(\SubE{\mcal{B}}^=)}
{\SubEntG{\mcal{AB}}(\SubE{\mcal{AB}})}
.
\end{equation}
}

%%%%%%%%%%%%%%%%%%%
\subsubsection{Boltzmann temperature fails to predict heat flow}
\label{sec:examples:power_law_dos:boltzmann_heat_flow}
%%%%%%%%%%%%%%%%%%%

The Boltzmann temperature of the compound system has a more complicated relation to the initial Boltzmann temperatures  $\SubTempBinit{\mcal{A}} = \SubTempB{\mcal{A}}(\SubEinit{\mcal{A}})$ and $\SubTempBinit{\mcal{B}}=\SubTempB{\mcal{B}}(\SubEinit{\mcal{B}})$.
If $\es{\mcal{A}}$, $\es{\mcal{B}}$, and  $\es{\mcal{A}} +\es{\mcal{B}} \neq 1$ (otherwise at least one of the involved temperatures is infinite), then
\begin{equation}
\begin{split}
\SubTempB{\mcal{AB}} &=  \frac{\SubE{\mcal{AB}}}{\es{\mcal{AB}}  - 1}
\\&
= 
\frac{(\es{\mcal{A}} - 1) \SubTempBinit{\mcal{A}} + (\es{\mcal{B}} - 1) \SubTempBinit{\mcal{B}}}{\es{\mcal{A}} +  \es{\mcal{B}} - 1}
.
 \end{split}
\end{equation} 
This implies in particular, that when two power-law systems with equal Boltzmann temperature $\SubTempBinit{\mcal{A}} = \SubTempBinit{\mcal{B}}$ are brought into thermal contact, the compound system Boltzmann temperature differs from the initial temperatures, $\SubTempB{\mcal{AB}} \neq \SubTempBinit{\mcal{A}} = \SubTempBinit{\mcal{B}}$. Moreover, even the signs of the temperatures may differ. For example, if $\es{\mcal{A}} < 1$ and $\es{\mcal{B}} < 1$, but $\es{\mcal{A}} + \es{\mcal{B}} > 1$, then $\SubTempBinit{\mcal{A}} < 0$ and $\SubTempBinit{\mcal{B}} < 0$, but $\SubTempB{\mcal{AB}} > 0$.

\par
The ordering of the initial Boltzmann temperatures does not fully determine the direction of the net energy flow during thermalization. In particular, heat may flow from an initially colder system to a hotter system. If for example, system $\mcal{A}$ has a power-law DoS with index $\es{\mcal{A}} = 3/2 $ and initial energy $\SubEinit{\mcal{A}} = 3 \Es{}$, and system $\mcal{B}$ has index $\es{\mcal{B}} = 2 $ and initial energy $\SubEinit{\mcal{B}} = 5 \Es{}$, then the initial Boltzmann temperature $\SubTempBinit{\mcal{A}} = 6 \Es{}$ is higher than $\SubTempBinit{\mcal{B}} = 5 \Es{}$.
However, the final energy $\EV{\SubE{\mcal{A}}}_{\SubE{\mcal{AB}}} = 24/7 \Es{} > 3 \Es{}$. Thus, the initially hotter system $\mcal{A}$ gains energy during thermal contact.

\par
Morever, equal initial Boltzmann temperatures do not preclude heat flow at contact (i.e., do not imply \lq{}potential\rq{} thermal equilibrium). If, for example, $\es{\mcal{A}} = 3/2 $, $\SubEinit{\mcal{A}} = 3 \Es{}$, $\es{\mcal{B}} = 2 $ and $\SubEinit{\mcal{B}} = 6 \Es{}$, then  $\SubTempBinit{\mcal{A}} = \SubTempBinit{\mcal{B}} = 6 \Es{}$. However, $\EV{\SubE{\mcal{A}}'}_{\SubE{\mcal{AB}}} = 27/7 \Es{} > 3 \Es{}$. Thus, system $\mcal{A}$ gains energy through thermal contact with a system initially at the same Boltzmann temperature.

%%%%%%%%%%%%%%%%%%%
\subsubsection{Boltzmann temperature violates the zeroth law}
\label{sec:examples:power_law_dos:boltzmann_zeroth_law}
%%%%%%%%%%%%%%%%%%%

As already mentioned in Section~\ref{sec:zeroth_law}, the Boltzmann temperature may violate the zeroth law~\eqref{eq:zeroth_law_ii}. Here we show this explicitly for systems with a power-law DoS:
\begin{equation}
\begin{split}
	\EV{\SubTempB{i}(\SubE{i}')}_{\SubE{\mcal{AB}}} 
	&= 
	\frac{\es{\mcal{AB}} - 1}{\es{\mcal{AB}}} \frac{\es{i}}{\es{i}-1}  \SubTempB{\mcal{AB}}(\SubE{\mcal{AB}})
	\\&\neq
	\SubTempB{\mcal{AB}}(\SubE{\mcal{AB}}).
\end{split}
\end{equation}
In terms of Boltzmann temperature, subsystems are hotter than their parent system. In particular, the smaller the index $\es{i}$ (often implying a smaller system), the hotter is system $i$ compared to the compound system. Thus any two systems with different power-law index do not have the same Boltzmann temperature in thermal equilibrium. Moreover, for systems permitting different decompositions into subsystems with power-law densities, such as an ideal gas with several particles, the subsystems temperatures depend on the particular decomposition.

\par
In Section~\ref{sec:zeroth_law}, we also mentioned that the inverse Boltzmann temperature satisfies a relation similar to Eq.~\eqref{eq:zeroth_law_ii} for certain systems. If $\es{i}>1$, then one finds indeed
\begin{equation}
\label{eq:examples:power_law_dos:mean_inverse_subsystem_Boltzmann_temperature}
\begin{split}
	\EV{\SubTempB{i}^{-1}(\SubE{i}')}_{\SubE{\mcal{AB}}} 
	&= 
	\frac{\es{\mcal{AB}} - 1}{\SubE{\mcal{AB}}}
	=
	\SubTempB{\mcal{AB}}^{-1}(\SubE{\mcal{AB}})
,
\end{split}
\end{equation}
either through direct application of Eq.~\eqref{eq:zeroth_law_ii_inverse_Boltzmann}, or by calculation of the integral~\eqref{eq:mean_for_functions_of_subsystem_energy}.

\par
If however, $\es{i}<1$, then Eqs.~\eqref{eq:zeroth_law_ii_inverse_Boltzmann} and \eqref{eq:examples:power_law_dos:mean_inverse_subsystem_Boltzmann_temperature} do not hold. Instead, $\SubTempB{i}^{-1}(\SubE{i}') < 0$ for all $\SubE{i}'>0$, and the integral~\eqref{eq:mean_for_functions_of_subsystem_energy} diverges for the mean of inverse subsystem Boltzmann temperature,  $\EV{\SubTempB{i}^{-1}(\SubE{i}')}_{\SubE{\mcal{AB}}} = -\infty$. In contrast, the inverse compound system Boltzmann temperature $\SubTempB{\mcal{AB}}^{-1}(\SubE{\mcal{AB}})$ is finite (and positive for $\es{\mcal{AB}} > 1$) for all $\SubE{\mcal{AB}}>0$.

\par
As also mentioned in Section~\ref{sec:zeroth_law}, the Boltzmann temperatures at the most likely energy partition agree for certain systems in equilibrium. If both $\es{\mcal{A}} >1$ and $\es{\mcal{B}} > 1$, then the energy distribution $\pdfE{i}(\SubE{i}'|\SubE{\mcal{AB}})$ is maximal for $\SubE{i}' = \SubEML{i} = \SubE{\mcal{AB}} (\es{i} - 1)/(\es{\mcal{AB}} - 2)$, yielding
\begin{equation}
\label{eq:examples:power_law_dos_ML_T_B}
\begin{split}
\SubTempB{\mcal{A}}(\SubEML{\mcal{A}}) &= \SubTempB{\mcal{B}}(\SubEML{\mcal{B}})
=
\frac{\SubE{\mcal{AB}}}{\! \es{\mcal{AB}} - 2 \!}
> 
\SubTempB{\mcal{AB}}(\SubE{\mcal{AB}})
.
\end{split}
\end{equation}
Thus, the thereby defined subsystem temperatures agree, and their value is even independent of the particular decomposition of the compound system (which is usually not true for more general systems), but these subsystem Boltzmann temperatures are always larger than the  Boltzmann temperature of the compound system.

\par
If $\es{\mcal{A}} < 1$ and/or $\es{\mcal{B}} < 1$, the energy distribution~\eqref{eq:examples:power_law_dos:energy_pdf} becomes maximal for $\SubE{\mcal{A}}' = 0$ and/or $\SubE{\mcal{A}}' = \SubE{\mcal{AB}}$. There, one of the Boltzmann temperatures vanishes, whereas for the other system, $\SubTempB{i} = \SubE{\cal{AB}}/(\es{i}-1)$. Thus both subsystem temperatures differ from each other, and also from the temperature of the compound systems.

%%%%%%%%%%%%%%%%%%%%%%%%%%%%%%%%%%%%%%%%%%%%%%%%%
\subsection{Polynomial densities}
\label{sec:examples:polynomial_dos}
%%%%%%%%%%%%%%%%%%%%%%%%%%%%%%%%%%%%%%%%%%%%%%%%%

%================================================
\begin{figure}
%\centerline{\includegraphics[width=1\linewidth]{Gibbs_subsystem_temperature}}
\centerline{\includegraphics[width=1\linewidth]{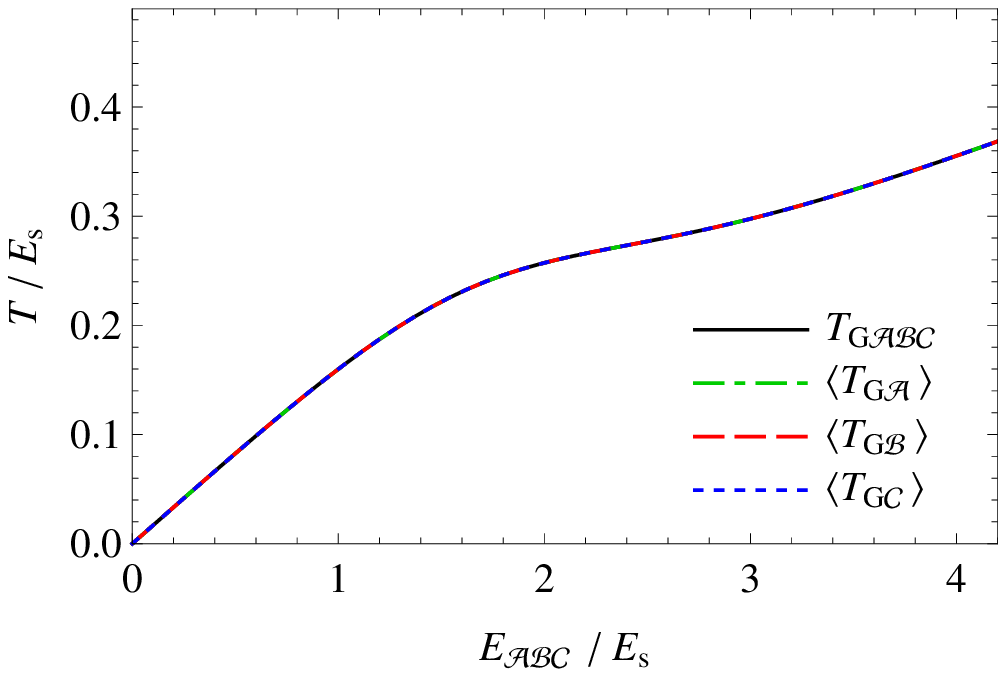}}
\caption{
\label{fig:Gibbs_subsystem_temperature}
Comparison of the subsystem Gibbs temperatures $\EV{\SubTempG{i}}_{\SubE{\mcal{ABC}}}$ and the compound system Gibbs temperature $\SubTempG{\mcal{ABC}}(\SubE{\mcal{ABC}})$ as function of compound system energy $\SubE{\mcal{ABC}}$ for systems with DoS~\eqref{eq:examples:polynomial_dos}. Note that for any given energy, all these temperatures agree. 
}
\end{figure}
%================================================

%================================================
\begin{figure}
%\centerline{\includegraphics[width=1\linewidth]{Boltzmann_subsystem_temperature}}
\centerline{\includegraphics[width=1\linewidth]{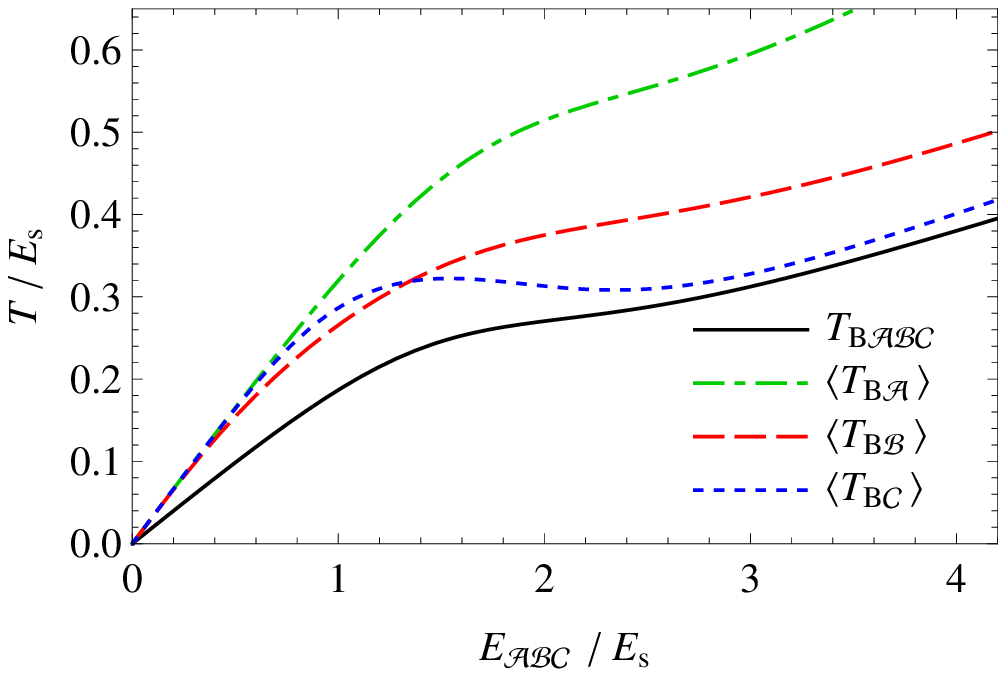}}
\caption{
\label{fig:Boltzmann_subsystem_temperature}
Comparison of the subsystem Boltzmann temperatures $\EV{\SubTempB{i}}_{\SubE{\mcal{ABC}}}$ and the compound system Boltzmann temperature $\SubTempB{\mcal{ABC}}(\SubE{\mcal{ABC}})$ as function of compound system energy $\SubE{\mcal{ABC}}$ for systems with DoS~\eqref{eq:examples:polynomial_dos}. Note that for almost all energies, these temperatures disagree. 
}
\end{figure}
%================================================

%================================================
\begin{figure}
%\centerline{\includegraphics[width=1\linewidth]{most_likely_Boltzmann_subsystem_temperature}}
\centerline{\includegraphics[width=1\linewidth]{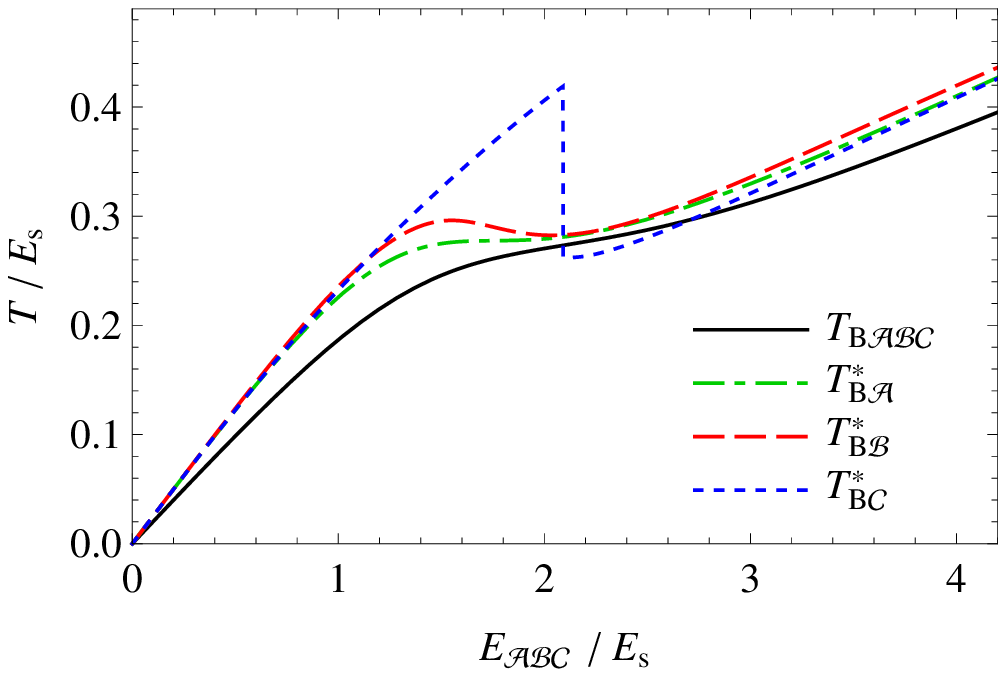}}
\caption{
\label{fig:most_likely_Boltzmann_subsystem_temperature}
Comparison of the subsystem Boltzmann temperatures $\SubTempBML{i}$ at the most likely subsystem energy and the compound system Boltzmann temperature $\SubTempB{\mcal{ABC}}(\SubE{\mcal{ABC}})$ as a function of the total energy $\SubE{\mcal{ABC}}$ for systems with DoS~\eqref{eq:examples:polynomial_dos}. Note that for almost all energies, these temperatures disagree. For energies $\SubE{\mcal{ABC}}\approx 2\Es{}$,  
the energy distribution $\pdfE{\mcal{C}}(\SubE{\mcal{C}}'|\SubE{\mcal{ABC}})$ is bimodal, with the order of the two peaks heights changing at $\SubE{\mcal{ABC}} = 2.09\Es{}$, causing a discontinuity in the most likely Boltzmann temperature $\SubTempBML{\mcal{C}}$.
}
\end{figure}
%================================================

\par
Systems with a pure power-law DoS exhibit relatively simple relations between the compound system Boltzmann temperature and the most likely subsystem Boltzmann temperatures, see Eq.~\eqref{eq:examples:power_law_dos_ML_T_B}.  Models with polynomial DoS present a straightforward generalization of a power-law DoS but exhibit a richer picture with regard to the decomposition dependence of subsystem Boltzmann temperatures. For coupled systems with pure power-law DoS, the most likely subsystem Boltzmann temperatures, although different from the compound system's Boltzmann temperature, all have the same value. This is not always the case for compositions of systems with more general polynomial DoS, as we will show next.

\par
For definiteness, consider three systems $\mcal{A}$, $\mcal{B}$, and $\mcal{C}$ with densities
\bse
\label{eq:examples:polynomial_dos}
\begin{align}
\SubDoS{\mcal{A}}(\SubE{\mcal{A}}) &= 
  \frac{\IDoSs{\mcal{A}}}{\Es{}}
  \begin{cases}
    \dfrac{\SubE{\mcal{A}}}{\Es{}}, & 0 < \SubE{\mcal{A}}, \\
    0, & \text{otherwise,}
  \end{cases}
\\
\SubDoS{\mcal{B}}(\SubE{\mcal{B}}) &= 
  \frac{\IDoSs{\mcal{B}}}{\Es{}}
  \begin{cases}
    \dfrac{\SubE{\mcal{B}}}{\Es{}} +\dfrac{\SubE{\mcal{B}}^{3}}{\Es{}^{3}}, &   0 < \SubE{\mcal{B}}, \\
    0 ,& \text{otherwise,}
  \end{cases}
\\
\SubDoS{\mcal{C}}(\SubE{\mcal{C}}) &= 
  \frac{\IDoSs{\mcal{C}}}{\Es{}}
  \begin{cases}
    \dfrac{\SubE{\mcal{C}}}{\Es{}} + \dfrac{\SubE{\mcal{C}}^{6}}{\Es{}^{6}}, &  0 < \SubE{\mcal{C}}, \\
    0, & \text{otherwise.}
  \end{cases}
\end{align}
\ese
Here, $\Es{}>0$ again defines an energy scale, and $\IDoSs{i}>0$ denotes an amplitude.

\par
When the three systems are thermally coupled to form an isolated compound system $\mcal{ABC}$ with energy $\SubE{\mcal{ABC}}$, the subsystem Gibbs temperatures $\EV{\SubTempG{i}}$ always agree with the  Gibbs temperature $\SubTempG{\mcal{ABC}}$  of the compound system (Fig.~\ref{fig:Gibbs_subsystem_temperature}). In contrast, the subsystem Boltzmann temperatures $\EV{\SubTempB{i}}$ almost never agree with each other or with the compound system Boltzmann temperature  $\SubTempB{\mcal{ABC}}$, differing by factors of almost two in some cases  (Fig.~\ref{fig:Boltzmann_subsystem_temperature}). Using the Boltzmann temperature $\SubTempBML{i} = \SubTempB{i}(\SubEML{i})$ of the subsystem $i$ at its most likely energy $\SubEML{i}$ as indicator of the subsystem temperature yields 
a very discordant result (Fig.~\ref{fig:most_likely_Boltzmann_subsystem_temperature}).

%================================================
\begin{figure*}
\centerline{
\includegraphics[height=4.75cm]{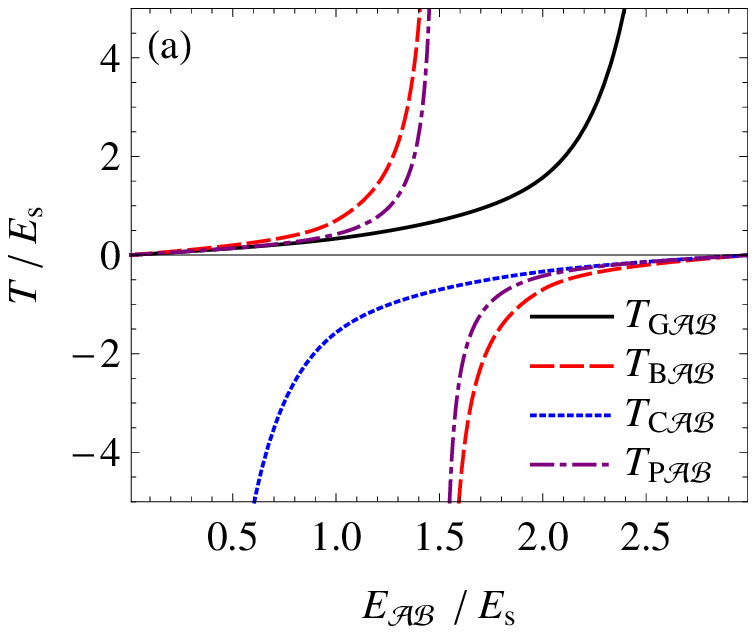}
\hfill
\includegraphics[height=4.75cm]{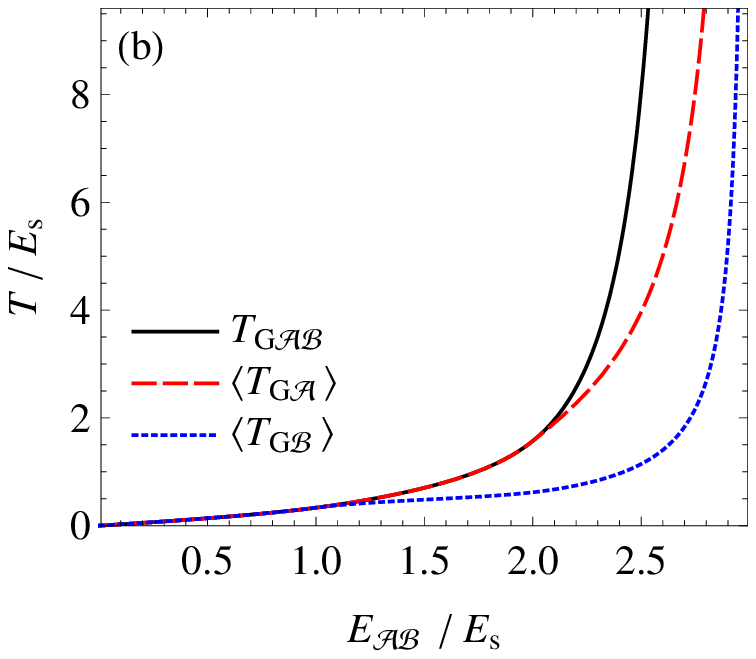}
\hfill
\includegraphics[height=4.75cm]{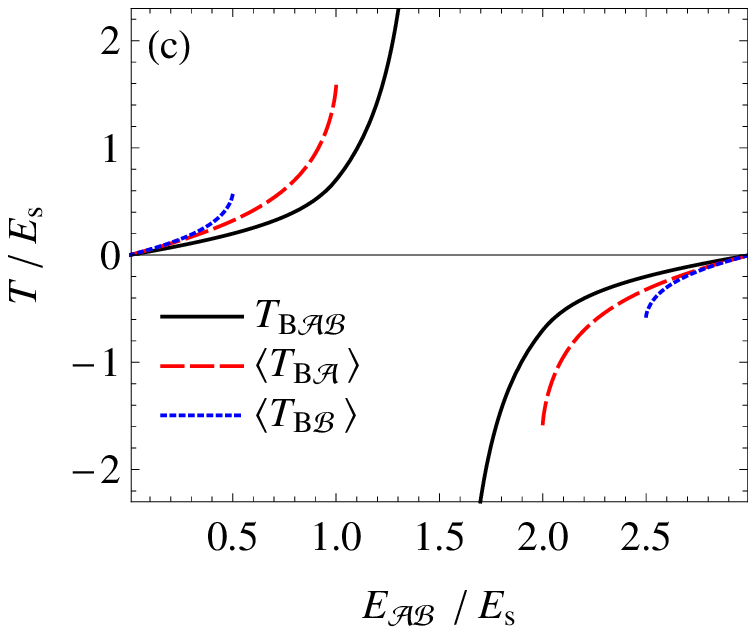}
}
\centerline{
\includegraphics[height=4.75cm]{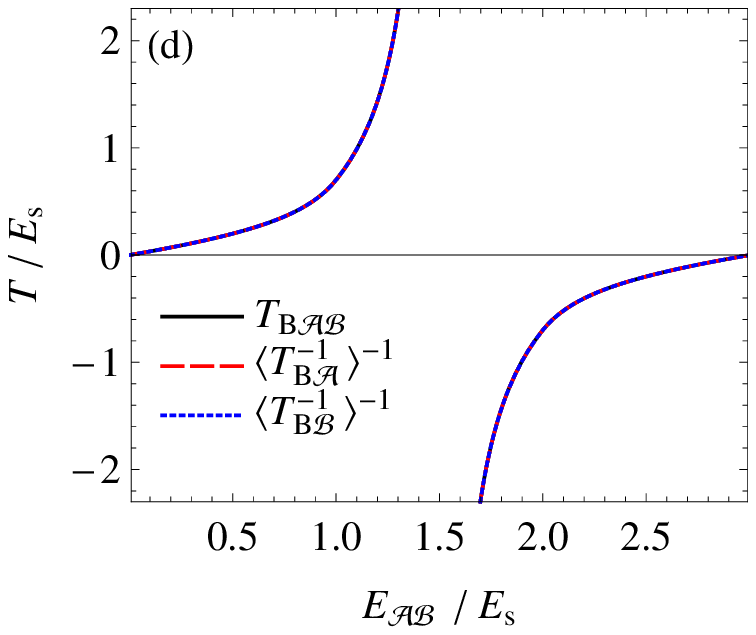}
\hfill
\includegraphics[height=4.75cm]{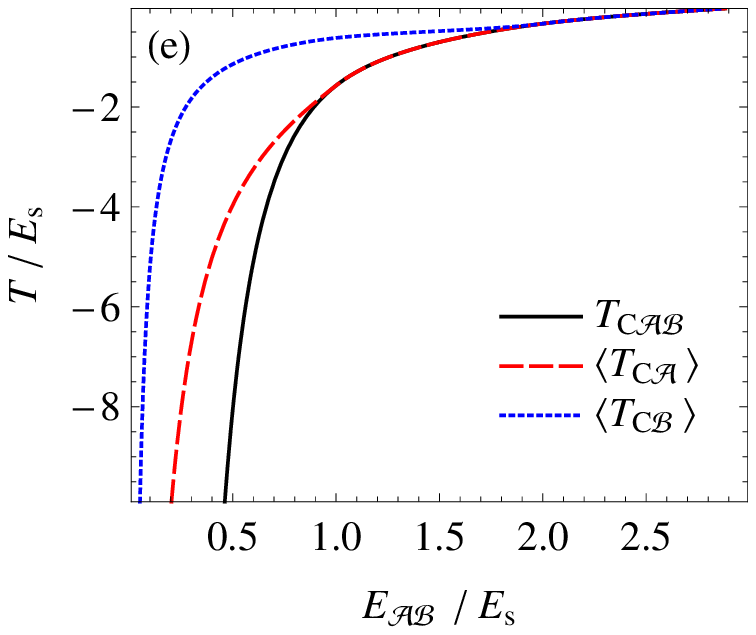}
\hfill
\includegraphics[height=4.75cm]{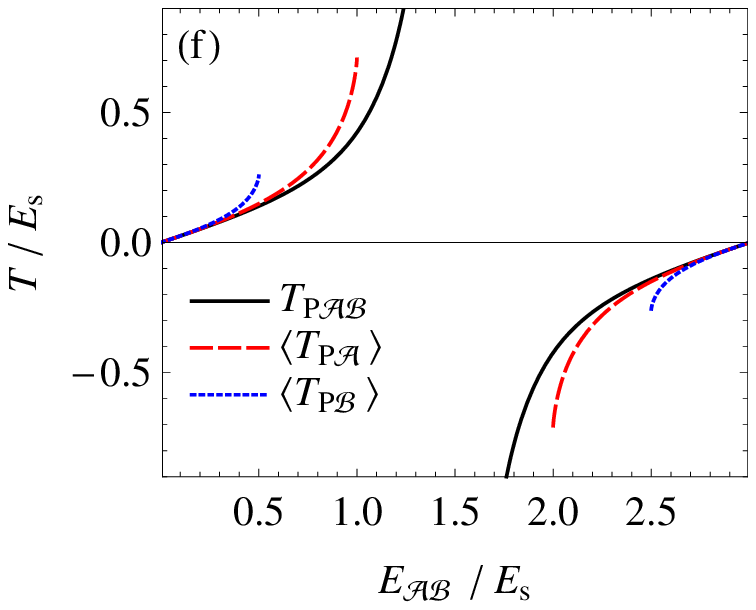}
}\caption{
\label{fig:two_bounded_dos}
Temperatures of two thermally coupled systems with bounded DoS~\eqref{eq:examples:simple_bounded_dos} (with $\Es{\mcal{A}} = 2\Es{}$ and $\Es{\mcal{B}} = 1\Es{}$, see main text for details) as functions of total system energy $\SubE{\mcal{AB}}$: (a) comparison of different definitions for the compound system temperature, (b)-(f) comparison between compound system temperature and subsystem temperatures.
}
\end{figure*}
%================================================

%%%%%%%%%%%%%%%%%%%%%%%%%%%%%%%%%%%%%%%%%%%%%%%%%
\subsection{Bounded densities}
\label{sec:examples:bounded_dos}
%%%%%%%%%%%%%%%%%%%%%%%%%%%%%%%%%%%%%%%%%%%%%%%%%

We  now consider thermal contact between systems that have an upper energy bound and a finite volume of states $\IDoSinf{}$. This general definition covers, among others, systems of weakly coupled localised magnetic moments (paramagnetic \lq{}spins\rq{}) in an external magnetic field. Restricting the considerations to systems with finite $\IDoSinf{}$ allows us to discuss of the complementary Gibbs entropy and the alternative entropy~$\EntP$, in addition to Gibbs and Boltzmann entropy.

\par
To keep the algebra simple, we consider systems $i=\mcal{A},\mcal{B}$, with energies $\SubE{i}$ and integrated DoS, DoS, and differential DoS given by
\bse
\label{eq:examples:simple_bounded_idos_dos_and_ddos}
\begin{align}
\label{eq:examples:simple_bounded_idos}
\SubIDoS{i}(\SubE{i}) &= 
	\frac{\! \IDoSs{i}\!}{\Es{i}^3}
	\!\begin{cases}
		0 ,& \SubE{i} < 0, \\
	 	\SubE{i}^2\! \left(3\Es{i} - 2 \SubE{i} \right), \!\!\!\! &  0 \leq \SubE{i} \leq \Es{i},\!\!\!\! \\
		\Es{i}^3 ,&  \Es{i} < \SubE{i},
	\end{cases}
\\
\label{eq:examples:simple_bounded_dos}
\SubDoS{i}(\SubE{i})	&= 
	\frac{\!6\IDoSs{i}\!}{\Es{i}^3}
	\begin{cases}
	 	\SubE{i} \left(\Es{i} - \SubE{i} \right),  &  0 \leq \SubE{i} \leq \Es{i}, \\
		0,  & \text{otherwise,}
	\end{cases}
\\
\label{eq:examples:simple_bounded_ddos}
\SubDDoS{i}(\SubE{i}) &= 
	\frac{\!6\IDoSs{i}\!}{\Es{}^3}
	\begin{cases}
	 	\left(\Es{i} - 2 \SubE{i} \right), &   \quad0 \leq \SubE{i} \leq \Es{i}, \\
		0 , &  \quad\text{otherwise.}
	\end{cases}
\end{align}
\ese
Here, $\Es{i} > 0$ defines the energy bandwidth, and $\IDoSs{i} = \SubIDoS{i}(\Es{i})= \SubIDoS{i}(\infty)$ denotes the total number of states of system $i$. 
%We assume for the bandwidths that $\Es{\mcal{A}} \geq \Es{\mcal{B}}$.

\par
Note that the DoS~\eqref{eq:examples:simple_bounded_dos} can be generalized to \mbox{$\SubDoS{i}\propto \SubE{i}^{\es{i}}(\Es{i} - \SubE{i})^{\es{i}}$}, producing a DoS with rectangular shape for $\es{i}=0$ (the DoS for a single classical spin in an external magnetic field), a semicircle for $\es{i}=1/2$, an inverted parabola for $\es{i}=1$ (considered here), and a bell shape for $\es{i}>1$. However, these generalizations would merely complicate the algebra without providing qualitatively different results.

%%%%%%%%%%%%%%%%%%%
\subsubsection{Boltzmann entropy violates second law}
\label{sec:examples:bounded_dos:boltzmann_second_law}
%%%%%%%%%%%%%%%%%%%

It is straightforward to construct a simple example where the Boltzmann entropy violates the Planck version of the second law. Consider two systems $\mcal{A}$ and $\mcal{B}$ with DoS~\eqref{eq:examples:simple_bounded_dos}, $\Es{\mcal{A}} = \Es{\mcal{B}} = \Es{}$, $\IDoSs{\mcal{A}} = \IDoSs{\mcal{B}} = \IDoSs{}$, and system energies $\SubEinit{\mcal{A}} = \SubEinit{\mcal{B}} = \Es{}/2$, before coupling. The sum of the Boltzmann entropies before coupling is given by $\SubEntB{\mcal{A}}(\SubEinit{\mcal{A}}) + \SubEntB{\mcal{B}}(\SubEinit{\mcal{B}}) = \ln\bigl[ (9 \epsilon^2 \IDoSs{}^2)/(4 \Es{}^2)\bigr]$. The Boltzmann entropy of the coupled system is obtained as $\SubEntB{\mcal{AB}}(\SubEinit{\mcal{A}} + \SubEinit{\mcal{B}}) = \ln\bigl[ (6 \epsilon \IDoSs{}^2)/(5 \Es{})\bigr]$. Hence, 
\begin{equation}
  \SubEntB{\mcal{AB}}(\SubEinit{\mcal{A}} + \SubEinit{\mcal{B}})
  <
  \SubEntB{\mcal{A}}(\SubEinit{\mcal{A}}) + \SubEntB{\mcal{B}}(\SubEinit{\mcal{B}})
\end{equation}
for $\epsilon > 8 \Es{} / 15$. The only way to avoid this problem in general is to always use an infinitesimal $\epsilon$ and a DoS that exactly gives the number of states at the energy in question (i.e. the exact degeneracy) devoid of any energy coarse-graining -- but this leads to other severe conceptual problems~(see SI of Ref~\cite{2014DuHi_NatPhys}).

%%%%%%%%%%%%%%%%%%%
\subsubsection{Temperatures and the zeroth law}
\label{sec:examples:bounded_dos:zeroth_law}
%%%%%%%%%%%%%%%%%%%

For a slightly more general discussion of thermal contact, we still consider two systems $\mcal{A}$ and $\mcal{B}$ with DoS~\eqref{eq:examples:simple_bounded_dos} with parameters $\Es{\mcal{A}} = 2\Es{}$, $\Es{\mcal{B}} = 1\Es{}$, $\IDoSs{\mcal{A}} = \IDoSs{\mcal{B}} = \IDoSs{}$, and system energies $0 \leq \SubEinit{\mcal{A}} \leq \Es{}$ and $0 \leq \SubEinit{\mcal{B}} \leq 2 \Es{}$ before coupling.
The energy dependence of the compound system and subsystem temperatures after coupling are shown in Fig.~\ref{fig:two_bounded_dos} for the various entropy definitions. For total energies $0 < \SubE{\mcal{AB}} < 3 \Es{}$, i.e. within the admissible range, the compound system temperature is always positive for the case of the Gibbs entropy, and always negative for the case of the complementary Gibbs entropy. For both the  Boltzmann and the Gibbs definition, the compound system temperature is positive for low energies, diverges at $\SubE{\mcal{AB}} = (3/2) \Es{}$, and is negative for larger energies.

\par
 When $\EntB$ or $\EntP$ are used, the mean temperatures of the subsystems never agree with each other or with the compound system temperature. Moreover, the mean subsystem temperature $\lan \SubTempB{i}\ran$  and $\lan \SubTempP{i}\ran$  are only well-defined for energies close to the smallest or largest possible energy. For intermediate total energies, subsystem energies in the range where the Boltzmann temperature $\TempB$ and the alternative temperature~$\TempP$ diverge, have a non-zero probability density, so that  $\lan \SubTempB{i}\ran$  and $\lan \SubTempP{i}\ran$ become ill-defined.

\par
The mean  Gibbs temperatures of the subsystems agree with the  Gibbs temperature of the compound system for low energies. For high energies, however, the subsystem Gibbs temperatures are lower than the compound system temperature. The complementary Gibbs temperature shows the opposite behavior. Only the mean \emph{inverse} Boltzmann temperature shows agreement between the subsystems and the compound system for all energies.

%%%%%%%%%%%%%%%%%%%
\subsubsection{Temperatures and heat flow}
\label{sec:examples:bounded_dos:heat_flow}
%%%%%%%%%%%%%%%%%%%

%================================================
\begin{figure}
\centerline{
\includegraphics[width=1\linewidth]{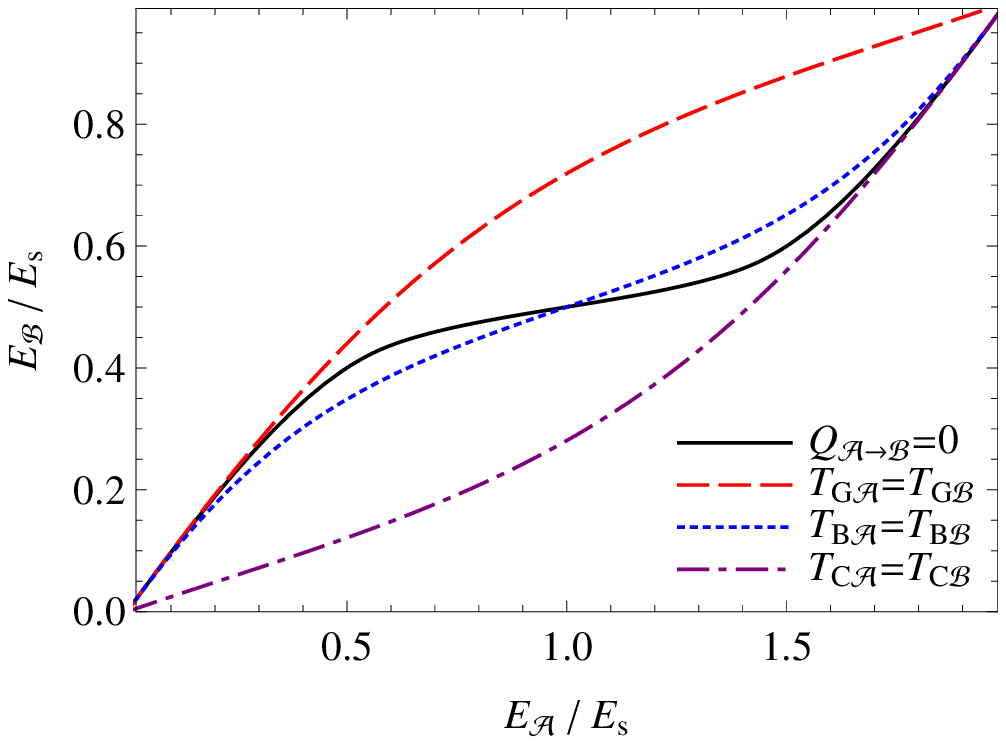}
}
\caption{
\label{fig:two_bounded_dos_heat_and_temperature}
Combinations of initial energies $\SubEinit{\mcal{A}}$ and $\SubEinit{\mcal{B}}$ for two systems $\mcal{A}$ and $\mcal{B}$ with bounded DoS~\eqref{eq:examples:simple_bounded_dos} (with $\Es{\mcal{A}} = 2\Es{}$ and $\Es{\mcal{B}} = 1\Es{}$) for which the heat transfer $\Heat_{\mcal{A}\to\mcal{B}}$ during thermalization vanishes (black solid line). For initial energies below/above the solid line, heat is transferred from system $\mcal{A}$/$\mcal{B}$ to the other system. Also shown are the energies for equal initial Gibbs (red dashed), Boltzmann (blue dotted), or complementary Gibbs (purple dash-dotted line) temperatures (line of equal $\TempP$ is not shown, which are very close to line for equal Boltzmann temperatures). Below/above the lines, system $\mcal{A}$/$\mcal{B}$ is initially hotter than the other system.
}
\end{figure}
%================================================

As Fig.~\ref{fig:two_bounded_dos_heat_and_temperature} shows, for each of the considered temperature definitions, there are combinations of initial energies for which the heat transfer during thermalization does not agree with the naive expectation from the ordering of the initial temperatures. That is, none of these temperatures can be used to correctly predict the direction of heat flow from the ordering of temperatures in all possible cases.

%%%%%%%%%%%%%%%%%%%%%%%%%%%%%%%%%%%%%%%%%%%%%%%%%
\subsection{Classical Hamiltonians with bounded spectrum}
\label{sec:examples:classical_bounded}
%%%%%%%%%%%%%%%%%%%%%%%%%%%%%%%%%%%%%%%%%%%%%%%%%

We discuss two classical non-standard Hamiltonian systems, where the equipartition formula~\eqref{eq:mce:equipartition_theorem}  for the Gibbs temperature holds even for a bounded spectrum with partially negative Boltzmann temperature.

%%%%%%%%%%%%%%%%%%%%%%%%%%%%%%%%%%%%%%%%%%%%%%%%%
\subsubsection{Kinetic energy band}
\label{sec:examples:cos_potential}
%%%%%%%%%%%%%%%%%%%%%%%%%%%%%%%%%%%%%%%%%%%%%%%%%
Consider the simple band Hamiltonian
\be
H(p)=\eps_*[1-\cos(p/p_*)]
\ee
where  $\eps_*>0$ is an energy scale, $p_*>0$ a momentum scale, and the momentum coordinate is restricted the first Brillouin zone $p/p_*\in [-\pi,\pi]$.
Adopting units $\eps_*=1$ and $p_*=1$, the DoS is given by
\bse
\be
\DoS(E)%=\f{2}{|\sin[\arccos(1-E)] |} 
=\f{2}{\sqrt{(E-2)E}},
\ee
and the integrated DoS by
\be
\IDoS(E)
%=4\arcsin(\sqrt{E/2})
=2\arccos(1-E)
\ee
\ese
where $E\in[0,2]$. Noting that $p(\p H/\p p) =p\sin p>0$ and that there are exactly two possible momentum values per energy $p(E)=\pm \arccos(1-E)$, one finds in units $\kB=1$
\be
\left\lan p\f{\p H}{\p p}\right\ran_E= \f{\IDoS}{\DoS}=\TempG.
\ee

%%%%%%%%%%%%%%%%%%%%%%%%%%%%%%%%%%%%%%%%%%%%%%%%%
\subsubsection{Anharmonic oscillator}
\label{sec:examples:anharmonic_oscillator}
%%%%%%%%%%%%%%%%%%%%%%%%%%%%%%%%%%%%%%%%%%%%%%%%%

%================================================
\begin{figure}
\centerline{
\includegraphics[width=1\linewidth]{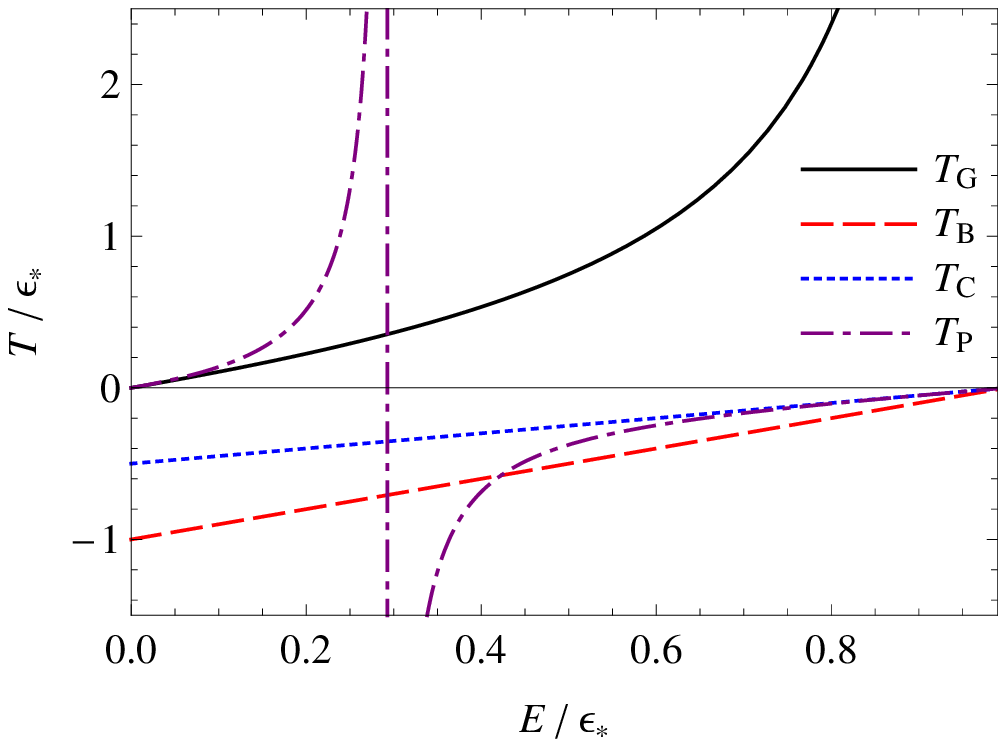}
}
\caption{
\label{fig:anharmonic_oscillator_temperatures}
Gibbs temperature~$\TempG$ (black solid), Boltzmann temperature~$\TempB$ (red dashed), complementary Gibbs temperature~$\TempC$ (blue dotted), and temperature~$\TempP$ (purple dash-dotted line) for the anharmonic oscillator with Hamiltonian~\eqref{eq:oscillator_hamiltonian}.}
\end{figure}
%================================================

Another simple ergodic system with bounded energy spectrum is an anharmonic oscillator described by the Hamiltonian 
\begin{equation}
\label{eq:oscillator_hamiltonian}
H(p,q) = \eps_*\left[1 - \sqrt{1-\left(p/p_*\right)^2-\left(q/q_*\right)^2}\right],
\end{equation}
where the parameters $\eps_*,p_*,q_*>0$ define the characteristic energy, momentum and length scales, 
$p \in[-p_*,p_*]$ is the oscillator momentum, and $q \in[-q_*,q_*]$ the position of the oscillator.
The energy of this system is bounded by $0$ and $\eps_*$. In the low energy limit, corresponding to initial conditions $(p_0,q_0)$ such that~\mbox{$(p/p_*)^2+(q_0/q_*)^2\ll 1$}, the Hamiltonian dynamics reduces to that of an ordinary harmonic oscillator. Fixing mass, length and time units such that~\mbox{$\eps_*=p_*=q_*=1$}, the Hamiltonian takes the simple dimensionless form 
\bse
\begin{equation}
H(p,q) = 1 - \sqrt{1-p^2-q^2}=E,
\end{equation}
with $|p| \leq 1$ and $|q| \leq 1$, and the Hamilton equations of motion read 
\be
\f{dq}{dt}=\f{p}{\sqrt{1-p^2-q^2}},
\qquad
\f{dp}{dt}=\f{-q}{\sqrt{1-p^2-q^2}}.
\ee 
These can be rewritten as
\be
\f{dq}{dt}=\f{p}{1-E},
\qquad
\f{dp}{dt}=\f{-q}{1-E},
\ee
\ese 
yielding the solution
\bse\label{eq:oscillator_solutions}
\be
q(t)=q_0 \cos \left(\frac{t}{E-1}\right) - p_0 \sin \left(\frac{t}{E-1}\right), \\
p(t)=p_0 \cos \left(\frac{t}{E-1}\right)+{q_0} \sin \left(\frac{t}{E-1}\right).
\ee
\ese
Using two-dimensional spherical coordinates, one finds the integrated DoS
\bse
\be
\IDoS(E)=
\begin{cases}
0, & E<0 \\
\pi(2E-E^2), &0\le E\le 1 \\
\pi, &E>1,
\end{cases}
\ee
corresponding to $\IDoSinf{}=\pi$. The DoS reads
\be
\DoS(E)=
\begin{cases}
2\pi(1-E), &0\le E\le 1\\
0, &\textrm{otherwise}.
\end{cases}
\ee
and its derivative 
\be
\nu(E)=
\begin{cases}
-2\pi&0\le E\le 1\\
0, &\textrm{otherwise}.
\end{cases}
\ee
\ese
This gives the temperatures (Fig.~\ref{fig:anharmonic_oscillator_temperatures}) 
\bse
\be
\TempG&=&
\frac{E(E-2)}{2 (E-1)},
\\
\TempB&=&E-1,
\\
\TempC&=&
\frac{E-1}{2},
\\
\TempP&=&
\frac{E(E-2) (E-1)}{4 (E-2) E+2}.
\ee
\ese

As illustrated in Fig.~\ref{fig:anharmonic_oscillator_temperatures}, only the Gibbs temperature~$\TempG$ is always positive in the permissible energy range $0<E<1$ and satisfies the equipartition theorem
\be\label{eq:anharm_equi}
\left\lan q\f{\p H}{\p q}\right\ran_E=\left\lan p\f{\p H}{\p p}\right\ran_E= \TempG.
\ee
Note that, due to the trivial ergodicity of the oscillator, one can replace the microcanonical averages in Eq.~\eqref{eq:anharm_equi} by time-averages, which can be computed directly from the time-dependent solutions~\eqref{eq:oscillator_solutions}.

%================================================
\begin{table*}[t]
\center
\begin{tabular}{l | l |c| c |c |c }
\hline
\hline 
Entropy              & $S(E)$                                                          & equipartition & zeroth law                   & first law                         & second law                  \\
                     &                                                                 &               & Eq.~\eqref{eq:zeroth_law_ii} & Eq.~\eqref{eq:consistency_general} & Eq.~\eqref{eq:second_law} \\
\hline 
Gibbs                & $\ln [\IDoS(E)]$                                                & +             & +                            & +                                  & +                               \\
Complementary Gibbs  & $\ln [\IDoSinf{} - \IDoS(E) \bigr]$                             & --            & --                           & +                                  & +                               \\
Alternative (Penrose)& $\ln [\IDoS(E)] + \ln [\IDoSinf{} - \IDoS(E) ] - \ln\IDoSinf{}$ & --            & --                           & +                                  & +                               \\
Modified Boltzmann   & $\ln [\IDoS(E + \epsilon) - \IDoS(E) ]$                         & --            & --                           & --                                 & --                               \\
Boltzmann            & $\ln [\epsilon \DoS(E) ]$                                       & --            & --                           & --                                 & --                              \\
\hline
\end{tabular}
\caption{
\label{sec:summary:tab:comparison}
The different microcanonical entropy candidates from Sec.~\ref{sec:mce:entropy_candidates} and whether they satisfy the laws of thermodynamics. These results hold under the specifications given in the text. The equipartition statement refers to classical Hamiltonian systems with sufficiently regular phase space topology (see Sec.~\ref{sec:mce:Gibbs_entropy_and_temperature}). The validation of the zeroth law considers part (Z1), and assumes that the DoS of the considered subsystems is strictly positive (but not necessarily monotonic) from $0$ up to the total system energy $E$ (see Sec.~\ref{sec:zeroth_law}). The validation of the first law is based on the consistency relations~\eqref{eq:consistency_general} (see Sec.~\ref{sec:first_law}). The validation of the second law merely assumes a non-negative DoS for energies larger than the groundstate energy $E=0$ (see Sec.~\ref{sec:second_law}).
Additional notes regarding the zeroth law: When the subsystem DoS is not strictly positive for all positive subsystem energies, the Gibbs temperature $\TempG$ may fail the zeroth law for large total energies.
For systems with upper energy bound, the complementary  Gibbs temperature $\TempC$ may satisfy the zeroth law instead for energies close to the maximal energy.
Under suitable conditions, the inverse Boltzmann temperature $1/\TempB{}$ satisfies the \lq{}inverse\rq{} zeroth law~\eqref{eq:zeroth_law_ii_inverse_Boltzmann}.
}
\end{table*}
%================================================

%%%%%%%%%%%%%%%%%%%%%%%%%%%%%%%%%%%%%%%%%%%%%%%%%
\section{Discrete spectra}
\label{sec:extensions:discrete}
%%%%%%%%%%%%%%%%%%%%%%%%%%%%%%%%%%%%%%%%%%%%%%%%%

Having focussed exclusively on exact statements up to this point, we would still like to address certain more speculative issues that, in our opinion, deserve future study. The above derivations relied on the technical assumption that the integrated DoS $\Omega$ is continuous and piecewise differentiable. This condition is satisfied for systems that exhibit a continuous spectrum, including the majority of classical Hamiltonians.  Interestingly, however, the analysis of simple quantum models suggests that at least some of the above results extend to discrete spectra~\cite{2014DuHi_NatPhys} -- if one considers analytic continuations of the discrete level counting function  (DLCF) $\DLCF(E_n)$. The DLCF is the discrete counterpart of the integrated DoS but is \textit{a priori} only defined on the discrete set of eigenvalues $\{E_n\}$. We briefly illustrate the heuristic procedure for three basic examples.

\par
 For the quantum harmonic oscillator with spectrum $E_n=\hbar\varpi(n+1/2)$, $n=0,1,\ldots$, the DLCF $\DLCF(E_n)$ is obtained by inverting the spectral formula, which yields $\DLCF(E_n)=1+n=(1/2)+ E_n/(\hbar\varpi)$ for all $n=0,1,\ldots$, and has the analytic continuation $\IDoS(E)=(1/2)+ E/(\hbar\varpi)$, now defined for all $E\in\R$ with $E\geq (1/2) \hbar \varpi$. From the associated Gibbs entropy $\EntG=\ln \IDoS$, one can compute the heat capacity $\HeatCG(E) = (\p \TempG(E)/\p E)^{-1}=\kB$, which agrees with the heat capacity of a classical  oscillator.

\par
By the same procedure, one can analyze the thermodynamic properties of a hydrogen atom with discrete spectrum $E_n= - \ER/n^2<0$,  where $n=1,2\ldots$ and $\ER$ the Rydberg energy. Inversion of the spectral formula and accounting for the degeneracy per level, $g_n=n^2$, combined with analytic continuation yields $\IDoS(E)=f(\sqrt{-\ER/E})$ with \mbox{$f(n)=(1/6)n(1+n)(1+2n)$} and \mbox{$E\in [-\ER, 0)$}. One can then compute the energy-dependent heat capacity $\HeatCG$ in a straightforward manner and finds that $\HeatCG(E)$  is negative, as expected for attractive $1/r$-potentials\footnote{Intuitively, as the energy is lowered through the release of photons (heat), the kinetic energy of the electron increases. This is analogous to a planet moving in the gravitational field of a star.}, approaching $\HeatCG(E) = -(169/173)\kB$ for $E\to -\ER$ and $\HeatCG(E) = -(3/2)\kB$ for $E\to 0$.

\par
As the third and final example, consider a quantum particle in a one-dimensional box potential of width~$L$, with spectrum $E_n= (\hbar\pi)^2 n^2/(2mL^2)$, where $n=1,2\ldots$ In this case, the analytical continuation $\IDoS(E)=(2mE)^{1/2}L/(\hbar\pi)$  yields the heat capacity $\HeatCG(E)=\kB/2$. More interestingly, the Gibbs prediction for the pressure, $\PressG(E)=\TempG (\p \EntG/\p L)=2E_n/L$, is in perfect agreement with the mechanical pressure obtained from the spectrum $\Press(E_n)= - \p E_n /\p L$. Thus, for all three  examples, one obtains reasonable predictions for the heat capacities\footnote{It is easy to check that the Boltzmann entropy fails to produce reasonable results in all three cases.} and, for the box potential, even the correct pressure law. We find this remarkable, given that the analytic continuation covers \lq quasi-energies\rq{} that are not part of the conventional spectrum.

\par
It is sometimes argued that thermodynamics should only be studied for infinite systems. We consider such a dogmatic view artificially restrictive, since the very active and successful field of finite-system thermodynamics has contributed substantially to our understanding of physical computation limits~\cite{1961Landauer,2012Berut_Nature,2014Roldan_NPhys}, macromolecules~\cite{2001Hummer,PhysRevLett.97.218103,2008Junghans_JChemPhys,PhysRevE.76.046110} and thermodynamic concepts in general~\cite{PhysRevLett.96.070603,2001MuStBo,2001BoMuHa,2006HiDu,2006Cleuren_PRL,2009CaTaHa_PRL,Campisi2011RvMP,2012Campisi_PRL,Campisi13JPCB117,2013Chetrite_PRL,2014Gelbwaser_PRE,2008Deffner,2000Kastner,Talkner2008,Talkner2013} over the last decades.  A scientifically more fruitful approach might be to explore \emph{when} and \emph{why} the analytic continuation method\footnote{This approach can, in principle, be applied to any discrete spectrum although it will in general be difficult to provide explicit analytical formulas for $\IDoS(E)$.}  produces reasonable predictions for heat capacities, pressure, and similar thermodynamic quantities.\footnote{A possible explanation may be that, if one wants to preserve the continuous differential structure of thermodynamics, the analytical continuation simply presents, in a certain sense, the most natural continuous approximation to discrete finite energy differences. In this case, the continuation method does not yield fundamentally new insights but can still be of practical use as a parameter-free approximation technique to estimate heat capacities and other relevant quantities.}

%%%%%%%%%%%%%%%%%%%%%%%%%%%%%%%%%%%%%%%%%%%%%%%%%
%\newpage 
\section{Conclusions}
\label{sec:summary}
%%%%%%%%%%%%%%%%%%%%%%%%%%%%%%%%%%%%%%%%%%%%%%%%%

We have presented a detailed comparison of the most frequently encountered 
microcanonical entropy candidates. After reviewing the various entropy definitions, we first showed that, regardless of which definition is chosen, the microcanonical temperature of an isolated system can be a non-monotonic, oscillating function of the energy (Sec.~\ref{sec:mce:non-unique_T}). This fact implies that, contrary to claims in the recent literature~\cite{2014VilarRubi,2014FrenkelWarren,2014Schneider_Comment}, naive temperature-based heat-flow arguments cannot be used to judge the various entropy candidates. Any objective evaluation should be based on whether or not a given thermostatistical entropy definition is compatible with the laws of thermodynamics.

\par
Focussing on exact results that hold for a broad class of densities of states, we found that only the Gibbs entropy simultaneously satisfies the zeroth, first and second law of thermodynamics (as well as the classical equipartition theorem). If one accepts the results in Table~\ref{sec:summary:tab:comparison} as mathematically correct facts, then there remains little choice but to conclude that the thermodynamic characterization of an isolated systems should build on the Gibbs entropy, implying a strictly non-negative absolute temperature and Carnot efficiencies not larger than 1 for system with or without upper energy bounds.

\par
It is sometimes argued~\cite{2014Schneider_Comment} that, if the spectrum of a given Hamiltonian is symmetric under an energy reflection $E \to - E$, the thermodynamic entropy must exhibit the symmetry $S(E) \to S(-E)$.  However, the fact that certain (artificially truncated) Hamiltonians possess symmetric spectra does \emph{not} imply that low energy states and high energy states are physically or thermodynamically equivalent. Within conventional thermodynamics the internal energy is a \emph{directed} (ordered) quantity:  The assumption that a higher energy state is, in principle, physically distinguishable from a lower energy state is a basic axiom of thermodynamics -- without this empirically supported fact it would not be possible to distinguish whether a thermodynamic cycle produces work or absorbs work.  Even for \lq truly\rq{} symmetric Hamiltonians, states with $E$ and $-E$ are not thermodynamically equivalent: A system occupying the groundstate can be easily heated (e.g., by injecting a photon) whereas it is impossible to add a photon to a system in the highest energy state. The Gibbs entropy reflects this asymmetry in a natural manner.

\par
Arguments in favor of the Boltzmann entropy, such as the symmetry argument above, are sometimes motivated by the idea that information entropy and thermodynamic entropy must be equivalent. It is conceptually pleasing if, in some cases, the thermodynamic entropy of a certain ensemble can be identified with one of the numerous information entropies~\cite{1960Renyi,1978Wehrl,1991Wehrl} -- but there exists no fundamental reason why a physical quantity should always coincide with a particular abstract information measure\footnote{In fact, one could try to use empirically established physical laws to constrain information measures, rather than the other way round.}. If, however, one desired such a formal connection, one could equally well add the Gibbs entropy to the long list~\cite{1960Renyi,1978Wehrl,1991Wehrl} of information measures.

\par
Some earlier~\cite{1991Montgomery} objections against the Gibbs entropy~\cite{2014FrenkelWarren,2014Schneider_Comment} purport that the Gibbs temperature does not adequately capture relevant statistical details of large population-inverted systems -- and that it should therefore be replaced by the Boltzmann entropy. We consider arguments of this type inappropriate and misleading, for they intermingle two questions that should be treated separately. The first question relates to whether a certain entropy definition is thermodynamically consistent. This problem is well-posed and can be answered unambiguously, as shown in Table~\ref{sec:summary:tab:comparison},  for the majority of physically relevant systems. The second question is whether there exist other types of characterizations of many-particle systems that add to the partial information encoded in the thermodynamic variables. Clearly, even if the popular Boltzmann entropy is in conflict with the thermodynamic laws for practically all finite systems, this quantity still encodes valuable information about the statistical properties of certain physical systems. In particular, the Boltzmann temperature can provide a useful effective description of spin systems, lasers and other population-inverted systems. But one should be cautious before inserting the Boltzmann temperature (or any other effective temperature) into the Carnot efficiency formula or thermodynamic equations of state - especially if the resulting formulas would suggest the possibility of perpetual motion.

%%%%%%%%%%%%%%%%%%%%%%%%%%%%%%%%%%%%%%%%%%%%%%%%%
\section{Outlook}
\label{sec:outlook}
%%%%%%%%%%%%%%%%%%%%%%%%%%%%%%%%%%%%%%%%%%%%%%%%%

The exact results summarized in this paper apply to systems with a continuous energy spectrum. In Sec.~\ref{sec:extensions:discrete}, we discussed analytic continuation of the discrete level counting function as a possible way to extend the results to systems with discrete spectra. Future studies may reveal when and why the analytic continuation method
yields reasonable predictions for the thermodynamic properties of such systems.

\par
As discussed in Sec.~\ref{sec:zeroth_law}, for classical systems, the validity of the extension (Z1) of the zeroth law for the subsystem Gibbs temperatures is closely related the the validity of the equipartition theorem. In its simplest form, the equipartition theorem is valid for classical systems with a simple phase space $\R^N$ and standard Hamiltonians~\cite[][]{Khinchin}. The examples in Sec.~\ref{sec:examples:classical_bounded} show that equipartition may hold even for certain system with a more complex phase space. This raises the question which conditions exactly have to be imposed on phase space geometry and Hamiltonians for equipartition to hold (completely or least partially) or fail in classical systems.

\par
The consistency conditions arising from the first law (Sec.~\ref{sec:first_law}) require that the microcanonical entropy must be of the form $S(E, Z)=f\bigl[\IDoS(E,Z)\bigr]$, where $\IDoS(E,Z)$ is the integrated density of states and $f$ a differentiable function. \highlightchange{Demanding additivity of $S$ under factorization of $\IDoS(E,Z)$ or, alternatively, that the related temperature is consistent with the value measured by a gas thermometer, fixes $f=\ln$ and thereby singles out the Gibbs entropy. If one abandons the operational connection with the gas thermometer then, in addition to the Gibbs entropy, one obtains a large class of possible entropy candidates $S(E, Z)=f\bigl[\IDoS(E,Z)\bigr]$  that are consistent with the first law. Here, we only briefly studied two such alternatives and one could construct other interesting entropy candidates, study their properties (in particular whether they satisfy the zeroth and second law), and possibly devise a corresponding thermometer.
}
\par
The example in Sec.~\ref{sec:examples:bounded_dos} illustrates that all of the considered temperature definitions may fail to predict the direction of heat flow when two previously isolated systems are brought into thermal contact to form a combined isolated system. The discussion in Sec.~\ref{sec:mce:non-unique_T} suggests that it will be difficult to identify a microcanonical temperature that correctly predicts heat flow directions between arbitrary microcanonical systems. It remains, however, an open question whether or not one could in principle define such a temperature.

\par
Finally, the main assumption adopted here was that thermostatistical properties of an isolated system are well described by the standard microcanonical density operator~\eqref{eq:mce:pdf_mce}. With regard to astrophysical applications, it may be worthwhile to study how the above results can be adapted to isolated systems that possess additional integrals of motion~\cite{2002Gross_Full} or exhibit otherwise limited mixing dynamics that prevents them from reaching full thermodynamic equilibrium.

\begin{acknowledgments}
We thank D.~Frenkel, O.~Penrose, U.~Schneider, J.-S. Wang, and P.~Warren for helpful discussions, and we are very grateful to T. Ensslin, K. Sakmann and P. Talkner for thoughtful comments.
\end{acknowledgments}

%%%%%%%%%%%%%%%%%%%%%%%%%%%%%%%%%%%%%%%%%%%%%%%%%%%%%%%
\newpage

%%%%%%%%%%%%%%%%%%%%%%%%%%%%%%%%%%%%%%%%%%%%%%%%%%%%%%%
\appendix
\newpage

%%%%%%%%%%%%%%%%%%%%%%%%%%%%%%%%%%%%%%%%%%%%%%%%%%%%%%%

\begin{widetext}

\section{Compound systems}
\label{app:sec:compound_system_density}

Here, we present brief derivations of certain equations for compound systems. We assume that all energies,  densities, etc. for the compound system $\mcal{AB}$ and its subsystems $\mcal{A}$ and $\mcal{B}$ are defined as stated in the main text (see Sec.~\ref{sec:zeroth_law}). Then, Eq.~\eqref{eq:convolution_dos} can be derived as follows: 
\begin{equation}
\label{a-e:convolution_dos}
\begin{split}
\DoS(E)
&=
  \Tr\left[ \DiracDelta(E - H) \right]
\\&=
  \Tr_{\mcal{A}}
	\left\{
	\Tr_{\mcal{B}}
	\left[
	\DiracDelta(E - \SubH{\mcal{A}} - \SubH{\mcal{B}}) 
	\right]
	\right\}
\\&=
	\Tr_{\mcal{A}}
	\left\{	
	\Tr_{\mcal{B}}
	\left[
  \int_{-\infty}^{\infty}\idiff{\SubE{\mcal{A}}'}\,\DiracDelta(\SubE{\mcal{A}}' - \SubH{\mcal{A}})
  \int_{-\infty}^{\infty}\idiff{\SubE{\mcal{B}}'}\,\DiracDelta(\SubE{\mcal{B}}' - \SubH{\mcal{B}})
	\DiracDelta(E - \SubH{\mcal{A}} - \SubH{\mcal{B}})
	\right]
	\right\}
\\&=
	\Tr_{\mcal{A}}
	\left\{	
	\Tr_{\mcal{B}}
	\left[
  \int_{-\infty}^{\infty}\idiff{\SubE{\mcal{A}}'}\,\DiracDelta(\SubE{\mcal{A}}' - \SubH{\mcal{A}})
  \int_{-\infty}^{\infty}\idiff{\SubE{\mcal{B}}'}\,\DiracDelta(\SubE{\mcal{B}}' - \SubH{\mcal{B}})
	\DiracDelta(E - \SubE{\mcal{A}}' - \SubE{\mcal{B}}')
	\right]
	\right\}
\\&=
  \int_{-\infty}^{\infty}\idiff{\SubE{\mcal{A}}'}\, \Tr_{\mcal{A}}\left[\DiracDelta(\SubE{\mcal{A}}' - \SubH{\mcal{A}}) \right]
  \int_{-\infty}^{\infty}\idiff{\SubE{\mcal{B}}'}\,	\Tr_{\mcal{B}}\left[\DiracDelta(\SubE{\mcal{B}}' - \SubH{\mcal{B}}) \right]
	\DiracDelta(E - \SubE{\mcal{A}}' - \SubE{\mcal{B}}')
\\&=
  \int_{-\infty}^{\infty}\idiff{\SubE{\mcal{A}}'}\, \SubDoS{\mcal{A}}(\SubE{\mcal{A}}')
  \int_{-\infty}^{\infty}\idiff{\SubE{\mcal{B}}'}\,	\SubDoS{\mcal{B}}(\SubE{\mcal{B}}')
	\DiracDelta(E - \SubE{\mcal{A}}' - \SubE{\mcal{B}}')
\\&=
  \int_{0}^{\infty}\idiff{\SubE{\mcal{A}}'} 
  \int_{0}^{\infty}\idiff{\SubE{\mcal{B}}'}\,
  \SubDoS{\mcal{A}}(\SubE{\mcal{A}}') \SubDoS{\mcal{B}}(\SubE{\mcal{B}}')
	\DiracDelta(E - \SubE{\mcal{A}}' - \SubE{\mcal{B}}')
\\&=
  \int_{0}^{E}\idiff{\SubE{\mcal{A}}'}\,
	\SubDoS{\mcal{A}}(\SubE{\mcal{A}}')
  \SubDoS{\mcal{B}}(E - \SubE{\mcal{A}}')
.
\end{split}
\end{equation}
Here, we exploited that $\SubDoS{i}(\SubE{i}) = 0$ for $\SubE{i} < 0$ to restrict the integral boundaries in some of the latter steps.

A very similar calculation leads to the first expression in Eq.~\eqref{eq:convolution_idos} for the integrated DoS:
\begin{equation}
\label{a-e:convolution_phase_volume}
\begin{split}
\IDoS(E)
&=
  \Tr\left[ \Heaviside(E - H) \right]
\\&=
  \int_{0}^{\infty}\idiff{\SubE{\mcal{A}}'} 
  \int_{0}^{\infty}\idiff{\SubE{\mcal{B}}'}\,
  \SubDoS{\mcal{A}}(\SubE{\mcal{A}}') \SubDoS{\mcal{B}}(\SubE{\mcal{B}}')
	\Heaviside(E - \SubE{\mcal{A}}' - \SubE{\mcal{B}}')
.
\end{split}
\end{equation}
To obtain the alternative expression for $\IDoS(E)$ with only one integral requires a few simple additional steps:
\begin{equation}
\begin{split}
\IDoS(E)
&=
  \int_{0}^{\infty}\idiff{\SubE{\mcal{B}}'}
	\left[
  \int_{0}^{\infty}\idiff{\SubE{\mcal{A}}'} 
	\SubDoS{\mcal{A}}(\SubE{\mcal{A}}') 
	\Heaviside(E - \SubE{\mcal{B}}' - \SubE{\mcal{A}}')
	\right]
  \SubDoS{\mcal{B}}(\SubE{\mcal{B}}')
\\&=
  \int_{0}^{\infty}\idiff{\SubE{\mcal{B}}'}\,
	\SubIDoS{\mcal{A}}(E - \SubE{\mcal{B}}')
  \SubDoS{\mcal{B}}(\SubE{\mcal{B}}')
\\&=
  \int_{0}^{E}\idiff{\SubE{\mcal{A}}'}\,
	\SubIDoS{\mcal{A}}(\SubE{\mcal{A}}')
  \SubDoS{\mcal{B}}(E - \SubE{\mcal{A}}')
.
\end{split}
\end{equation}

To derive the expression~\eqref{eq:convolution_ddos} for derivative of the DoS, we start by applying the general definition~\eqref{eq:mce:ddos_mce} to the second expression for the DoS of the compound system in Eq.~\eqref{eq:convolution_dos}. Assuming that $\SubDoS{\mcal{A}}(\SubE{\mcal{A}}')$ is sufficiently well-behaved near $\SubE{\mcal{A}}' = 0$, we obtain:
\begin{equation}
\label{a-e:convolution_ddos}
\begin{split}
\DDoS(E)
&=
 %\totder{\DoS(E)}{E} 
%\\&=
\totder{}{E}
  \int_{0}^{E}\idiff{\SubE{\mcal{A}}'}\,
	\SubDoS{\mcal{A}}(\SubE{\mcal{A}}')
  \SubDoS{\mcal{B}}(E - \SubE{\mcal{A}}')
\\&=
\totder{}{E}
  \int_{0}^{E}\idiff{\SubE{\mcal{B}}'}\,
	\SubDoS{\mcal{A}}(E - \SubE{\mcal{B}}')
  \SubDoS{\mcal{B}}(\SubE{\mcal{B}}')
\\&=
	\SubDoS{\mcal{A}}(0^+)
  \SubDoS{\mcal{B}}(E)
	+
  \int_{0}^{E}\idiff{\SubE{\mcal{B}}'}\,
	\SubDDoS{\mcal{A}}(E - \SubE{\mcal{B}}')
  \SubDoS{\mcal{B}}(\SubE{\mcal{B}}')
\\&=
  \int_{0}^{E}\idiff{\SubE{\mcal{A}}'}\,
	\SubDDoS{\mcal{A}}(\SubE{\mcal{A}}')
  \SubDoS{\mcal{B}}(E - \SubE{\mcal{A}}')
	+
	\SubDoS{\mcal{A}}(0^+)
  \SubDoS{\mcal{B}}(E)
	.
\end{split}
\end{equation}

Using the same approach as in the derivation~\eqref{a-e:convolution_dos}, expression~\eqref{eq:energy_dist_sub_A} for the probability density~\eqref{eq:energy_dist_sub} of the energy for subsystem $\mcal{A}$ can be derived as follows:
\begin{equation}
\label{a-e:pdfE_A}
\begin{split}
\pdfE{\mcal{A}}(\SubE{\mcal{A}}'|E)
&=
  \Tr\left[\DOp{}\,\DiracDelta(\SubE{\mcal{A}}' - \SubH{\mcal{A}}) \right]
\\&=
  \Tr\left[ \frac{\DiracDelta(E - \SubH{\mcal{A}} - \SubH{\mcal{B}})}{\DoS(E)} \DiracDelta(\SubE{\mcal{A}}' - \SubH{\mcal{A}})  \right]
\\&=
  \int_{-\infty}^{\infty}\idiff{\SubE{\mcal{A}}''}\, \SubDoS{\mcal{A}}(\SubE{\mcal{A}}'')
  \int_{-\infty}^{\infty}\idiff{\SubE{\mcal{B}}''}\,	\SubDoS{\mcal{B}}(\SubE{\mcal{B}}'')
	\frac{\DiracDelta(E - \SubE{\mcal{A}}'' - \SubE{\mcal{B}}'')}{\DoS(E)} \DiracDelta(\SubE{\mcal{A}}' - \SubE{\mcal{A}}'')
\\&=
  \frac{\SubDoS{\mcal{A}}(\SubE{\mcal{A}}')\;\SubDoS{\mcal{B}}(E - \SubE{\mcal{A}}')}{\DoS(E)}
.
\end{split}
\end{equation}

Equation~\eqref{eq:mean_for_functions_of_subsystem_energy} can be verified by observing:
\begin{equation}
\label{a-e:ev_F_H_i}
\begin{split}
  \EV{F(\SubH{i})}_E
&=
  \Tr\left[\DOp{}\, F(\SubH{i}) \right]
\\&=
  \int_{-\infty}^{\infty}\idiff{\SubE{i}'}\,
	  \Tr\left[\DOp{}\, \DiracDelta(\SubE{i}' - \SubH{i}) F(\SubH{i}) \right]
\\&=
  \int_{-\infty}^{\infty}\idiff{\SubE{i}'}\,
	  \Tr\left[\DOp{}\, \DiracDelta(\SubE{i}' - \SubH{i}) \right] F(\SubE{i})
\\&=
  \int_{0}^{E}\idiff{\SubE{i}'}\,
	\pdfE{i}(\SubE{i}'|E)  F(\SubE{i})
.
\end{split}
\end{equation}

%%%%%%%%%%%%%%%%%%%%%%%%%%%%
\section{Zeroth law and equipartition}
\label{app:sec:zero_equi}
%%%%%%%%%%%%%%%%%%%%%%%%%%%%

We assume $H(\bs \xi)=H_{\mcal{A}}(\bs \xi_\mcal{A})+H_{\mcal{B}}(\bs \xi_\mcal{B})$. To prove Eq.~\eqref{eq:zeroth_law_equi}, we use the same approach as in the derivation~\eqref{a-e:convolution_dos}:
\bse
\be
\TempG(E)
&=& \EV{\pq_i \f{\p H_\mcal{A}}{\p \pq_i}}_{E}
\notag\\ 
&=& \Tr \left[\left(\pq_i \f{\p H_\mcal{A}}{\p \pq_i}\right)\; \f{\gd(E -  H_\mcal{A}- H_\mcal{B})}{\go(E)}\right]
\notag\\ 
&=&\f{1}{\go(E)} \Tr_\mcal{A}\Tr_\mcal{B} \left[\left(\pq_i \f{\p H_\mcal{A}}{\p \pq_i}\right)\;  \gd(E -  H_\mcal{A}- H_\mcal{B})\right]
\notag\\ 
&=&
\f{1}{\go(E)} 
  \int_{-\infty}^{\infty}\idiff{\SubE{\mcal{A}}'}\, \Tr_{\mcal{A}}\left[\DiracDelta(\SubE{\mcal{A}}' - \SubH{\mcal{A}}) \left(\pq_i \f{\p H_\mcal{A}}{\p \pq_i}\right)\;  \right]
  \int_{-\infty}^{\infty}\idiff{\SubE{\mcal{B}}'}\,	\Tr_{\mcal{B}}\left[\DiracDelta(\SubE{\mcal{B}}' - \SubH{\mcal{B}}) \right]
	\DiracDelta(E - \SubE{\mcal{A}}' - \SubE{\mcal{B}}')
\notag\\ 
&=&
  \int_{-\infty}^{\infty}\idiff{\SubE{\mcal{A}}'}\, \Tr_{\mcal{A}}\left[\DiracDelta(\SubE{\mcal{A}}' - \SubH{\mcal{A}}) \left(\pq_i \f{\p H_\mcal{A}}{\p \pq_i}\right)\;  \right]
  \f{1}{\go(E)} 
\Tr_{\mcal{B}}	\left[\DiracDelta(E - \SubE{\mcal{A}}' - \SubH{\mcal{B}}) \right]
\notag\\ 
&=&
  \int_{-\infty}^{\infty}\idiff{\SubE{\mcal{A}}'}\, \Tr_{\mcal{A}}\left[\DiracDelta(\SubE{\mcal{A}}' - \SubH{\mcal{A}})
   \left(\pq_i \f{\p H_\mcal{A}}{\p \pq_i}\right)\;  \right]
  \f{\go_{\mcal{B}}(E - \SubE{\mcal{A}}' )}{\go(E)} 
\notag\\ 
&=&
  \int_{-\infty}^{\infty}\idiff{\SubE{\mcal{A}}'}\,   \f{\go(\SubE{\mcal{A}}')\;\go_{\mcal{B}}(E - \SubE{\mcal{A}}' ) }{\go(E)} \;\Tr_{\mcal{A}}\left[\f{\DiracDelta(\SubE{\mcal{A}}' - \SubH{\mcal{A}})  }{\go(\SubE{\mcal{A}}')}
  \left(\pq_i \f{\p H_\mcal{A}}{\p \pq_i}\right)\;  \right]
\notag\\ 
&=&
  \int_{-\infty}^{\infty}\idiff{\SubE{\mcal{A}}'}\,   \pdfE{\mcal{A}}(\SubE{\mcal{A}}'|E)\;\EV{\pq_i \f{\p H_\mcal{A}}{\p \pq_i}}_{H_\mcal{A}=\SubE{\mcal{A}}'}.
  \ee
\ese

%%%%%%%%%%%%%%%%%%%%%%%%%%%%
\pagebreak
\section{Second law}
\label{app:sec:second}
%%%%%%%%%%%%%%%%%%%%%%%%%%%%

\paragraph*{Complementary Gibbs entropy.} 
To verify that $\EntC=\ln [\Omega_\infty -\Omega(E)]$ complies with Planck's second law~\eqref{eq:second_law}, we need to show that
\be
\label{app:sec:second_compG}
\Omega_\infty -\Omega(E_\mcal{A}+ E_\mcal{B}) 
\ge
\left[\Omega_{\mcal{A}\infty} -\Omega_\mcal{A}(\SubE{\mcal{A}}) \right] 
\left[\Omega_{\mcal{B}\infty} -\Omega_\mcal{B}(\SubE{\mcal{B}}) \right] 
.
\ee
%or, equivalently,
%\be
%\Omega_\infty -\Omega(E_\mcal{A}+ E_\mcal{B}) 
%\ge
%\Omega_{\mcal{A}\infty} \Omega_{\mcal{B}\infty} -
%\Omega_\mcal{A}(\SubE{\mcal{A}})  \Omega_{\mcal{B}\infty}  -
%\Omega_\mcal{B}(\SubE{\mcal{B}})  \Omega_{\mcal{A}\infty} +
%\Omega_\mcal{A}(\SubE{\mcal{A}})\Omega_\mcal{B}(\SubE{\mcal{B}})
%\ee
Assuming upper spectral  bounds $E_\mcal{A}^+$ and $E_\mcal{B}^+$ for the subsystems, the energy of the compound system $\mcal{AB}$ has an upper bound $E^+=E_\mcal{A}^+ +E_\mcal{B}^+$ and, therefore,
\be
\IDoS_\infty
\;=\;
\IDoS(E^+)
 &=&
\int_0^{E_\mcal{A}^+} \!\!\diff{\SubE{\mcal{A}}'}
\int_0^{E_\mcal{B}^+} \!\!\diff{\SubE{\mcal{B}}'}
\,
\SubDoS{\mcal{A}}(\SubE{\mcal{A}}')\;\SubDoS{\mcal{B}}(\SubE{\mcal{B}}')\;
\Heaviside(\SubEinit{\mcal{A}}^+ +\SubEinit{\mcal{B}}^+ - \SubE{\mcal{A}}' - \SubE{\mcal{B}}')
 \notag\\
 &=&
\int_0^{E_\mcal{A}^+} \!\!\diff{\SubE{\mcal{A}}'}
\int_0^{E_\mcal{B}^+} \!\!\diff{\SubE{\mcal{B}}'}
\,
\SubDoS{\mcal{A}}(\SubE{\mcal{A}}')\;\SubDoS{\mcal{B}}(\SubE{\mcal{B}}')
 \notag\\
 &=&
\Omega_{\mcal{A}\infty} \Omega_{\mcal{B}\infty}
.
\ee
It then remains to show that
\be
\Omega_\mcal{A}(\SubE{\mcal{A}})  \Omega_{\mcal{B}\infty}  +
\Omega_\mcal{B}(\SubE{\mcal{B}})  \Omega_{\mcal{A}\infty} 
\ge
\Omega(E_\mcal{A}+ E_\mcal{B}) + \Omega_\mcal{A}(\SubE{\mcal{A}})\Omega_\mcal{B}(\SubE{\mcal{B}})
.
\ee
Each summand corresponds to an integral over the same function $f(\SubE{\mcal{A}}',\SubE{\mcal{B}}')=\SubDoS{\mcal{A}}(\SubE{\mcal{A}}')\SubDoS{\mcal{B}}(\SubE{\mcal{B}}')\ge 0$.  Using a graphical representation of the underlying integration regions, similar to that in Fig.~\ref{fig:Gibbs_second_law}, one indeed finds that this inequality is always fulfilled.

\paragraph*{Penrose entropy.} 
For $\EntP(E) = \ln \IDoS(E) + \ln [ \IDoS_\infty - \IDoS(E) ] - \ln \IDoS_\infty$ from Eq.~\eqref{sec:mce:eq:df_Penrose_entropy}, we have to verify that
\be
\f{\IDoS(E_\mcal{A}+ E_\mcal{B})\;[\IDoS_\infty - \IDoS(E_\mcal{A}+ E_\mcal{B}) ]}{\IDoS_\infty} 
\ge
\f{\Omega_\mcal{A}(\SubE{\mcal{A}})\;[ \Omega_{\mcal{A}\infty}- \Omega_\mcal{A}(\SubE{\mcal{A}})]}{  \Omega_{\mcal{A}\infty}} 
\f{\Omega_\mcal{B}(\SubE{\mcal{B}})\;[ \Omega_{\mcal{B}\infty} - \Omega_\mcal{B}(\SubE{\mcal{B}}) ]}{ \Omega_{\mcal{B}\infty}} 
.
\ee
Using $\Omega_{\infty}=\Omega_{\mcal{A}\infty} \Omega_{\mcal{B}\infty}$, this simplifies to
\be
\IDoS(E_\mcal{A}+ E_\mcal{B})\;[\IDoS_\infty - \IDoS(E_\mcal{A}+ E_\mcal{B}) ] 
\ge
\Omega_\mcal{A}(\SubE{\mcal{A}})\;\Omega_\mcal{B}(\SubE{\mcal{B}})\;
[ \Omega_{\mcal{A}\infty}- \Omega_\mcal{A}(\SubE{\mcal{A}})]
[ \Omega_{\mcal{B}\infty} - \Omega_\mcal{B}(\SubE{\mcal{B}}) ]
.
\ee
This inequality holds by virtue of Eqs.~\eqref{eq:Gibbs_second_law} and~\eqref{app:sec:second_compG}.

%
%
%\begin{equation}
%\begin{split}
%&
%\IDoS(\SubEinit{\mcal{A}}+\SubEinit{\mcal{B}})
% \\&=
%\int_0^{\infty} \!\!\diff{\SubE{\mcal{A}}'}
%\int_0^{\infty} \!\!\diff{\SubE{\mcal{B}}'}
%\,
%\SubDoS{\mcal{A}}(\SubE{\mcal{A}}')\SubDoS{\mcal{B}}(\SubE{\mcal{B}}')
%\\&\quad\times
%\Heaviside(\SubEinit{\mcal{A}}+\SubEinit{\mcal{B}} - \SubE{\mcal{A}}' - \SubE{\mcal{B}}')
% \\&\geq
%\int_0^{\infty} \!\!\diff{\SubE{\mcal{A}}'}
%\int_0^{\infty} \!\!\diff{\SubE{\mcal{B}}'}
%\,
%\SubDoS{\mcal{A}}(\SubE{\mcal{A}}')\SubDoS{\mcal{B}}(\SubE{\mcal{B}}')
%\\&\quad\times
%\Heaviside(\SubEinit{\mcal{A}} - \SubE{\mcal{A}}')
%\Heaviside(\SubEinit{\mcal{B}} - \SubE{\mcal{B}}')
%\\&=
%\SubIDoS{\mcal{A}} (\SubE{\mcal{A}}) \,\SubIDoS{\mcal{B}} (\SubE{\mcal{B}})
%.
%\end{split}
%\end{equation}

\end{widetext}

%%%%%%%%%%%%%%%%%%%%%%%%%%%%%%%%%%%%%%%%%%%%%%%%%%%%%%%%%%%%%%%%%%%%%%%
\end{document}